%% file: HudsonPhDThesis.tex
\pdfoutput=1
\documentclass[12pt, double, phd]{osudiss-2}

\usepackage{amsmath, amssymb, amsthm,mathtools}
\usepackage{blkarray}
\usepackage{mathdots}
\usepackage{graphicx} 
\usepackage{mwe}
\usepackage{float}
\usepackage{rotating} % for rotating figures/tables
\usepackage{multirow} % for multirow figures
\usepackage{caption}
\usepackage{subcaption}
\usepackage{bm,bbm} % for bold math---useful
\usepackage{booktabs} % for better looking tables
\usepackage{dsfont}
\usepackage{bbm}
\usepackage{bookmark} % bookmarks in pdf file
\hypersetup{colorlinks=true,linkcolor=blue, linktoc=page} 
\usepackage{cite} % for backreferenced citations - supercedes \usepackage[numbers, sort&compress]{natbib} 
\input{Misc/macros.tex}
\title{Resonant Floquet scattering\\ of ultracold atoms}
\author{D. Hudson Smith}
\advisorname{Eric Braaten}
\degree{Doctor of Philosophy} % Default value
\member{Louis DiMauro}
\member{Ilya Gruzberg}
\member{Ulrich Heinz}
\authordegrees{B.S, B.A.}
\graduationyear{2016}
\unit{Graduate Program in Physics\vspace{-1em}} % defaults to ``Graduate Program in Physics''

\begin{document}
\frontmatter

\include{Frontmatter/Hudson_Abs}
\dedication{To my wife and children.}
\include{Frontmatter/Hudson_Ack}
\include{Frontmatter/Hudson_Vita}

\tableofcontents 
\newpage
\listoffigures 
\newpage
\listoftables 

% \clearpage
%\PrintListofAbbreviations{List of Abbreviations}
% \phantomsection
% \printglossary[type=\acronymtype,style=long] 

%print glossary - comment out if you don't want this. Make sure you also add \glsdisablehyper if you don't want to print a glossary, but do use the glossaries package to keep track of acronyms
%\clearpage %remove if you don't want a page break before list of abbreviations
%\PrintListofAbbreviations{List of Abbreviations} %Title is in { } - change if desired

\mainmatter
% \startdoublespace
% \startsinglespace

\include{Chapters/Intro/Intro}
\include{Chapters/Experiment/Experiment}

\include{Chapters/ScatteringTheory/ScatteringTheory}
\include{Chapters/Floquet/Floquet}
\include{Chapters/Formalism/Formalism}

\include{Chapters/PeriodicPotentials/PeriodicPotentials}
\include{Chapters/Applications/Applications}
\include{Chapters/Conclusion/Conclusion}

% Note: GS 2010 requires bibliography/references _before_ the appendix
% if you believe their guidelines; however, conversations with GS
% staff suggests _they don't care_. Go figure. So do what you like.
\backmatter

% We use BIBTeX for the bibliography---you don't have to
% \bibliographystyle{natureURL} % use your favorite BIBTeX style
 \bibliographystyle{h-physrev3-dk} % use your favorite BIBTeX style
% \bibliographystyle{rmp} % use your favorite BIBTeX style
% \bibliographystyle{opcit} % use your favorite BIBTeX style
%\renewcommand\bibname{Master list of references} % rename bibliography 
% "You may decide how this section should be titled. The terms References or Bibliography are the most commonly chosen titles."
%\nocite{*} % To display all refs, even uncited refs (useful when editting)
\bibliography{thesisbib}
% can use multiple .bib files, but no white space allowed for multiple included references, and ".bib" cannot be included

% If for some reason you are anti-BIBTeX, then you would use the% following instead of the above:
%\begin{thebibliography}{99}
% ...
%\end{thebibliography}

\appendix

\include{Appendices/PartialWaveIntegralEqn}
\include{Appendices/SquareWellWF}
\include{Appendices/MultiChannel}
\include{Appendices/AnalyticParametrization}
% \include{Appendices/Appendix_ERC}
% \include{Appendices/Appendix_Unstable}

\end{document}

%% file: Misc/macros.tex
\usepackage{mathtools}

\newcommand{\abar}{\bar a}
\newcommand{\atil}{\tilde a}
\newcommand{\Vbar}{\bar{V}}
\newcommand{\Vtil}{\widetilde{V}}
\newcommand{\Bbar}{\bar{B}}
\newcommand{\Btil}{\widetilde{B}}

\newcommand{\mbf}[1]{\bm{#1}}
\newcommand{\VvdW}{V_\mathrm{vdW}}
\newcommand{\RvdW}{R_{\mathrm{vdW}}}
\newcommand{\EvdW}{E_\mathrm{vdW}}

\newcommand{\ketbra}[2]{\ket{#1}\bra{#2}}
\DeclarePairedDelimiter\bra{\langle}{\rvert}
\DeclarePairedDelimiter\ket{\lvert}{\rangle}
\DeclarePairedDelimiterX\matel[3]{\langle}{\rangle}{#1 \delimsize\vert #3 \delimsize\vert #2}
\DeclarePairedDelimiterX\braket[2]{\langle}{\rangle}{#1 \delimsize\vert #2}

\newcommand{\mr}[1]{\mathrm{#1}}

\newcommand{\rD}{\hspace{-0.5 px}{\rm D}}

\newcommand{\beq}{\begin{equation}} 
\newcommand{\eeq}{\end{equation}} 

\newcommand{\imineq}[2]{\vcenter{\hbox{\includegraphics[height=#2ex]{#1}}}}

%% file: Frontmatter/Hudson_Abs.tex
% !TEX root = HudsonPhDThesis.tex
\begin{abstract}
  In systems of ultracold atoms, pairwise interactions are resonantly enhanced by the application of an oscillating magnetic field that is parallel to the spin-quantization axis of the atoms. The resonance occurs when the frequency of the applied field is precisely tuned near the transition frequency between the scattering atoms and a diatomic molecule. The resulting cross section can be made more than two orders of magnitude larger than the cross section in the absence of the oscillating field. The low momentum resonance properties have a universal description that is independent of the atomic species. To arrive at these conclusions, we first develop a formal extension of Floquet theory to describe scattering of atoms with time-periodic, short-range interaction potentials. We then calculate the atomic scattering properties by modeling the atomic interactions with a square well potential with oscillating depth and then explicitly solving the time-dependent Schr{\"o}dinger equation. We then apply the Floquet formalism to the case of atoms scattering with a contact interaction described by a time-periodic scattering length, obtaining analytic results that agree with those obtained by solving the time-dependent Schr{\"o}dinger equation.
\end{abstract}

%%% Local Variables:
%%% mode: latex
%%% TeX-master: "../HudsonPhDThesis"
%%% End:

%% file: Frontmatter/Hudson_Ack.tex
% !TEX root = HudsonPhDThesis.tex
\begin{acknowledgments}

  I express my deep gratitude to Eric Braaten, my research advisor and mentor. Over the past five years, he has pursued my development as a physicist by tirelessly answering my questions, by continuously offering well-reasoned feedback and advice, and by generously providing me with many opportunities to attend scientific conferences and workshops. I owe much of what I am as a physicist to Eric.

  I owe more than I can express to my wife. Her love and support keep me afloat. I thank my mom, dad, and brothers for their continued prayers and encouragement. I also thank Rich Schelp for his friendship, mentorship, and for helping me take my first steps as a physicist.
  
  I also thank the local community of graduate students and postdocs. In particular, I thank Christian Langmack, Chuck Bryant, Sushant More, Abhishek Mohapatra, Hong Zhang, Xiangyu Yin, and Shaun Hampton for their friendship and for making the OSU physics department a fun and intellectually stimulating environment. Without them and many others, life as a graduate student would have been very dull indeed. I must also thank the physics administrative staff. I am particularly grateful to Kris Dunlap whose patience with me appears to be never-ending.

\end{acknowledgments}

%% file: Frontmatter/Hudson_Vita.tex
% !TEX root = ../HudsonPhDThesis.tex
\begin{vita}
% \small
\dateitem{September 2007--May 2011}{BS in Physics and BA in Mathematics, Erskine College, Due West, South Carolina}
\dateitem{September 2011--August 2016}{PhD in Physics, Ohio State University, Columbus, Ohio}
%\dateitem{July 2013}{M.S. in Physics, The Ohio State University, Columbus, Ohio}
\begin{publist}
 {\it Avalanche mechanism for atom loss near an atom-dimer Efimov resonance}
  \\Christian Langmack, D. Hudson Smith, and Eric Braaten
  \\Phys. Rev. A {\bf 86}, 022718 (2012) [\texttt{arXiv:1205.2683}] \\\\
 {\it Avalanche mechanism for the enhanced loss of ultracold atoms}
  \\Christian Langmack, D. Hudson Smith, and Eric Braaten
  \\Phys. Rev. A {\bf 87}, 023620 (2013) [\texttt{arXiv:1209.4912}] \\\\
 {\it Atom loss resonances in a Bose-Einstein condensate}
  \\Christian Langmack, D. Hudson Smith, and Eric Braaten
  \\Phys. Rev. Lett. {\bf 111}, 023003 (2013) [\texttt{arXiv:1302.5925}] \\\\
 {\it Two-body and three-body contacts for identical bosons near unitarity} 
  \\D. Hudson Smith, Eric Braaten, Daekyoung Kang, and Lucas Platter
  \\Phys. Rev. Lett. {\bf 112}, 110402 (2014) [\texttt{arXiv:1309.6922}]\\\\
 {\it Selection rules for hadronic transitions of XYZ mesons}
  \\Eric Braaten, Christian Langmack, and D. Hudson Smith
  \\Phys. Rev. Lett. {\bf 112}, 222001 (2014) [\texttt{arXiv:1401.7351}]\\\\\\\\
 {\it Born-Oppenheimer approximation for the XYZ mesons}
  \\Eric Braaten, Christian Langmack, and D. Hudson Smith
  \\Phys. Rev. D {\bf 90}, 014044 (2014) [\texttt{arXiv:1402.0438}]\\\\
 {\it Association of atoms into universal dimers using an oscillating magnetic field}
  \\Christian Langmack, D. Hudson Smith, and Eric Braaten
  \\Phys. Rev. Lett. {\bf 114}, 103002 (2015) [\texttt{arXiv:1406.7313}]\\\\
 {\it Inducing resonant interactions in ultracold atoms with a modulated magnetic field}
  \\D. Hudson Smith
  \\Phys. Rev. Lett. {\bf 115}, 193002 (2015) [\texttt{arXiv:1503.02688}]\\\\
 {\it Induced two-body scattering resonances from a square-well potential \\with oscillating depth}
  \\D. Hudson Smith
  \\EPJ Web of Conf. \textbf{113}, 02005 (2015)\\\\
 {\it Dynamics of small trapped one-dimensional Fermi gas\\ under oscillating magnetic field}
 \\X. Y. Yin, Yangqian Yan, and D. Hudson Smith
 \\Phys. Rev. A {\bf 94}, 043639 (2016) [\texttt{arXiv:1608.06966}]

\end{publist}
\begin{fieldsstudy}
\majorfield{Physics}
%\onestudy{Atomic phisics: Efimov physics in ultracold atoms}{Eric Braaten} % optional
\begin{studieslist}
\studyitem{Atomic physics: Universal aspects of ultracold atomic gases}{} % optional
%\studyitem{High energy physics: QCD phenomenology and heavy quarkonium}{} % optional
\end{studieslist}
% Alternatively you can do:
% \begin{studieslist}
% \studyitem{Topic 1}{Professor 1}
% \studyitem{Topic 2}{Professor 2}
% \studyitem{Topic 3}{Professor 3}
% \end{studieslist}
\end{fieldsstudy}
\end{vita}

%%% Local Variables:
%%% mode: latex
%%% TeX-master: "../HudsonPhDThesis"
%%% End:

%% file: Chapters/Intro/Intro.tex
% !TEX root = ../HudsonPhDThesis.tex
\cleardoublepage

\chapter{Introduction}
One of the great achievements of modern physics is the quantum mechanical description of matter. On a microscopic level, matter behaves in exciting and counterintuitive ways. Particles behave not like points in space but instead like waves that extend over a region of space. Particle energies can take on discrete values. Particles come in two basic types called \textit{bosons} and \textit{fermions}, which have dramatically different quantum behavior. The quantum description of matter has led to accurate, systematic explanations for diverse phenomena, ranging from the chemical properties of elements in the periodic table to superconductivity in metals at very low temperatures. More recently, novel experiments with ultracold trapped atoms have begun to unveil the intricacies of the microscopic, quantum-mechanical world in unprecedented ways. These experiments study trapped gases of atoms that have been cooled to such low temperatures that a particle’s quantum mechanical nature takes over. 

A unique feature of cold atomic gases is the ability to experimentally adjust the strength of interactions between particles. This tunability makes it possible to measure the properties of cold atomic gases all the way from the non-interacting regime, where the particles have no influence on one another, to the regime with infinitely strong interactions. This control over the quantum nature of atoms has led to unprecedented breakthroughs in few- and many-body physics. 

For ultracold atoms, a convenient measure of particle interactions is the $s$-wave scattering length $a$. Roughly speaking, the magnitude of $a$ corresponds to the distance over which particles feel each other, and the sign of $a$ indicates whether the interaction is attractive or repulsive. At very low energy the 2-body elastic cross section for distinguishable particles is simply
%------------------------------
\begin{equation}
\sigma = 4\pi a^2.
\end{equation}
%------------------------------
For large $|a|$, the cross section becomes large, signaling strong atomic interactions. The experimental techniques for controlling the strength of atomic interactions involve the manipulation of the effective scattering length $a$. By the careful application of external magnetic or optical fields, the scattering length $a$ can be made resonantly large, positive or negative. 

\begin{figure}[!t]
\centering
\begin{subfigure}[t]{.48\textwidth}
\includegraphics[width=\textwidth]{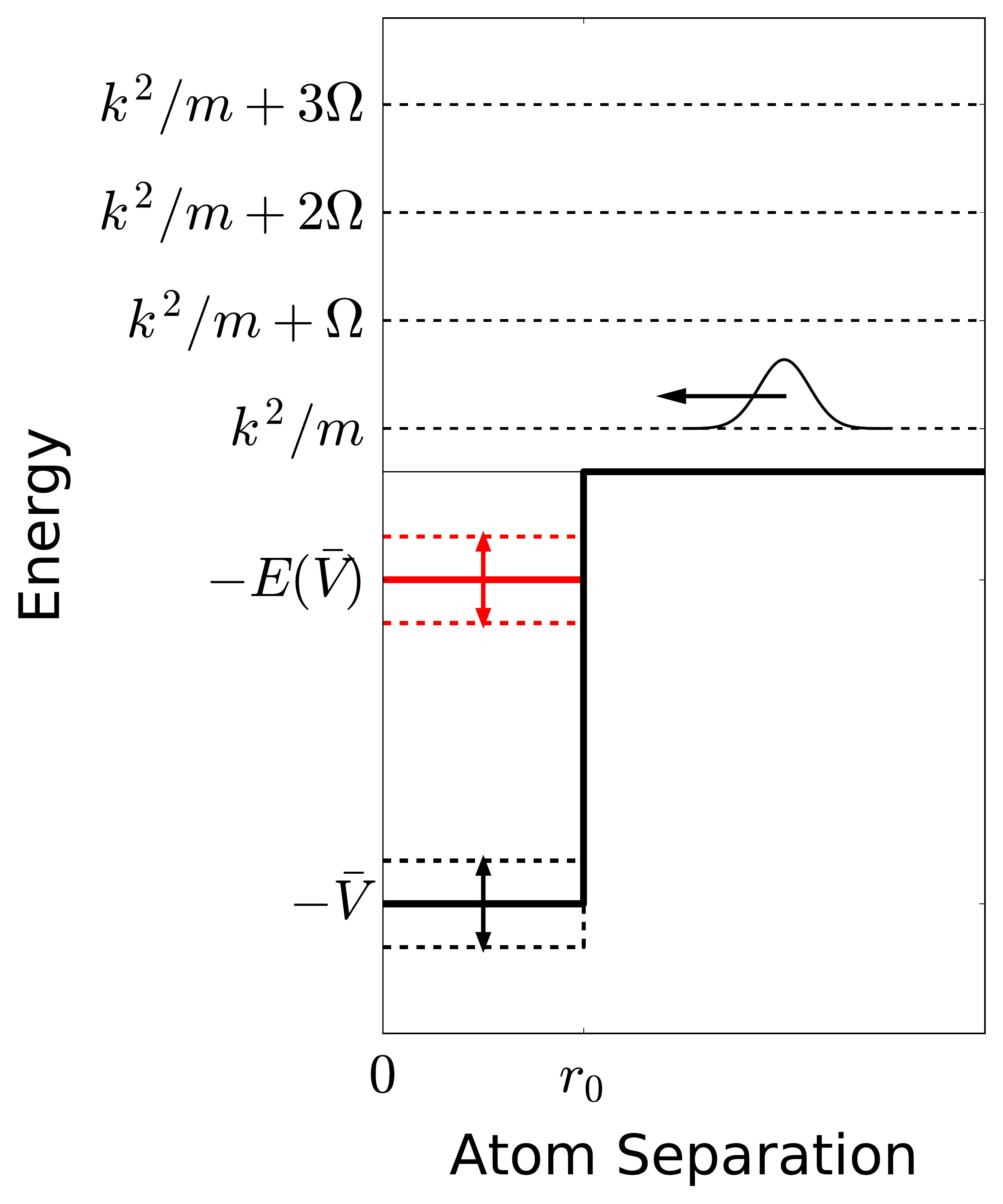}
\caption{The incoming wave.}
\end{subfigure}%
\hfill
\begin{subfigure}[t]{.48\textwidth}
\includegraphics[width=\textwidth]{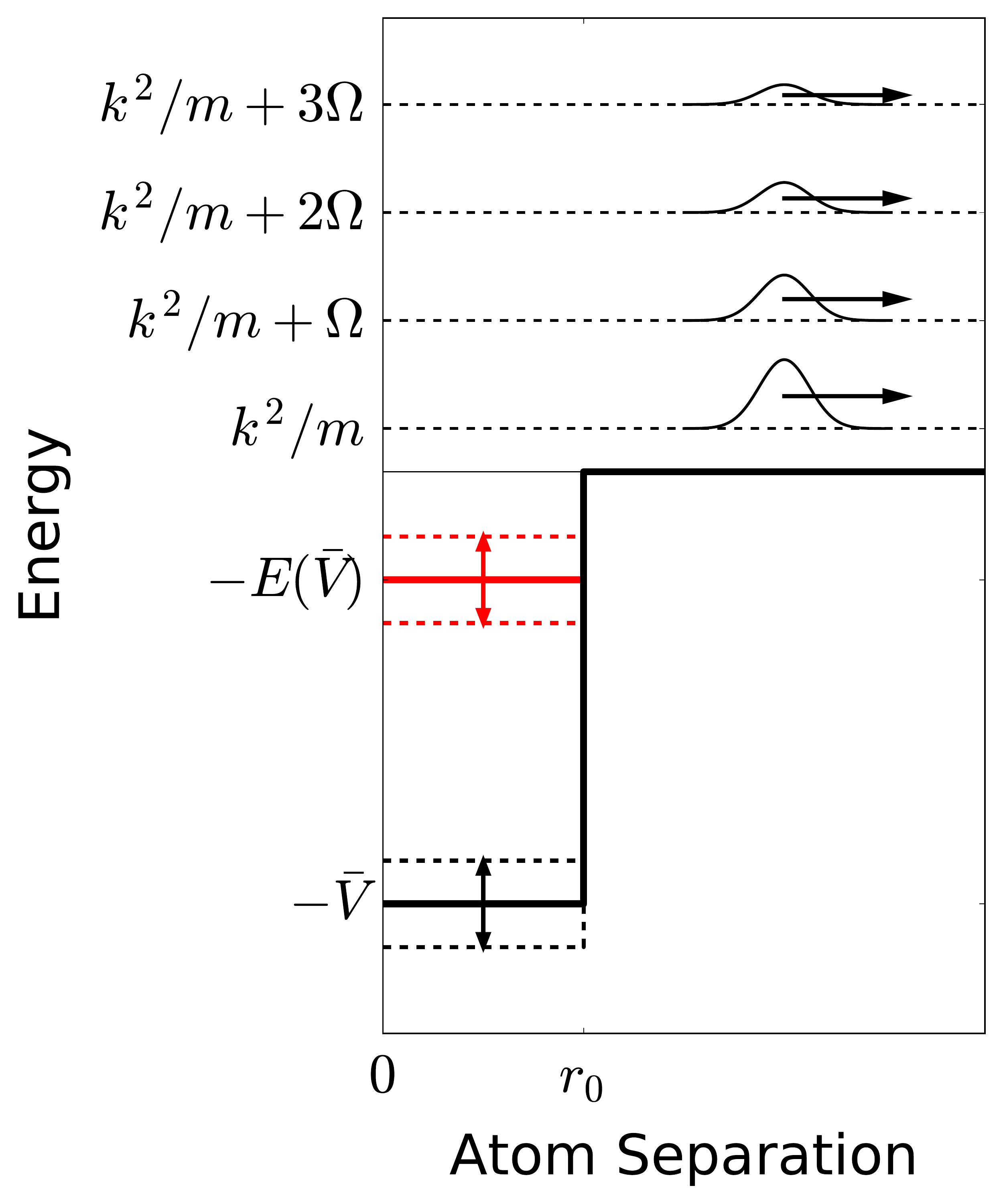}
\caption{The outgoing wave.}
\end{subfigure}
\caption[A time-periodic interaction potential splits an incoming wave into an outgoing wave multiple energy components]{A time-periodic interaction potential splits an incoming wave (panel (a)) into multiple, outgoing waves (panel (b)) with evenly-spaced energies. The double-ended arrows indicate the oscillation range of the depth of the potential $\Vbar$ and the corresponding oscillation range of a bound state energy $-E(\Vbar)$.}
\label{fig:floquet_heuristic}
\end{figure}

In this thesis, we discuss a new technique for resonantly enhancing the effective interaction between atoms, which we term \textit{Modulated Magnetic Feshbach Resonance} (MMFR). This technique, involves the application of an oscillating magnetic field near a precise frequency. Microscopically, the effect of this field is to modulate the strength of the 2-atom interaction potential in time. This modulated potential can significantly alter the scattering properties of the 2-body system. The oscillating potential can inject or remove energy from the scattering particles while they are in the scattering region. The potential injects or removes energy in quanta of $\hbar\Omega$, where $\Omega$ is the oscillation frequency. This is more complicated than scattering from a time-independent potential, for which energy is conserved, but it is less complicated than scattering from a potential with arbitrary time dependence for which energy conservation can be violated by arbitrary amounts. Figure \ref{fig:floquet_heuristic} gives a heuristic picture of how the oscillating potential splits an incoming wave into multiple, evenly-spaced energy components.

%%%%%%%%%%%%%%%%%%%%%%%%%%%%%%%%%%%%%%%%%%%%%%%%%%%%%%
%%%%%%%%%%%%%%%%%%%%%%%%%%%%%%%%%%%%%%%%%%%%%%%%%%%%%%

This basic property of scattering from time-periodic potentials leads to the possibility of experimentally tuning the effective interactions between particles by tuning the oscillation frequency $\Omega$. In fact, as we will show, for potentials supporting bound states (as in Fig.~\ref{fig:floquet_heuristic}), a scattering resonances is induced if $\Omega$ is tuned near the transition frequency between the scattering state of the scattering particles and the bound state. Of course, as a prerequisite, the experimentalist must be able to modulate the effective interaction potential between particles. We will show in Chapter~\ref{Chap:Applications} that the effective interactions between cold, neutral atoms \textit{can} be modulated by the application of an oscillating magnetic field. We find that the effective cross section can be enhanced by several orders of magnitude.

This thesis is organized as follows: In Chapter~\ref{chap:exper}, we discuss the basic properties of ultracold atomic gases, providing the physical context for our later discussions. In Chapter \ref{chap:scat}, as a prerequisite for understanding scattering of atoms in the presence of an oscillating magnetic field, we review scattering in the absence of such a field. In Chapter \ref{Chap:Floquet}, we discuss Floquet theory, which provides a natural framework for dealing with time-periodic potentials. Chapter \ref{Chap:Formalism} begins the unique contributions of this thesis. There we derive an extension of Floquet theory to describe 2-body scattering with a short-range, time-periodic interaction potential.  In Chapter \ref{chap:atomic_scattering}, we derive the scattering properties of atoms with a short-range time-periodic potential modeled by a square-well with oscillating depth. In Chapter \ref{Chap:Applications}, we use the Floquet scattering formalism derived in Chapter \ref{Chap:Formalism} to calculate the scattering properties of atoms with an explicitly zero-range time-periodic potential. We conclude in Chapter \ref{Chap:Final}.

%%% Local Variables:
%%% mode: latex
%%% TeX-master: "../../HudsonPhDThesis"
%%% End:

%% file: Chapters/Experiment/Experiment.tex
% !TEX root = ../HudsonPhDThesis.tex
\cleardoublepage

\chapter{Ultracold atomic gases}
\label{chap:exper}

Modern experiments with ultracold gases have demonstrated the unprecedented ability to create and manipulate macroscopic samples of atoms whose properties are governed by the underlying quantum statistics of the particles. The creation of such quantum gases relies upon advanced technologies for trapping and cooling the atoms. Once created, the manipulation of ultracold gases can be achieved at the quantum level by tuning the effective interactions between the particles. This ability has led to many advances in our understanding of the role of interactions in quantum matter and has potentially many applications in the field of quantum engineering. This chapter discusses the basic properties of ultracold atomic gases, providing the physical context for our later discussions. Refs.~\cite{UBG_review,UFG_review} provide extensive reviews of the properties of cold Bose and Fermi gases as well as the associated experimental techniques.

In Section \ref{chap:exper::sec:alkali}, we review the properties of alkali atoms, which are the most common species used in experiments with ultracold atomic gases. In Section \ref{chap:exper::sec:low_temp}, we discuss the theoretical motivations and experimental procedures for cooling atoms. In Section \ref{chap:exper::sec:tuning}, we discuss techniques for tuning the effective interactions between atoms.

\section{Alkali atoms}
\label{chap:exper::sec:alkali}

The types of neutral atoms that are most easily cooled to ultralow temperatures and whose interactions can be most easily manipulated are the alkali atoms more massive than hydrogen (H) and less massive than francium (Fr). They are lithium (Li), sodium (Na), potassium (K), rubidium (Rb), and cesium (Cs). Hydrogen is difficult to cool due to its small mass, and francium is rarely used due to its relatively short half-life of $22$ minutes \cite{feshbach, francium}. The ability to cool and manipulate the alkali atoms largely arises from their simple electronic structure, which leads to a relatively simple response of these atoms to a magnetic field, as discussed in Section \ref{chap:exper::sec:alkali::sub:hyper}. At low energies, the interaction properties of the alkali atoms are governed by their van der Waals interaction potentials, as discussed in Section \ref{chap:exper::sec:alkali::sub:potentials}.

\subsection{Hyperfine spin states}
\label{chap:exper::sec:alkali::sub:hyper}

An alkali atom in its electronic ground state has multiple spin states. There are two contributions to its spin: the electronic spin $\bm{S}$ with quantum number $s={1\over2}$ and the nuclear spin $\bm{I}$ with quantum number $i$. The $2(2i+1)$ spin states can be labeled $|m_s,m_i \rangle$, where $m_s$ and $m_i$ specify the eigenvalues of $S_z$ and $I_z$. The Hamiltonian for a single atom includes a {\it hyperfine} term that can be expressed in the form
% ------------------------
\begin{eqnarray}
  H_{\rm hyperfine} =
  {2 E_{\rm hf} \over (2 i + 1)\hbar^2} \bm{I} \cdot \bm{S}.
  \label{Hhyperfine}
\end{eqnarray}
% ------------------------
This term splits the ground state of the atom into two hyperfine multiplets with energies differing by $E_{\rm hf}$. The sum of the electronic and nuclear spin is called the hyperfine spin $\bm{F} = \bm{I}+\bm{S}$. The eigenstates of $H_{\mathrm{hyperfine}}$ are labeled by the quantum numbers $f$ and $m_f$ which specify the eigenvalues of $\bm{F}^2$ and $F_z$. The eigenvalues of $H_{\rm hyperfine}$ are
% ------------------------
\begin{eqnarray}
  E_{f,m_f} =
  \frac{f(f+1) - i(i+1) - \frac{3}{4}}{2 i + 1} E_{\rm hf} .
\end{eqnarray}
% ------------------------
The two hyperfine multiplets of an alkali atom consist of $2i+2$ states with $f = i + {1\over2}$ and $2i$ states with $f = i - {1\over2}$. For example, a $^7$Li atom has nuclear spin quantum number $i={3\over2}$. The two hyperfine multiplets consist of five states with $f=2$ and three states with $f=1$. The $f=2$ multiplet is higher in energy by $E_{\rm hf}$. The frequency associated with the hyperfine splitting is $E_{\rm hf}/h \approx 803.504~$MHz \cite{ScatteringModelsBraaten, feshbach}.

In the presence of a magnetic field $\bm{B} = B \bm{\hat z}$,
the Hamiltonian for a single atom has a {\it magnetic} term.
The magnetic moment \mbox{\boldmath $\mu$} of the atom is dominated by the
term proportional to the spin of the electron:
\mbox{\boldmath $\mu$} = $\mu \, \bm{S}/({1\over2}\hbar)$.
The magnetic moment $\mu$ of an alkali atom such as Li
is approximately that of the single electron in the outermost shell:
$\mu \approx-2\mu_B$, where $\mu_B$ is the Bohr magneton.
The magnetic term in the Hamiltonian can be expressed in the form
% ------------------------
\begin{eqnarray}
  H_{\rm magnetic} =
  - {2 \mu \over \hbar} \bm{S} \cdot \bm{B} .
  \label{Hmagnetic}
\end{eqnarray}
% ------------------------
If $B \neq 0$, this term splits the two hyperfine multiplets of an alkali atom into $2(2i+1)$ hyperfine states. In a weak magnetic field satisfying $\mu B \ll E_{\rm hf}$, each hyperfine multiplet is split into $2f+1$ equally-spaced Zeeman levels $| f, m_f \rangle$. In a strong magnetic field satisfying $\mu B \gg E_{\rm hf}$, the states are split into a set of $2i+1$ states with $m_s=+{1\over2}$ whose energies increase linearly with $B$ and a set of $2i+1$ states with $m_s=-{1\over2}$ whose energies decrease linearly with $B$. Each of those states is the continuation in $B$ of a specific hyperfine state $| f, m_f \rangle$ at small $B$. It is convenient to label the states by the hyperfine quantum numbers $f$ and $m_f$ for general $B$, in spite of the fact 
that those states are not eigenstates of $\bm{F}^2$ if $B \neq 0$. We denote the eigenstates of $H_{\rm hyperfine} + H_{\rm magnetic}$ by $| f, m_f ; B \rangle$ and their eigenvalues by $E_{f,m_f}(B)$. The two eigenstates with the maximal value of $|m_f|$ are independent of $B$:
% ------------------------
\begin{eqnarray}
  \left| f=i+\mbox{$1\over2$}, m_f= \pm (i+\mbox{$1\over2$}) ; B \right\rangle
  = \left| m_s = \pm \mbox{$1\over2$}, m_i = \pm i \right\rangle.
  \label{hfmax}
\end{eqnarray}
% ------------------------
Their eigenvalues are exactly linear in $B$:
% ------------------------
\begin{eqnarray}
  E_{f,m_f}(B) = \frac{i}{2i+1} E_{\rm hf} \mp \mu B.
\end{eqnarray}
% ------------------------
If $B \neq 0$, each of the other eigenstates $| f, m_f ; B \rangle$
is a linear superposition of the two states 
$| f= i -{1 \over 2}, m_f \rangle$
and $| f= i + {1 \over 2}, m_f \rangle$.

%%%%%%%%%%%%%%%%%%%%%%%%%%%%%%%%%%%%%%%%%%%%%%%%%%%%%%%%%% 
\begin{figure}[tb]	%\centerline{\includegraphics*[width=1.0\columnwidth,angle=0,clip=true]{hyper-split.ps}}
\centerline{\includegraphics*[width=1.0\columnwidth,angle=0,clip=true]{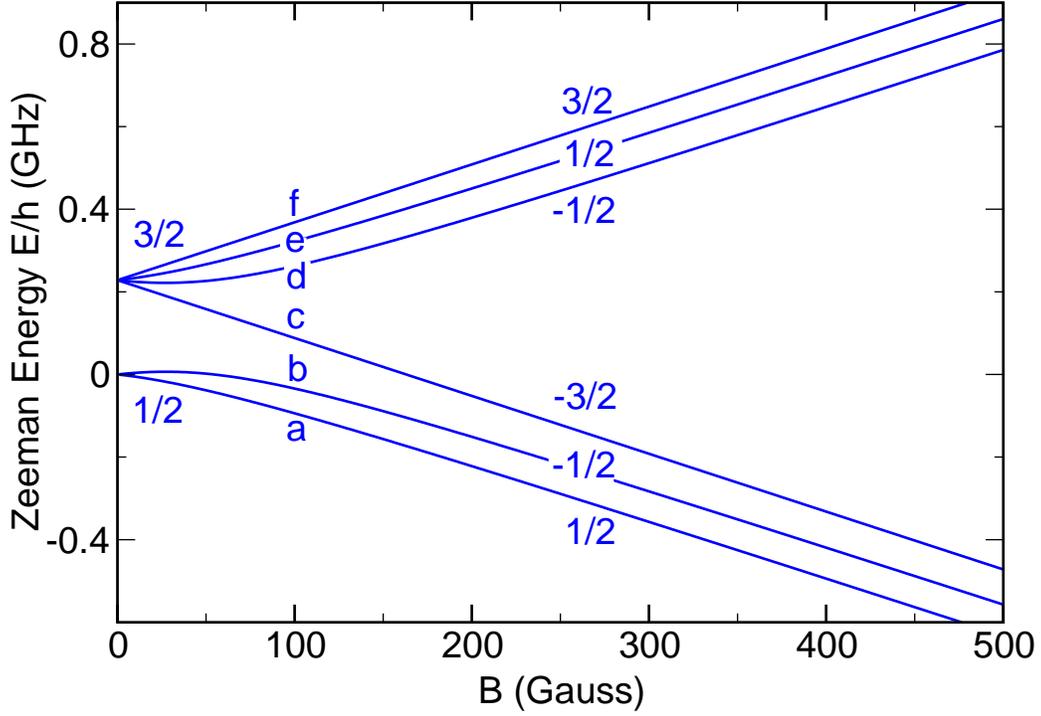}}
% \vspace*{-1.0cm}
\caption[Hyperfine energy levels of $^{7}$Li atoms]
{The hyperfine energy levels
  $^{7}$Li atoms as a function of the magnetic field \cite{feshbach}.}
% Figure from Ref.~\cite{Heidelberg}}
\label{fig:hyperfine}
\end{figure}
%%%%%%%%%%%%%%%%%%%%%%%%%%%%%%%%%%%%%%%%%%%%%%%%%%%%%%%%%% 

The dependence of the hyperfine energy levels of $^7$Li atoms
on the magnetic field is illustrated in Fig.~\ref{fig:hyperfine}.
At $B=0$, the hyperfine multiplets with $f=2$ and $f=1$
are split by $E_{\rm hf}$.
The magnetic energy scale $\mu B$ is comparable to the hyperfine splitting
$E_{\rm hf}$ when $B$ is about 287 Gauss.
At higher magnetic fields, the four $m_s=-\frac12$ states decrease linearly
with $B$, while the four $m_s=+\frac12$ states increase linearly.

\subsection{Interaction potentials}
\label{chap:exper::sec:alkali::sub:potentials}

Due to the large separation of mass scales between the atomic nucleus and the electrons, the atomic interactions are very accurately characterized by the Born-Oppenheimer (BO) potentials for the atoms. Each potential can be labeled by the combination of quantum numbers $^{2s+1}\Gamma_{g/u}$, where $s$ is the total electronic spin quantum number, $\Gamma=\Sigma,\, \Pi,\, \Delta,\,\ldots$ (or 1, 2, 3, $\ldots$) specifies the total orbital angular momentum of the atom pair, and $g/u$ (or {\it gerade/ungerade}) specifies the electronic inversion symmetry (only present for identical atoms). An electronic configuration is {\it gerade} ({\it ungerade}) if the phase of the wavefunction is even (odd) with respect to inversion through the molecular center of mass. The BO potential for an atom pair in the configuration $^{2s+1}\Gamma_{g/u}$ equals the potential energy of the atom pair in that configuration as a function of the separation of the atomic nuclei \cite{feshbach}.

\begin{figure}[tb]
  \centerline{\includegraphics*[width=1.0\columnwidth,angle=0,clip=true]{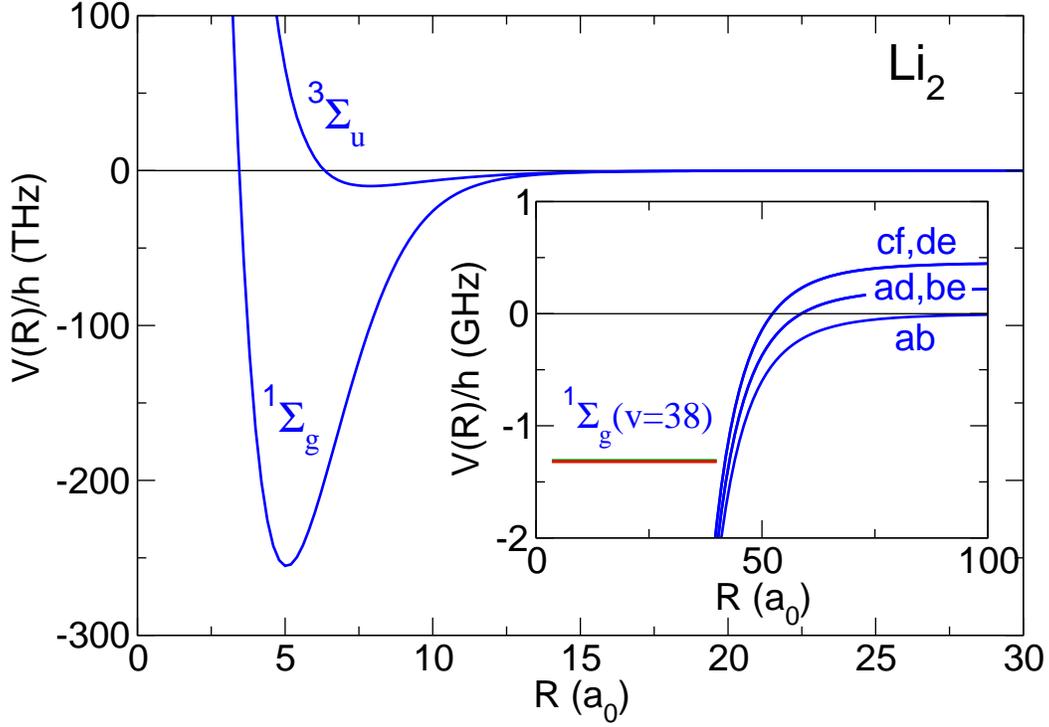}}
  % \vspace*{-1.0cm}
  \caption[Born-Oppenheimer potentials for $^6$Li]
  {The Born-Oppenheimer potentials $^{1}\Sigma_{g}$ and $^{3}\Sigma_{u}$ for $^6$Li plotted as functions of the nuclear separation in units of Bohr radii \cite{feshbach}. The inset gives an enlarged view of the BO potentials at large $R$ for the five hyperfine spin configurations with zero total $z$-projection of hyperfine spin (See Fig.~\ref{fig:hyperfine}).}
  % Figure from Ref.~\cite{Heidelberg}}
  \label{fig:BO_potentials}
\end{figure}

We consider alkali atoms in their ground state for which $\Gamma=\Sigma$. In that case all of the Born-Oppenheimer potentials are isotropic: they only depend on the nuclear separation $R$. Two separated alkali atoms in their ground state individually have zero orbital angular momentum and electron spin of 1/2. The total electronic spin can therefore be $0$, in which case the system is {\it gerade}, or $1$, in which case the system is {\it ungerade}. Thus, the relevant BO potentials for scattering alkali atoms are $^{1}\Sigma_{g}$ and $^{3}\Sigma_{u}$. These BO potentials for the $^{6}$Li system are plotted in Fig.~\ref{fig:BO_potentials}. 

%%%%%%%%%%%%%%%%%%%%%%%%%%%%%%%%%%%%%%%%%%%%%%%%%%%%%%%%%% 
%%%%%%%%%%%%%%%%%%%%%%%%%%%%%%%%%%%%%%%%%%%%%%%%%%%%%%%%%% 

For $s$-state atoms, the leading term in the BO potential at large $R$ are independent of the hyperfine spin configuration. The large-$R$ potentials can then be expressed in terms of a single parameter, $C_6$, which depends on the choice of atomic species, but not on the hyperfine spins of the atoms:
%------------------------------
\begin{equation}
V(R)\longrightarrow \VvdW(R)=-\frac{C_6}{R^6}.
\label{eq:vanderwaals}
\end{equation}
%------------------------------
$\VvdW(R)$ is the {\it van der Waals} potential. The natural length scale $R_\mathrm{vdW}$ for this potential is 
%------------------------------
\begin{equation}
\RvdW=\left(\frac{2\mu C_6}{\hbar^2}\right)^{1/4},
\label{eq:RvdW}
\end{equation}
%------------------------------
where $\mu$ is the reduced mass of the atom pair. The corresponding energy scale is
%------------------------------
\begin{equation}
\EvdW=\VvdW(\RvdW)=\frac{\hbar^2}{2\mu \RvdW^2}=\frac{\hbar^3}{\sqrt{8\mu^{3}C_{6}}}.
\label{eq:EvdW}
\end{equation}
%------------------------------
Table~\ref{tab:vdw} gives $\RvdW$ and $\EvdW$ for several alkali atoms. In units of Boltzmann's constant $k_B$, the typical van der Waals energy scale is $\EvdW\sim 1$mK, which is three orders of magnitude larger than the typical collision energy in an ultracold gas experiment $E\sim 1\mu$K. As a result, the atomic scattering properties are only sensitive to the low-energy (long-distance) details of the interatomic potential. Moreover, if the BO potential supports a bound state with binding energy much smaller than $\EvdW$, the properties of such a bound state are only sensitive to the long range van der Waals potential \cite{ScatteringModelsBraaten, feshbach, Gao1998,Gao2000}. 

\begin{table}
\centering
\caption[van der Waals scales for several atomic species]
{$\RvdW$ and $\EvdW$ for several atomic species. (1 amu = 1/12 mass of a
$^{12}$C atom, $a_0$= 0.0529177 nm) }\label{tab:vdw}
\begin{tabular}{rcccccl}
\hline\hline
species 	& mass			& $\RvdW$ 	& $\EvdW/k_B$   & $\EvdW/h$ & Ref.\\
    		& (amu)			& ($a_0$) 	& (mK) 		& (MHz) & \\
\hline
${^6}$Li 	& 6.0151223 	 	& 31.26 	& 29.47  	& 614.1  & \cite{Yan1996}\\
${^{23}}$Na 	& 22.9897680  	  	& 44.93  	& 3.732   	& 77.77  & \cite{Derevianko1999}\\
${^{40}}$K 	& 39.9639987  	  	& 64.90  	& 1.029   	& 21.44  & \cite{Derevianko1999}\\
$^{40}$Ca 	& 39.962591 	 		& 56.39 	& 1.363  	& 28.40  & \cite{Porsev2002}\\
${^{87}}$Rb 	& 86.909187  	  	& 82.58  	& 0.2922  	& 6.089  & \cite{vanKempen2002}\\
${^{133}}$Cs 	& 132.905429  	  	& 101.0  	& 0.1279  	& 2.666  & \cite{Chin2004}\\
	\hline\hline

\end{tabular}
\end{table}

\section{Dilute, cold gases}
\label{chap:exper::sec:low_temp}

In this section we discuss the importance of the density and temperature of atomic gases. We also describe some of the techniques used to create dilute, cold gases.

\subsection{The low-density regime}
\label{chap:exper::sec:dilute}
Consider an atomic gas with particle number density $n$. The diluteness of the gas is measured by the dimensionless parameter $n r_0^3$ where $r_0$ is the range of the 2-body interaction potential between the two different spin states. When $n r_0^3\ll 1$, the spacing between particles is larger than the range of the potential, and, under ordinary circumstances, the particles interact very weakly. Such a gas is approximately ``ideal'' and the effects of interactions can be incorporated perturbatively. When $n r_0^3\gtrsim 1$, the spacing between particles is comparable to or less than the range of the potential, and the effects of interactions cannot generally be incorporated perturbatively. If the system has a resonant $s$-wave scattering length $a$ such that $|a|\gg r_0$, then $a$ replaces $r_0$ as the relevant length scale for measuring diluteness. In that case, dilute gases satisfy $n |a|^3\ll 1$ \cite{pethicbec, Braaten:2001ay}.

\subsection{The low-temperature regime}
A particle with momentum $p$ has de Broglie wavelength $\lambda=2\pi\hbar/p$.  $\lambda$ describes the spacial extent of the wave function of a particle with momentum $p$. The typical de Broglie wavelength for a particle with mass $m$ in a gas at temperature $T$ is of the order of the thermal de Broglie wavelength, 
%------------------------------
\begin{equation}\label{eq:thermal_wav}
\lambda_T\equiv\sqrt{\frac{2\pi\hbar^2}{mk_\mathrm{B}T}}.
\end{equation}
%------------------------------
$\lambda_T$ can be regarded as the uncertainty in the position of a typical particle in a gas at temperature $T$. At high temperatures, $\lambda_T$ is small, and the constituents of the gas behave like classical, point-like particles. At low temperatures, particles have larger spacial extent. When this extent becomes on the same order as the inter-particle spacing ($\lambda_T\sim n^{-1/3}$, where $n$ is the particle density), the quantum mechanical nature of the particles modifies the properties of the system. For $\lambda_T\ll n^{-1/3}$, the particle behavior is completely governed by either Bose-Einstein statistics in the case of bosons or Fermi-Dirac statistics in the case of fermions. Thus, the pursuit of low temperatures is partly motivated by a desire to isolate and study the unique bosonic or fermionic statistical properties of matter.

Also, as discussed in Sec.~\ref{chap:exper::sec:alkali::sub:potentials}, the typical collision energy is of the same order as the temperature of the gas multiplied by $k_B$. If that energy scale is much less than the van der Waals energy scale $\EvdW$, the physical properties of the gas are only sensitive to the long-distance details of the scattering potential. This leads to a dramatic simplification in the theoretical description of ultracold gases, requiring only the van der Waals coefficient $C_6$ to describe the low-energy physics.

\subsection{Trapping and cooling ultracold gases}
\label{chap:exper::sec:trapping}
We will discuss a standard approach to creating and trapping an ultracold atomic gas. There are numerous variants on the general pattern discussed here. For a more thorough review, see Refs.~\cite{UBG_review, UFG_review}. Cooling occurs in several stages, over which the temperature of the gas can be reduced by about $9$ orders of magnitude. An initial beam of highly energetic atoms ($\sim 500~\text{K}$) passes through a Zeeman slower, which uses doppler cooling to reduce the energy to $\sim 1~\text{K}$. The atoms are then transferred to a magneto-optical trap (MOT), where a combination of typically six intersecting lasers combined with a non-uniform magnetic field confine and cool the cloud. The MOT doppler-cools the cloud to $\sim 1~\text{mK}$. The atoms are then transferred into an optical trap, which uses a single, tightly-focused beam to confine the atoms by the interaction between the electric field-gradient and the electric dipole moment of the atom. The depth of the optical trap can then be incrementally lowered. This allows the high-momentum component of the trapped atoms to escape, lowering the average temperature of the remaining atoms. This technique, called \textit{evaporative cooling}, cools the atoms to $\sim500~\text{nK}$.

\section{Tuning interactions}
\label{chap:exper::sec:tuning}

Multiple techniques exist for manipulating the effective interactions between neutral alkali-metal atoms. These techniques all exploit a coupling between the scattering state of interest and a two-atom bound state, often called the {\it resonance state}. The differences between these techniques lie in the choice of resonance state and in how the coupling between the scattering and resonance state is achieved. Here we briefly review the established techniques for controlling the effective interactions. 

%%%%%%%%%%%%%%%%%%%%%%%%%%%%%%%%%%%%%%%%%%%%%%%%%
\begin{figure}[t]
\centering
\includegraphics*[width=0.75\linewidth,angle=0,clip=true]{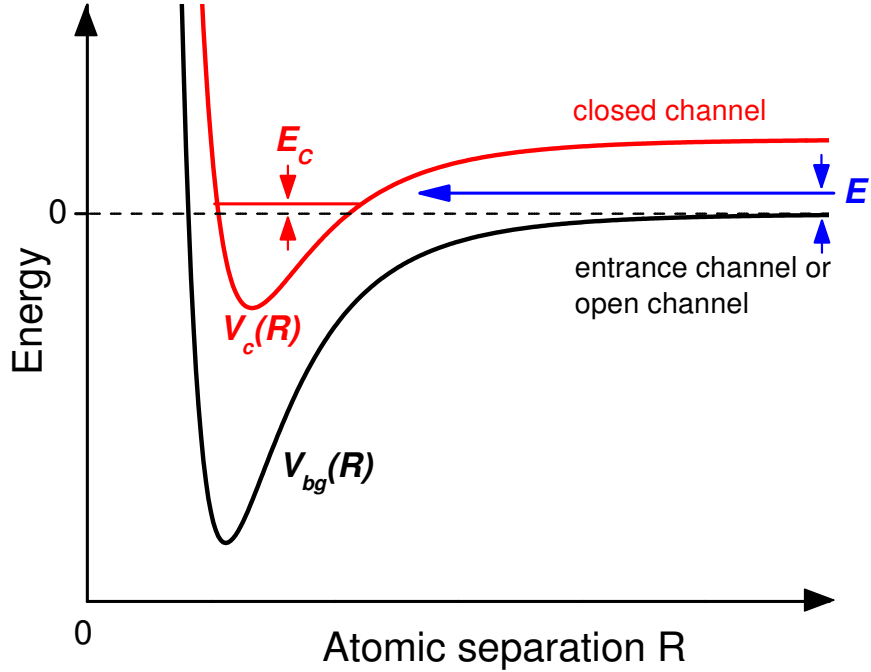}
\caption[Illustration of the magnetic Feshbach resonance mechanism]{Illustration of the magnetic Feshbach resonance (MFR) mechanism \cite{feshbach}. The scattering channel (lower black curve, sometimes called the ``entrance'' or ``open'' channel) is coupled to the bound-state channel (upper red curve, sometimes called the ``closed'' channel). A scattering resonance occurs when the energy difference $E_c$ between the bound-state energy and the scattering threshold approaches zero.}
\label{fig:Feshbach_1}
\end{figure}
%%%%%%%%%%%%%%%%%%%%%%%%%%%%%%%%%%%%%%%%%%%%%%%%%
%%%%%%%%%%%%%%%%%%%%%%%%%%%%%%%%%%%%%%%%%%%%%%%%%
\begin{figure}[t]
\centering
\includegraphics*[width=0.75\linewidth,angle=0,clip=true]{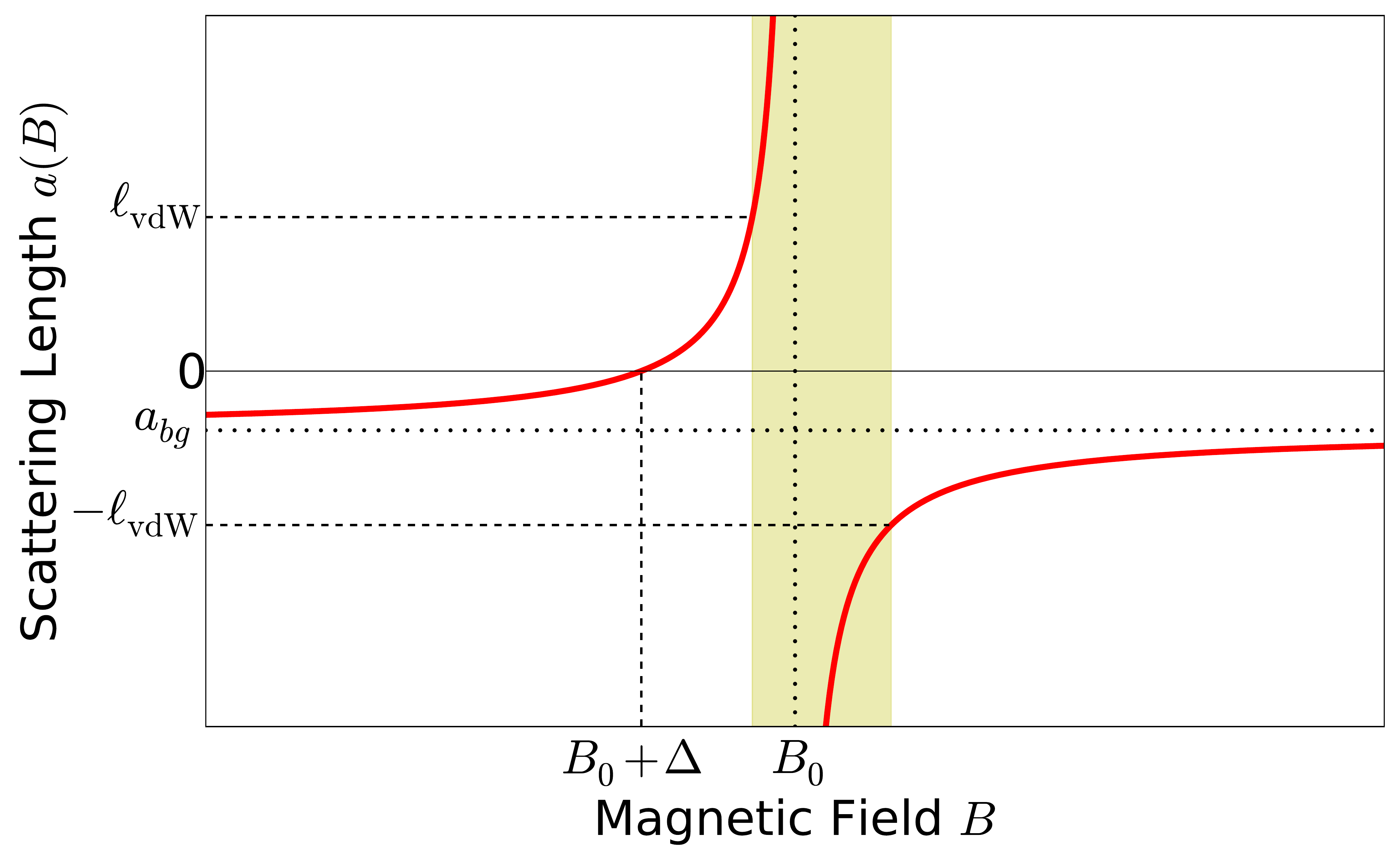}
\caption[Magnetic-field dependence of the scattering length near a magnetic Feshbach resonance]
{Dependence of the scattering length, $a$, on an applied magnetic field $B$ near a magnetic Feshbach resonance. The shaded region indicates the universal region where $|a| > \RvdW$. $B_0$ is the value of $B$ where $a$ is infinitely large. $B_0 + \Delta$ is the value of $B$ at the zero crossing of $a$, and $a_{bg}$ is the value of $a$ far from resonance \cite{LangmackThesis}. }
\label{fig:Feshbach}
\end{figure}
%%%%%%%%%%%%%%%%%%%%%%%%%%%%%%%%%%%%%%%%%%%%%%%%%%
\paragraph{Magnetic Feshbach resonance.}
\textit{Magnetic Feshbach resonance} (MFR) \cite{feshbach} is the most powerful and versatile method to date for manipulating the effective interactions between atoms. In MFR a constant magnetic field applied along the spin quantization axis of the atoms is used to couple the scattering state to a molecular state in a second hyperfine spin channel. Figure \ref{fig:Feshbach_1} illustrates the interaction potentials for the scattering channel and the channel containing the bound state. The energy difference between the scattering threshold and the bound state is modified by the application of a magnetic field $B$, because the magnetic moment of the bound state differs from the magnetic moment of the scattering state. A scattering resonance occurs at the magnetic field value $B_0$ where the energy of the bound level equals the scattering threshold energy. The corresponding $B$-dependent $s$-wave scattering length $a$ is shown in Fig.~\ref{fig:Feshbach}. Near $B_0$, $a$ is a simple function of $B$:
%--------------------
\begin{equation}\label{eq:feshbach}
\frac{1}{a(B)}=\frac{1}{a_{\mathrm{bg}}}\,\frac{B-B_0}{B-B_0-\Delta}+i\gamma,
\end{equation}
%--------------------
where $a_{\mathrm{bg}}$ is the background scattering length, $\Delta$ is the distance from the resonance to the nearest zero crossing, and $\gamma$ is non-zero only if the colliding atoms have a spin-relaxation scattering channel. 

While extremely useful in many scenarios, MFR is limited to atomic systems where scattering atoms can be brought to degeneracy with a bound state in a coupled channel by the application of a magnetic field of a reasonable magnitude. The properties of the associated resonance are then completely determined by the microscopic details of the interatomic potential and may not be favorable for experimental use. 

%%%%%%%%%%%%%%%%%%%%%%%%%%%%%%%%%%%%%%%%%%%%%%%%%
\begin{figure}[t!]
\centering
\includegraphics*[width=0.65\linewidth,angle=0,clip=true]{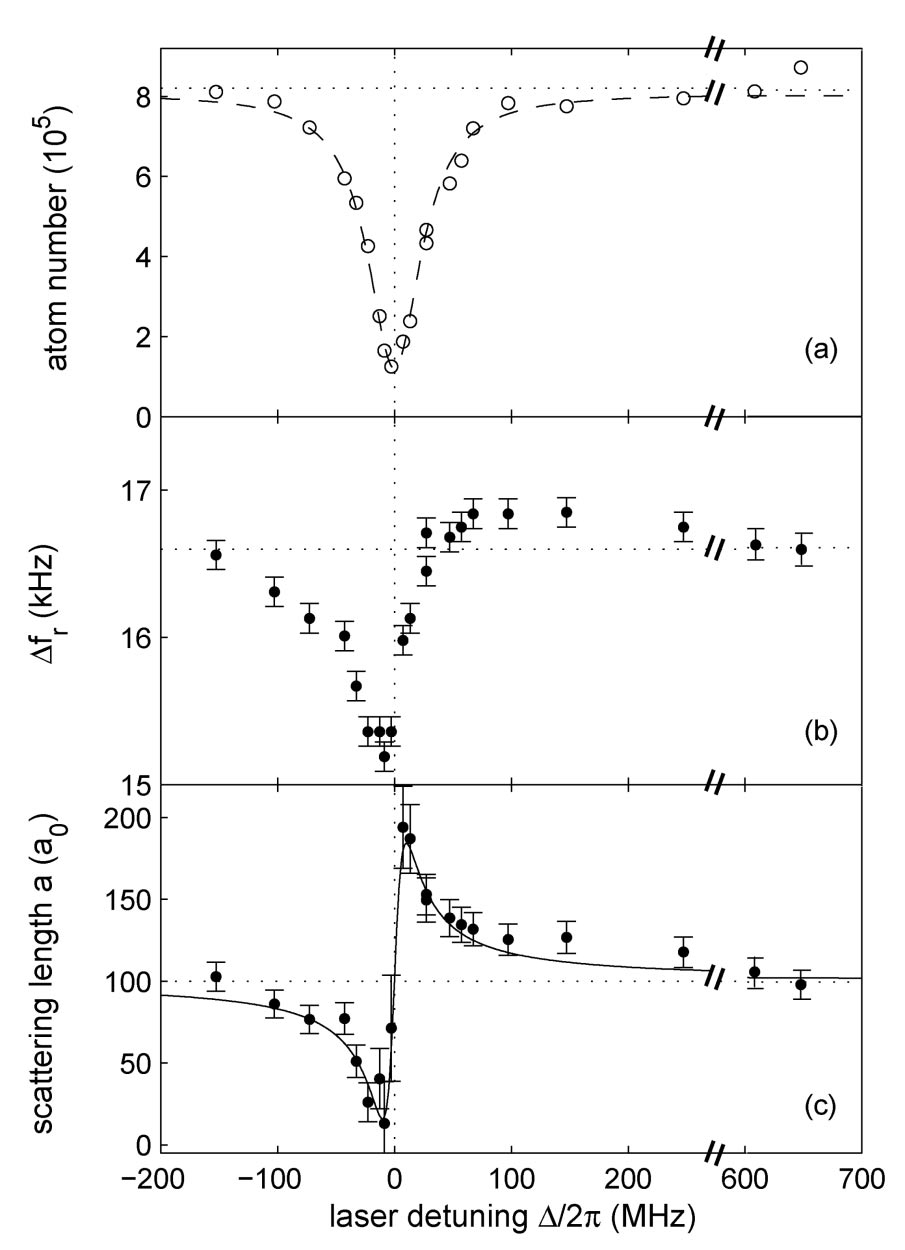}
\caption[Experimental results for an Optical Feshbach resonance in $^{87}$Rb]{Experimental results for an optical Feshbach resonance (OFR) in $^{87}$Rb \cite{Grimm_2004}.  (a) shows the number of atoms remaining after applying the oscillating field for a fixed interval of time as a function of the detuning. The decay of the $p$-wave molecule leads to atom loss. (c) shows the experimentally determined value of the effective scattering length as a function of the laser detuning from resonance. See Ref.~\cite{Grimm_2004} for a description of (b).   }
\label{fig:OFR}
\end{figure}
%%%%%%%%%%%%%%%%%%%%%%%%%%%%%%%%%%%%%%%%%%%%%%%%%%
\paragraph{Optical Feshbach Resonance}
In {\it optical Feshbach resonance} (OFR) \cite{Shlyapnikov,Julienne_1999,Lett_2000,Grimm_2004,Grimm_2005}, laser light that is slightly detuned from a transition between the scattering atoms and an electronically excited $p$-wave molecule induces a resonance in the scattering length. Figure \ref{fig:OFR} presents an experimental realization of OFR in $^{87}$Rb. One advantage of OFR over MFR is that laser light can be switched much more rapidly than electrostatic magnetic fields. In addition, the properties of an OFR depend upon the intensity of the laser, giving some control over the strength and width of the resonance. OFR has major limitations for alkali-metal atoms because the rapid spontaneous decay of the resonance molecule results in dramatic atom losses and severely limits the maximum value of the scattering length. This is demonstrated experimentally in panel (a) of Fig.~\ref{fig:OFR}.

%%%%%%%%%%%%%%%%%%%%%%%%%%%%%%%%%%%%%%%%%%%%%%%%%
\begin{figure}[h]
\centering
\includegraphics*[width=0.75\linewidth,angle=0,clip=true]{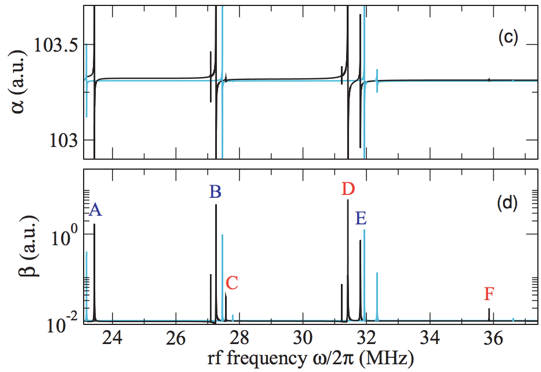}
\caption[Theoretical prediction for a radio-frequency Feshbach resonance in $^{87}$Rb]{Theoretical prediction for a radio-frequency Feshbach resonance (rfFR) in $^{87}$Rb \cite{Schmiedmayer_rfFR_2010}. The real (upper panel) and imaginary (lower panel) parts of the scattering length are shown as functions of the rf frequency. The blue and black curves correspond to different values of the constant magnetic field component parallel to the spin-quantization axis of the atoms.}
\label{fig:rffr}
\end{figure}
%%%%%%%%%%%%%%%%%%%%%%%%%%%%%%%%%%%%%%%%%%%%%%%%%%
\paragraph{Radio-frequency and Microwave Feshbach Resonance}
In {\it radio-frequency \\Feshbach resonance} (rfFR) and {\it microwave Feshbach resonance} (mwFR) \cite{Julienne_rfFR_2009,Schmiedmayer_rfFR_2010, Dalibard_mwFR_2010}, an oscillating magnetic field that is perpendicular to the spin-quantization axis of the atoms couples an atom pair to a molecule in another hyperfine channel, thereby modifying or inducing a resonance in the scattering length. These methods allow some control over the scattering length without introducing dramatic atom loss. One disadvantage of rf/mwFR is that the coupling between an atom pair and the resonance molecule tends to be very small, leading to very small enhancement of the real part of the scattering length. Also, it is difficult to produce large-amplitude rf and mw fields. Figure \ref{fig:rffr} shows a theoretical prediction for a rfFR in $^{87}$Rb.

%%%%%%%%%%%%%%%%%%%%%%%%%%%%%%%%%%%%%%%%%%%%%%%%%
\begin{figure}[h]
\centering
\includegraphics*[width=0.75\linewidth,angle=0,clip=true]{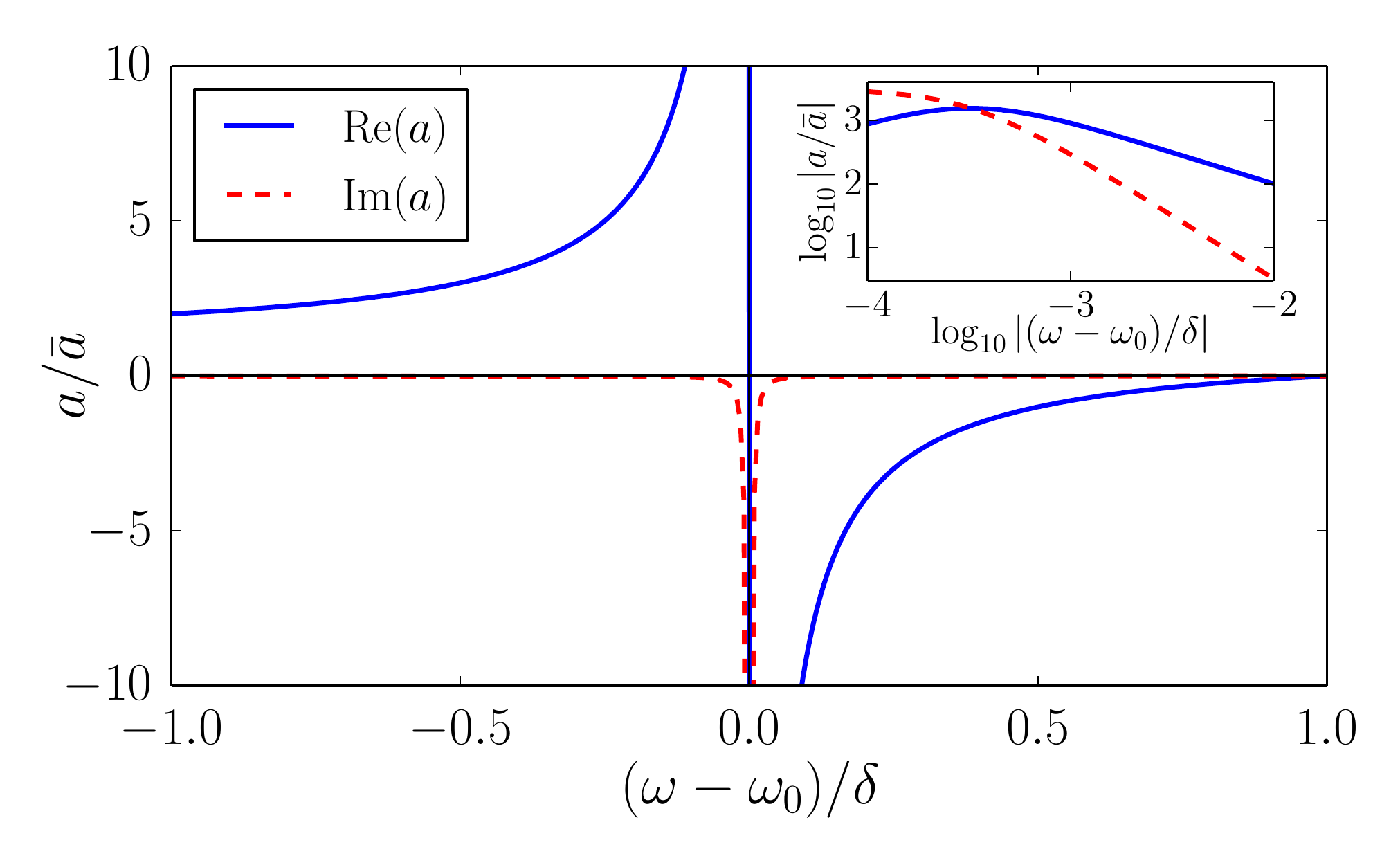}
\caption[Theoretical prediction for a modulated magnetic Feshbach resonance]{Theoretical prediction for a modulated magnetic Feshbach resonance (MMFR) \cite{Hudson_2015}. Shown are the real and imaginary parts of $a$ as functions of the oscillation frequency $\omega$. The inset shows the absolute values of the same quantities on a logarithmic scale. $\abar$, $\omega_0$, and $\delta$ are the value of the scattering length in the absence of the modulation, the position of the resonance, and the width of the resonance, respectively.}
\label{fig:MMFR}                %
\end{figure}
%%%%%%%%%%%%%%%%%%%%%%%%%%%%%%%%%%%%%%%%%%%%%%%%%%
\paragraph{Modulated Magnetic Feshbach Resonance}
We now discuss a new mechanism, {\it modulated-magnetic Feshbach resonance} (MMFR), for resonantly enhancing the scattering length in ultracold gases. MMFR was first introduced in Ref.~\cite{Hudson_2015}. This mechanism is related to {\it modulated-magnetic spectroscopy} or {\it wiggle spectroscopy}, which has been used to measure molecular binding energies and other properties for several alkali-metal atoms \cite{Wieman2005,PhysRevLett.95.190404,Inguscio0808,Chin_2009,Khaykovich1201,Hulet1302}. In wiggle spectroscopy and MMFR, an oscillating magnetic field 
%------------------------------
\begin{equation}
B(t) = \Bbar + \Btil \cos(\Omega t)
\end{equation}
%------------------------------
is applied parallel to the spin-quantization axis of the atoms and near the transition frequency between an atom pair and a shallow bound state in the scattering channel. The resulting time-dependent scattering length $a(t)$ is given by Eq.~\eqref{eq:feshbach} with $B$ replaced by $B(t)$. For sufficiently small values of $\Btil$, the resulting scattering length is linear in $\Btil$:
%------------------------------
\begin{equation}
a(t) = \abar + \atil \cos(\Omega t),
\label{eq:aoft_1}
\end{equation}
%------------------------------
where
%------------------------------
\begin{align}
\abar &= a(\Bbar),
\nonumber\\
\atil &= a'(\Bbar)\Btil.
\end{align}
%------------------------------

The time-dependent scattering length in Eq.~\eqref{eq:aoft_1} has a profound effect on the scattering properties of the atoms. Experiments in wiggle spectroscopy have demonstrated that the oscillating field induces transitions to the molecular state which, if unstable, will then decay, leading to measurable atom loss. What had not been realized prior to Ref.~\cite{Hudson_2015} was that under the circumstances of wiggle spectroscopy experiments, the oscillating field resonantly enhances the elastic scattering properties of the atoms, giving rise to the effective frequency-dependent scattering length
%------------------------------
\begin{equation}\label{eq:mmfeshbach}
\frac{1}{a(\Omega)}=\frac{1}{\abar}\,\frac{\Omega-\Omega_0}{\Omega-\Omega_0-\delta}+i\gamma.
\end{equation}
%------------------------------
This equation is a close parallel to Eq.~\eqref{eq:feshbach}. For $\Bbar$ near a MFR, the resonance parameters $\Omega_0$, $\delta$, and $\gamma$ in Eq.~\eqref{eq:mmfeshbach} are universal in a sense that will be made precise in \ref{Chap:Applications}. Figure \ref{fig:MMFR} shows predictions for the scattering length in the vicinity of a MMFR.

In Eq.~\eqref{eq:mmfeshbach} the imaginary part of $a$ arises from collisions in which a pair of low-energy atoms emits one or more quanta of frequency $\omega$ and forms a molecule. A complex $a$ also arises when controlling the scattering length of $^{85}$Rb with MFR because the only accessible broad resonance occurs in a hyperfine configuration with a spin-relaxation channel. However, this has not prevented pioneering studies of few- and many-body physics using $^{85}$Rb atoms \cite{Wieman2002, Wieman2005, Robins2011, Jin2012, Cornish2013, Jin2014}.

The frequencies needed for MMFR are much lower than for rf/mwFR. This allows for larger amplitudes of the oscillating magnetic field. Also, because of the parallel polarization of the oscillating field, the coupling between an atom pair and the bound state can be much stronger than for rf/mwFR. Larger oscillation amplitudes combined with stronger coupling results in greater enhancement of the scattering length. In Chapters \ref{chap:atomic_scattering} and \ref{Chap:Applications}, we calculate the scattering properties of atoms in the presence of a MMFR.

%%% Local Variables:
%%% mode: latex
%%% TeX-master: "../../HudsonPhDThesis"
%%% End:

%% file: Chapters/ScatteringTheory/ScatteringTheory.tex
% !TEX root = ../HudsonPhDThesis.tex
\cleardoublepage

\chapter{Scattering theory}
\label{chap:scat}

Before examining the scattering of neutral atoms in the presence of an
oscillating magnetic field, we must first understand scattering in the
absence of such a field. Here we aim to present the essential
details. For complete discussions see
Refs.~\cite{taylor_scattering_thy, newton_scattering_thy}. We first
review the basic concepts of scattering theory in
Sec.~\ref{chap:scat::sec:theory} and apply those concepts to the
low-energy scattering of alkali-metal atoms. In Section
\ref{chap:scat::sec:universality} we examine the scattering properties
of atoms that have an $s$-wave scattering length that is much larger
than all other intrinsic length scales. We conclude in
Sec.~\ref{chap:scat::sec:EFT} by showing that systems with $s$-wave
universality are succinctly described by an effective quantum field
theory. This field-theoretic description will be particularly useful
when discussing time-periodic, zero-range interactions in Chapter
\ref{Chap:Applications}.

\section{Basic scattering theory}
\label{chap:scat::sec:theory}

In this section, we briefly review the scattering of two distinguishable particles with short-range interaction potentials. Let $H_0$, $V$, and $|\Psi(t)\rangle$ be the kinetic energy operator, the short-range interaction potential, and the wavevector for the 2-body system, respectively. To describe the fate of two incident particles, we should solve the time-dependent Schr{\"o}dinger equation (TDSE):
%-------------------------------
\begin{equation}
(i\partial_t -H_0)|\Psi(t)\rangle =  V|\Psi(t)\rangle.
\label{eq:TDSE_ti}
\end{equation}
%-------------------------------
The solution for $\ket{\Psi(t)}$ must satisfy the asymptotic boundary condition:
%-------------------------------
\begin{equation}
\ket{\Psi(t)} \overset{t\rightarrow -\infty}{\longrightarrow} e^{-i(k^2/M) t}\ket{\bm{k}},
\label{eq:ScatBC_ti}
\end{equation}
%-------------------------------
where $\bm{k}$ is the relative momentum and $M$ is the mass. $\ket{\bm{k}}$ is an eigenstate of $H_0$:
%-------------------------------
\begin{equation}
H_0\ket{\bm{k}} =  \frac{k^2}{M}\ket{\bm{k}}.
\label{eq:H0k_ti}
\end{equation}
%-------------------------------
By using the boundary condition in Eq.~\eqref{eq:ScatBC_ti}, we are tacitly assuming that the desired solution does not depend upon the width of the incident physical wave packet. As discussed in many scattering theory texts, this is always the case for stationary potentials (see e.g. Ref.~\cite{taylor_scattering_thy}).

In the following sections, we will discuss the machinery for solving and understanding the TDSE, Eq.~\eqref{eq:TDSE_ti}, subject to the asymptotic boundary condition, Eq.~\eqref{eq:ScatBC_ti}. Our goal is not to re-derive the basic results of scattering theory in full, but rather to provide a basis for comparison once we move to the problem of scattering from time-periodic potentials. Many of the results in the latter case are highly analogous to the results that we will now discuss.

\subsection{Lippmann-Schwinger equation}
\label{chap:scat::sec:thy::subsec:LSE}

We begin our discussion of scattering theory by deriving the \textit{Lippmann-Schwinger equation} (LSE). The LSE fully encodes the physics of the TDSE, Eq.~\eqref{eq:TDSE_ti}, and the asymptotic boundary condition, Eq.~\eqref{eq:ScatBC_ti}, into one equation. This equation will prove very useful in the development of the scattering formalism and as a starting point for actual calculations.

Due to conservation of energy, we expect the time-dependent Schr{\"o}dinger equation in Eq.~\eqref{eq:TDSE_ti} to have solutions of the form
%-------------------------------
\begin{equation}
\ket{\Psi(t)} = e^{-i(k^2/M)t}\ket{\Phi},
\label{eq:psiAnsatz_ti}
\end{equation}
%-------------------------------
where $\ket{\Phi}$ is time independent. Inserting this into the TDSE and dropping the overall time-dependent phase gives
%-------------------------------
\begin{align}
(k^2/M - H_0)\ket{\Phi}=
V\ket{\Phi}.
\label{eq:SEomega_ti}
\end{align}
%-------------------------------
We can immediately write down a formal solution to Eq.~\eqref{eq:SEomega_ti} that also satisfies the boundary condition in Eq.~\eqref{eq:ScatBC_ti}:
%-------------------------------
\begin{equation}
\ket{\Phi} = \ket{\bm{k}} + G_0(k^2/M) V\ket{\Phi}.
\label{eq:LSE}
\end{equation}
%-------------------------------
$G_0(E)$ is the noninteracting Green's function:
%-------------------------------
\begin{equation}
G_0(E)=\frac{1}{E - H_0 + i0^+},
\label{eq:Green}
\end{equation}
%-------------------------------
where the $+i0^+$ prescription ensures that we include only causal solutions (i.e. the propagation of particles forward in time). By acting the operator $G_0^{-1}(k^2/M)$ on both sides of Eq.~\eqref{eq:LSE}, we recover Eq.~\eqref{eq:SEomega_ti}.

Equation \eqref{eq:LSE} is the Lippmann-Schwinger equation (LSE). The LSE fully encodes the information from both the time-dependent Sch{\"o}dinger equation and the boundary condition in the asymptotic past. It is not a solution in the practical sense because the wavevector $\ket{\Phi}$ depends upon itself operated upon by $G_0(k^2/M)V$. Nevertheless, the LSE naturally leads to methods for calculating the desired physical quantities. 

\subsection{Asymptotic wavefunction}
We can extract scattering observables from knowledge about the wavefunction in the region where the distance between the particles is much larger than the range of the potential. Using the LSE, we can extract the wavefunction $\Phi(\bm{r}) = \langle \bm{r}\ket{\Phi}$ in the region of large separation.

Projecting the Lippmann-Schwinger equation into position space gives
%-------------------------------
\begin{align}
\Phi(\bm{r}) 
&= e^{i\bm{k}\cdot\bm{r}} 
+ \int d^3r'G_0(\bm{r},\bm{r'};k^2/M) \langle\mbf{r'}|V\ket{\Phi}.
\label{eq:LSER}
\end{align}
%-------------------------------
$G_0(\bm{r},\bm{r'}; E)$ is the free position-space Green's function:
%-------------------------------
\begin{align}
G_0(\bm{r},\bm{r'}; E) = \matel*{\bm{r}}{\bm{r'}}{G_0(E)}
= -\frac{M}{4\pi} \frac{e^{i (M E)^{1/2} |\bm{r}-\bm{r}'|}}{|\bm{r}-\bm{r}'|},
\label{eq:G0r}
\end{align}
%-------------------------------
where $G_0(E)$ is defined in Eq.~\eqref{eq:Green}. In the region where the particle separation $r=|\bm{r}|$ is much greater than the range of the potential, Eq.~\eqref{eq:LSER} takes the form
%-------------------------------
\begin{align}
\Phi(\bm{r}) 
&\underset{r\rightarrow\infty}{\longrightarrow} e^{i\bm{k}\cdot\bm{r}}  -\frac{M}{4\pi}\frac{e^{ik r}}{r} \int d^3r'e^{-i k\bm{\hat{r}}\cdot\bm{r}'} \bra{\bm{r'}}V|\Phi\rangle \nonumber \\
&= e^{i\bm{k}\cdot\bm{r}} -\frac{M}{4\pi}\frac{e^{ ik r}}{r} \langle k\mbf{\hat{r}}|V|\Phi\rangle.
\label{eq:asymptoticphi_ti}
\end{align}
%-------------------------------
To obtain Eq.~\eqref{eq:asymptoticphi_ti} we have dropped all terms that fall faster than $1/r$. This equation manifestly satisfies the asymptotic boundary condition in Eq.~\eqref{eq:ScatBC_ti}, since the only incoming plane wave component equals the one specified by the boundary condition. All other components are spherical outgoing waves.

Defining the scattering amplitude
%-------------------------------
\begin{equation}
f(\bm{p},\mbf{k})=-\frac{M}{4\pi} \bra{\bm{p}}V|\Phi\rangle,
\label{eq:amplitude_deff_ti}
\end{equation}
%-------------------------------
we arrive at the standard form of the asymptotic wavefunction $\Phi_\infty(\bm{r})$:
%-------------------------------
\begin{align}
\Phi(\bm{r}) 
&\underset{r\rightarrow\infty}{\longrightarrow} \Phi_\mathrm{\infty}(\bm{r})  = e^{i\bm{k}\cdot\bm{r}} 
+f(k\bm{\hat r}, \bm{k} )\frac{e^{ ik r}}{r}.
\label{eq:asymptoticphi_ti_std}
\end{align}
%-------------------------------
In Eq.~\eqref{eq:amplitude_deff_ti}, the dependence on the incident relative momentum $\bm{k}$ is hidden in $\ket{\Phi}$. $f(\bm{p},\mbf{k})$ is the scattering amplitude for incident scatterers with relative momentum $\bm{k}$ and center of mass energy $k^2/M$ to exit with momentum $\bm{p}$ and energy $p^2/M$. Note that the asymptotic wavefunctions only depend on amplitudes that satisfy the on-shell condition $p^2/M = k^2/M$.

\subsection{Scattering cross section}
The quantum mechanical scattering from a short-range potential is fully characterized by the scattering amplitude $f(k\bm{\hat r},\mbf{k})$ defined in Eq.~\eqref{eq:amplitude_deff_ti}. In particular, we are interested in the differential cross section $d\sigma(k)/d\Omega$ for the incoming scatterers with momentum $\bm{k}$ to exit into the solid angle $d\Omega$. The differential cross section is:
%-------------------------------
\begin{equation}
d\sigma = \frac{R\cdot d\Omega}{|\bm{j}_{\mathrm{inc}}|},
\label{eq:dsigmadomega_1_ti}
\end{equation}
%-------------------------------
where $R$ is the differential rate of probability flow, and $|\bm{j}_{\mathrm{inc}}|$ is the magnitude of the incident particle current density. $R\cdot d\Omega$ equals the scattered current density into the solid angle $d\Omega$, $\bm{j}_{\mathrm{sc}}$, dotted with the corresponding area element. Dividing out an overall factor of $d\Omega$ this gives
%-------------------------------
\begin{equation}
R = \bm{j}_{\mathrm{sc}}\cdot \bm{\hat r} r^2
\label{eq:R1_ti}
\end{equation}
%-------------------------------
The scattered current density $\bm{j}_{\mathrm{sc}}$ is determined by the scattered part $\Phi_{\mathrm{sc}}(\bm{r})$ of the asymptotic wavefunction $\Phi_\infty(\bm{r})$ defined in Eq.~\eqref{eq:asymptoticphi_ti_std}:
%-------------------------------
\begin{align}
 \Phi_{\mathrm{sc}}(\bm{r}) = f(k\bm{\hat r},\bm{k})\frac{e^{ ik r}}{r}.
\end{align}
%-------------------------------
The scattered current density at large separations $r$ is
%-------------------------------
\begin{align}
\bm{j}_{\mathrm{sc}}
&\underset{r\rightarrow\infty}{\longrightarrow} \frac{1}{iM}\big[{\Phi_{\mathrm{sc}}(\bm{r})}^*\nabla\Phi_{\mathrm{sc}}(\bm{r})-\Phi_{\mathrm{sc}}(\bm{r})\nabla{\Phi_{\mathrm{sc}}(\bm{r})}^*\big]
\nonumber\\
&=\frac{2k}{M}\frac{\bm{\hat r}}{r^2}\big|f(k\bm{\hat r},\bm{k})\big|^2 + \mathcal{O}\left(1/r^3\right).
\end{align}
%-------------------------------
Inserting this result into Eq.~\eqref{eq:R1_ti}, we find
%-------------------------------
\begin{equation}
R = \frac{2k}{M}\big|f(k\bm{\hat r},\mbf{k})\big|^2.
\label{eq:R2_ti}
\end{equation}
%-------------------------------
The magnitude of the incident current density is simply
%-------------------------------
\begin{equation} 
|\bm{j}_{\mathrm{inc}}|= \frac{2k}{M}.
\label{eq:jinc_t1}
\end{equation}
%-------------------------------
Inserting $R$ and $|\bm{j}_{\mathrm{inc}}|$ into Eq.~\eqref{eq:dsigmadomega_1_ti}, we obtain the differential cross section
%-------------------------------
\begin{equation}
\frac{d\sigma}{d\Omega} = \big|f(k\bm{\hat r},\mbf{k})\big|^2.
\label{eq:dsigmadomega_2_ti}
\end{equation}
%-------------------------------

\subsection{Integral equation for the scattering amplitude}

The scattering amplitudes defined in Eq.~\eqref{eq:amplitude_deff_ti} depend on the unknown wavevector $\ket{\Phi}$, which is fully determined by the Lippmann-Schwinger Equation, Eq.~\eqref{eq:LSE}. Since the scattering observables only depend on the amplitude, it is convenient to convert the LSE for the wavevector $\ket{\Phi}$ into a similar equation for $f(\bm{p},\mbf{k})$. By premultiplying Eq.~\eqref{eq:LSE} with $-(M/4\pi)\bra{\bm{p}}V$, we find an integral equation for the scattering amplitudes:
%-------------------------------
\begin{align}
f(\mbf{p},\mbf{k})
&= -\frac{M}{4\pi}\langle \mbf{p}|V|\mbf{k}\rangle -\frac{M}{4\pi}\bra{\bm{p}}VG_0(k^2/M)
V\ket{\Phi}\nonumber \\
&= -\frac{M}{4\pi}\langle \mbf{p}|V|\mbf{k}\rangle + \int \frac{d^3q}{(2\pi)^3}
\langle \bm{p}|V|\bm{q}\rangle G_0(q, k^2/M) f(\mbf{q},\mbf{k}).
\label{eq:LSEf}
\end{align}
%-------------------------------
$G_0(q, E)$ is the free momentum-space Green's function
%-------------------------------
\begin{align}
G_0(q, E) = \frac{1}{E-q^2/M+i0^+}.
\label{eq:Gmomentum}
\end{align}
%-------------------------------
Equation \eqref{eq:LSEf} can be used as the starting point for exact or approximate calculations of the scattering amplitudes. Once the amplitudes $f(\bm{p},\bm{q})$ are known, the on-shell amplitudes that appear in Eq.~\eqref{eq:asymptoticphi_ti} are obtained by setting $\mbf{p}=k\hat{\bm{r}}$, where $\bm{k}$ is the relative momentum of the incoming particles.

\subsection{Partial wave expansion}
\label{chap:scat::sec:theory::subsec:partialwave}
In ultra cold atomic physics, we are often interested in the scattering of two atoms at low center-of-mass energy $k^2/M$ compared to the van der Waals energy scale $E_\mathrm{vdW}$ (Eq.~\ref{eq:EvdW}). In that case, it is convenient to perform a partial wave expansion of the scattering amplitudes because higher partial waves are suppressed by powers of the center-of-mass energy. In practice only a small number of partial waves are needed.

We will assume in what follows that the interaction potential is \textit{central}, i.e. $V(\bm{r})=V(r)$. As we will see, this assumption is justified in our case since, to a very good approximation, the interaction potentials between cold, neutral alkali atoms are central. The matrix elements $\langle\mbf{p}|V|\bm{q}\rangle$ and the scattering amplitudes $f(\mbf{p},\mbf{q})$ then depend on $\bm{\hat p}\cdot\bm{\hat q}$ but not upon the rotation angle around the axis of the incoming relative momentum. The scattering amplitudes and potential matrix elements therefore have expansions of the form
%-------------------------------
\begin{align}
f(\mbf{p},\mbf{q})
 &= 4\pi\sum\limits_{l,m}f^{l}(p,q)Y_{lm}^*(\Omega_{\bm{p}})Y_{lm}(\Omega_{\bm{q}}),
\nonumber\\
\langle\mbf{p}|V|\bm{q}\rangle
 &=4\pi\sum\limits_{l,m}V^{l}(p,q)Y_{lm}^*(\Omega_{\bm{p}})Y_{lm}(\Omega_{\bm{q}}),
\label{eq:sphericalHarmonics_ti}
\end{align}
%-------------------------------
where the $l$ sum runs from $0$ to $\infty$ and the $m$ sum runs from $-l$ to $l$. We have assumed that the expansion coefficients $f^l(p,q)$ and $V^l(p,q)$ do not depend upon the $m$ quantum number. This is equivalent to our assumption that the interaction potential is spherically symmetric. With this assumption, the $m$ sum can be performed analytically using the addition theorem for spherical harmonics:
%-------------------------------
\begin{align}
\sum\limits_{m}Y_{lm}^*(\Omega_{\bm{p}})Y_{lm}(\Omega_{\bm{q}}) = \frac{2l+1}{4\pi}P_l(\bm{\hat p}\cdot\bm{\hat q}),
\label{eq:addition_theorem}
\end{align}
%-------------------------------
where $P_l(x)$ is the $l^\mathrm{th}$ Legendre polynomial. Using this identity, Eqs.~\eqref{eq:sphericalHarmonics_ti} simplify to 
%-------------------------------
\begin{align}
f(\mbf{p},\mbf{q})
 &= \sum\limits_{l}f^{l}(p,q)(2l+1)P_l(\bm{\hat p}\cdot\bm{\hat q}), 
\nonumber\\
\langle\mbf{p}|V|\bm{q}\rangle
 &=\sum\limits_{l}V^{l}(p,q)(2l+1)P_l(\bm{\hat p}\cdot\bm{\hat q}).
\label{eq:LegendreExapansion_ti}
\end{align}
%-------------------------------
The Legendre polynomials obey the orthogonality relation
%-------------------------------
\begin{align}
\int\limits_{-1}^1 dx P_m(x)P_n(x)=\frac{2}{2n+1}\delta_{mn}.
\label{eq:Pcompleteness}
\end{align}
%-------------------------------
Using this identity, the partial wave components are
%-------------------------------
\begin{align}
f^l(p,q)
 &=\frac{1}{2}\int\limits_{-1}^1 d(\bm{\hat p}\cdot\bm{\hat q}) P_l(\bm{\hat p}\cdot\bm{\hat q})f(\bm{p},\bm{q}),\nonumber\\
V^l(p,q)
 &=\frac{1}{2}\int\limits_{-1}^1 d(\bm{\hat p}\cdot\bm{\hat q}) P_l(\bm{\hat p}\cdot\bm{\hat q})\langle\mbf{p}|V|\mbf{k}\rangle.
\label{eq:partialWaveComponents_ti}
\end{align}
%-------------------------------
Notice that the $s$-wave component ($l=0$) is the simple angle average of the original function since $P_0(x)=1$. 

Under the assumption that $V(\bm{r})=V(r)$, the angular momentum components of the wavefunction scatter independently. This allows us to re-express the LSE in terms of the individual angular momentum components.  Inserting the spherical harmonic expansions of $f(\mbf{p},\mbf{q})$ and  $\langle\mbf{p}|V|\bm{q}\rangle$ in Eqs.~\eqref{eq:sphericalHarmonics_ti} into the LSE, Eq.~\eqref{eq:LSEf}, and projecting out the $l^{\mathrm{th}}$ angular momentum component, we obtain the LSE for the partial wave scattering amplitudes $f^l(p,k)$:
%-------------------------------
\begin{align}
f^l(p,k)
&= -\frac{M}{4\pi}V^l(p,k)
+\frac{1}{2\pi^2}\int\limits_0^\infty dqq^2
\frac{V^l(p,q)}{k^2/M - q^2/M + i0^+} f^l(q,k).
\label{eq:FLSEpartialwave_ti}
\end{align}
%-------------------------------
The left-hand side and the first term on the right-hand side follow directly using the orthogonality relation for Legendre polynomials, Eq.~\eqref{eq:Pcompleteness}. To arrive at the second term on the right hand side, we first used the orthogonality relation for the spherical harmonics,
%-------------------------------
\begin{align}
\int\limits d\Omega \, Y_{lm}(\Omega)Y_{l'm'}(\Omega)=\delta_{ll'}\delta_{mm'},
\label{eq:Ycompleteness}
\end{align}
%-------------------------------
to integrate over the $\bm{q}$ angles. We then apply the addition theorem in Eq.~\eqref{eq:addition_theorem} and finally the orthogonality relation for the Legendre polynomials, Eq.~\eqref{eq:Pcompleteness}. For the full derivation, see Appendix \ref{App:partial_wave}.

\subsection{Effective range expansion}
\label{chap:scat::sec:thy::subsec:ERE}
For low-energy scattering from short-range potentials with no power-law tail, it can be shown that the $s$-wave ($l=0$) scattering amplitudes defined in Eq.~\eqref{eq:partialWaveComponents_ti} can be expressed as an expansion in powers of $k^2$ using the so-called \textit{effective range expansion}:
%-------------------------------
\begin{equation}
f^0(k)^{-1}+ik = -\frac{1}{a} + \frac{1}{2}{r_s}k^2  + \mathcal{O}(k^4),
\label{eq:ERE_ti}
\end{equation}
%-------------------------------
where $a$ and $r_s$ are the $s$-wave scattering length and effective range. In practice, the first $N$ resonance parameters can be extracted by fitting a polynomial of order $N$ in $k^2$ to $f(k)^{-1}+ik$ at small values of $k$.  The higher order terms in the effective range expansion can be calculated in this fashion, but they are rarely needed.

The van der Waals length, $\RvdW$, defined in Eq.~\eqref{eq:RvdW} provides the natural length scale for the effective range coefficients for low-energy scattering of alkali atoms. By dimensional analysis, any effective range expansion parameter can be expressed as a power of $\RvdW$ multiplied by a dimensionless coefficient. In the absence of an enhancement mechanism, we expect the dimensionless coefficient to be order 1. However, if the low-energy scatterers are nearly degenerate with and coupled to a bound or resonance state, the dimensionless prefactor can be orders of magnitude larger or smaller than 1, leading to a resonance or dissonance in the scattering amplitude.

The benefit of the effective range expansion is that it makes the momentum dependence completely explicit in terms of a small number of effective range expansion coefficients $a$, $r_s$, \textit{etc}. At low momentum, the dependence on higher order terms is suppressed by higher powers of $k^2$. Thus, this parametrization allows one to specify the full momentum-dependent scattering observables in terms of a small number of constants. For example, the low-energy $s$-wave cross section is
%-------------------------------
\begin{equation}\label{eq:crosssec_ERE_ti}
\sigma^0(k) = \frac{4\pi|a|^2}{\left|1 + iak -\frac{1}{2}ar_sk^2+\mathcal{O}(k^4)\right|^2}.
\end{equation}
%-------------------------------
For sufficiently low momenta, the cross section is simply $4\pi |a|^2$.

So far, we have only considered the effective range expansion for the $s$-wave scattering amplitude. We did this under the assumption that the higher partial waves are suppressed by powers of the scattering energy. If the interaction potential has a long-range tail, higher partial waves are not necessarily suppressed at low energy. However, it can be shown that for potentials that scale as $1/R^6$ for large atomic separations $R$, the partial wave expansion and subsequent effective range expansion still hold up to order $k^2$, where $k$ is the relative momentum of the scatterers. In general, the contribution to the scattering amplitude at higher orders receives corrections from all partial waves.

\subsection{\textit{S}-matrix}
It is common in scattering theory to discuss the properties of the scattering matrix or $S$-matrix, whose matrix elements relate the amplitude of incoming states in the asymptotic past to outgoing states in the asymptotic future. We represent the asymptotic state with momentum $\bm{p}$ as $\ket{\bm{p}}$. The elements of the $S$-matrix have the form
%------------------------------
\begin{equation}
  \matel{\bm{p}}{\bm{q}}{S} = (2\pi)^3\delta(\bm{p}-\bm{q})+\frac{i}{2\pi M}\delta(q^2/M-p^2/M)f(\bm{p},\bm{q}),
  \label{eq:Sf1}
\end{equation}
%------------------------------
where $f(\bm{p}, \bm{q})$ is the scattering amplitude defined in Eq.~\eqref{eq:amplitude_deff_ti}. The relationship between the scattering amplitude and the $S$-matrix is
%------------------------------
\begin{equation}
\matel{\bm{p}}{\bm{q}}{S-1} = \frac{i}{2\pi M}\delta(q^2/M-p^2/M)f(\bm{p},\bm{q}).
\label{eq:Stof_ti}
\end{equation}
%------------------------------

Since for describing cold atoms we are often interested in the partial wave amplitudes $f^l(p,q)$, it is convenient to define the partial wave $S$-matrix element $S^l(E)$ as
%------------------------------
\begin{equation}
\matel{E'l'm'}{Elm}{S} = \delta(E'-E)\delta_{l'l}\delta_{m'm}S^l(E).
\label{eq:Spartial_ti}
\end{equation}
%------------------------------
The states ${\ket{Elm}}$ are the simultaneous eigenstates of $H_0$, $L^2$, and $L_z$. The momentum dependence of the left-hand and right-hand sides of Eq.~\eqref{eq:Stof_ti} can be expanded into partial wave components using 
%------------------------------
\begin{equation}
1 = \int dE \sum\limits_{lm}\, \ket{Elm}\bra{Elm}
\label{eq:Elmcompleteness}
\end{equation}
%------------------------------
along with
%------------------------------
\begin{equation}
\braket{\bm{p}}{Elm} =\delta(p^2/M-E)\frac{Y_{lm}(\Omega_{\bm{p}})}
{\sqrt{Mp}}.
\label{eq:Elmexpansion}
\end{equation}
%------------------------------
Using the definition of $S^l(E)$ in Eq.~\eqref{eq:Spartial_ti}, we extract the relationship between the partial-wave scattering amplitude and the partial-wave $S$-matrix element:
%------------------------------
\begin{equation}
f^l(p,q) = \frac{S^l(E)-1}{2i\sqrt{pq}}.
\label{eq:ftoS_partial_ti}
\end{equation}
%------------------------------
Furthermore, the on-shell condition enforced by the energy $\delta$-functions in Eqs.~\eqref{eq:Stof} and \eqref{eq:Spartial} requires that $p=q$. Thus, the notation $f^l(p,q)$ is not ideal for representing on-shell scattering amplitudes. Nevertheless, we retain this notation since the off-shell amplitudes appear in the LSE for the partial-wave scattering amplitudes, Eq.~\eqref{eq:FLSEpartialwave_ti}.

\section{Universality for atoms with large scattering length}
\label{chap:scat::sec:universality}
As discussed in Sec.~\ref{chap:scat::sec:thy::subsec:ERE}, low energy 2-body scattering can be described systematically using the effective range expansion in the scattering energy. Generically, the values of the effective range expansion coefficients are set by the van der Waals length $\RvdW$. We now consider the case of an unnaturally large scattering length $|a|\gg \RvdW$. In this context, we introduce the concept of universality.

Consider two atoms scattering with the unnaturally large scattering length $a$ ($|a|\gg \RvdW$) and with energy small compared to the van der Waals energy, $\EvdW$, defined in Eq.~\eqref{eq:EvdW}. The properties of such a system are completely determined by $a$. These systems are said to be {\it universal} in that the properties of two apparently disparate systems depend upon $a$ in the same way, irrespective of the microscopic details of the systems. The scattering amplitude for such a system follows from Eq.~\eqref{eq:ERE_ti} after setting $r_s=0$, giving the universal scattering amplitude
%------------------------
\begin{equation}
f^0(k) =\frac{1}{-1/a-ik}.
\label{fk}
\end{equation}
%------------------------
The cross section is then
%------------------------
\begin{equation}
\sigma^0(k) =\frac{4\pi a^2}{1+(ak)^2}.
\label{sigma}
\end{equation}
%------------------------
The momentum-dependent cross section is determined by a single parameter: $a$. 

If the scattering length $a$ is large and positive, the 2-atom spectrum includes a diatomic molecule (\textit{dimer}) whose properties are completely determined by $a$. The binding energy of the dimer is determined by the location of the pole in the scattering amplitude $f^0(k)$, Eq.~\eqref{fk}, for complex values of the momentum $k$. Generally, if a scattering amplitude has a pole at the complex momentum $i\kappa$ with $\kappa>0$, then the potential supports a bound state with binding energy $\kappa^2/M$. The scattering amplitude in Eq.~\eqref{fk} has a pole at $k=i/a$. For $a>0$, the potential supports a bound state (the dimer) with energy
%------------------------
\begin{equation}
E_{\rD} =-\frac{1}{Ma^2},\hspace*{0.06\columnwidth} a>0.
\label{E_D}
\end{equation}
%------------------------
The size of the dimer is roughly $a$. In addition to the dimer, there may also be diatomic molecules with binding energies of order $1/(M \RvdW^2)$ or larger. The properties of these {\it deep dimers} depend on the details of the scattering potential and are non-universal in the sense described above.

The limit of large scattering length $|a|\gg \RvdW$ can alternatively be expressed as the \textit{zero-range limit} $\RvdW\to 0$. Conceptually, the zero-range limit can be achieved by taking the range of the interaction potential to zero while simultaneously increasing its depth so that the scattering length remains fixed. This limit is independent of the shape of the potential. In the zero-range limit, the universal scattering amplitude in Eq.~\eqref{fk} becomes exact up to arbitrarily high energies. The universal expression for the binding energy in Eq.~\eqref{E_D} also becomes exact. If there are any deep dimers, their binding energies become infinitely large. Since the universal results are the same in both limits, we will sometimes use the phrases {\it large scattering length}, {\it zero range}, and {\it universal} interchangeably.

In the limit $a\to \pm \infty$, the universal cross section approaches $4\pi/k^2$, which is the maximum value allowed by unitarity. The limit $a\to\pm\infty$ is therefore called the {\it unitary limit}. In this limit, there is no length scale associated with the interactions. Thus the system has a symmetry under scaling the spatial coordinates by an arbitrary positive factor $\lambda$ and the time by a factor $\lambda^2$. This symmetry is called {\it scale invariance}. 

The scale invariance of the unitary limit manifests itself at finite scattering length by simple scaling behavior under simultaneous scaling of $a$ and kinematic variables. For example, when $a$ and the momentum variable $k$ are scaled by the factors $\lambda$ and $\lambda^{-1}$, the cross section in Eq.~\eqref{sigma} is changed by a factor $\lambda^2$: $\sigma(\lambda^{-1} k;\lambda a)=\lambda^2 \sigma(k;a)$. The binding energy in Eq.~\eqref{E_D} also shows the scaling behavior. When $a$ is scaled by $\lambda$, $E_{\rD}$ is changed by a factor $\lambda^{-2}$: $E_{\rD}(\lambda a)=\lambda^{-2} E_{\rD}(a)$. This scaling behavior is a general feature of the system with large scattering length.  It follows from the fact that the scattering length $a$ is the only relevant interaction parameter at low energies \cite{Braaten:2004rn}.

One reason that universality is important is because it relates phenomena in various fields of physics. There are examples of systems with large scattering lengths in nuclear physics and high energy physics as well as in atomic physics. In nuclear physics, the best example is the neutron, whose two spin states interact with a large negative scattering length. In high energy physics, a good example is the charm mesons $D^{*0}$ and $\bar{D}^0$, which form a very weakly bound state called the $X(3872)$ and therefore must have a large positive scattering length. A classic example in atomic physics is $^4$He atoms, whose scattering length is about $+200~a_0$, which is much larger than the van der Waals length scale $R_{\rm vdW}\approx 10~a_0$.

Universality is particularly important in atomic physics because it is possible to tune the scattering length experimentally. As we discussed in Sec.~\ref{chap:exper::sec:tuning}, this can be accomplished with various techniques that induce a resonant coupling between the scattering atoms and a bound state. These techniques all use some control field that can be tuned to bring the scattering atoms to resonance. Near the resonance, the magnitude of the scattering length can be made extremely large, effectively bringing the atoms into the universal regime.

\section{Effective field theory for atoms with large scattering length}
\label{chap:scat::sec:EFT}
We now briefly discuss a low-energy effective field theory (EFT) for
atoms with large scattering length. As with the rest of this thesis,
we limit our discussion to equal-mass, distinguishable fermions which
we now label as types $1$ and $2$. The generalization to different
particle statistics is straightforward. For a more thorough discussion
of this topic, see \cite{Braaten:2004rn}.

\subsection{Lagrangian density}
We label the operators which annihilate either an atom of type $1$ or $2$ at the time $t$ and the position $\mbf{r}$ as $\psi_1(t,\mbf{r})$ or $\psi_2(t,\mbf{r})$, respectively (time and position arguments are suppressed from now on). We assume that the only relevant interactions are binary contact interactions between atoms of type 1 and type 2. The Lagrangian density for such a system is
%-----------------------------
\begin{equation}
\mathcal{L}=
\sum\limits_{\sigma=1}^2\psi^\dagger_\sigma\left( i\partial_t\
+\frac{\nabla^2}{2M}\right)\psi_\sigma 
-\frac{g}{M}\psi^\dagger_2\psi^\dagger_1\psi_1\psi_2,
\label{eq:L}
\end{equation}
%-----------------------------
where $g$ is the bare coupling constant. If the atoms are fermions, the field operators $\psi_1$ and $\psi_2$ obey the equal-time anticommutation relations
%-----------------------------
\begin{subequations}
\begin{align}
\{\psi_\alpha(t,\mbf{r}),\psi_\beta(t,\mbf{r'})\} &= 0, \label{eq:anticomm_zero}\\
\{\psi_\alpha(t,\mbf{r}),\psi_\beta^\dagger(t,\mbf{r'})\} &= \delta_{\alpha,\beta}\,\delta^3(\mbf{r}-\mbf{r'}).
\label{eq:anticomm_delta}
\end{align}
\end{subequations}
%-----------------------------

\subsection{Feynman rules}
The physical properties of the atomic system are determined by summing over Feynman diagrams. These diagrams encapsulate the possible particle dynamics specified by the Lagrangian density $\mathcal{L}$, defined in Eq.~\eqref{eq:L}. The terms in $\mathcal{L}$ are represented as diagrammatic elements which can be combined with other elements to form a diagram representing some physical process. We divide these elements into two categories: \textit{propagators} and \textit{vertices}.

A propagator in a diagram represents the amplitude for particle propagation. The propagators for $\psi_1$ or $\psi_2$, which correspond to the terms under the summation in Eq.~\eqref{eq:L}, are simply the imaginary number $i$ multiplied by the free momentum space Green's function $G_0(q,E)$ defined in Eq.~\eqref{eq:Gmomentum} for a particle with momentum $q$ and energy $E$. In diagrams, a propagator is represented as a solid line. Technically, each line should have a label representing the fermion type. For our purposes, however, we can infer these labels from the context. The propagator rule is
%-----------------------------
\begin{align}
\imineq{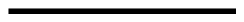}{0.4}\,\, =\,\,\frac{i}{E-q^2/M +i0^+}.
\end{align}
%-----------------------------

A vertex in a diagram represents the amplitude for particle interactions. $\mathcal{L}$ contains a single contact interaction vertex between particles of type 1 and 2. The rule for this vertex equals the imaginary number $i$ multiplied by the prefactor of the quartic term in $\mathcal{L}$. In diagrams, this vertex has two incoming lines of type 1 and type 2 and two outgoing lines of type 1 and type 2. The rule for the vertex is
%-----------------------------
\begin{align}
\imineq{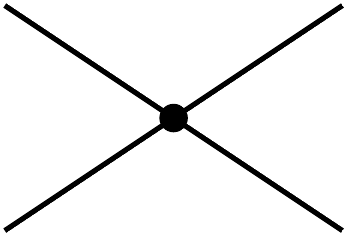}{10}\,\, = \,\,-\frac{ig}{M}.
\end{align}
%-----------------------------

\subsection{Lippmann-Schwinger equation}
%%%%%%%%%%%%%%%%%%%%%%%%%%%%%%%%%%%%%%%%%%%%%%%%%%%%%%
\begin{figure}[t]
\centering
\includegraphics[width=0.6\textwidth]{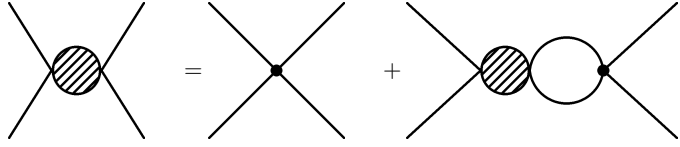}
\caption[Diagrammatic representation of the Lippmann Schwinger Equation]{Diagrammatic representation of the LSE for the scattering amplitude. The blob represents the imaginary number $i$ multiplied by the $s$-wave scattering amplitude $f^0(k)$, defined in Eq.~\eqref{eq:sphericalHarmonics_ti}.}
\label{fig:LSEDiagram}
\end{figure}
%%%%%%%%%%%%%%%%%%%%%%%%%%%%%%%%%%%%%%%%%%%%%%%%%%%%%%

The value of the bare coupling constant $g$ is chosen to reproduce the scattering length $a$. The scattering amplitude $f^0(k)$ can be calculated to all orders in $g$ by solving the Lippmann-Schwinger integral equation represented diagrammatically in Fig.~\ref{fig:LSEDiagram}. In the figure the gray blob represents  $(4\pi i / M)f^0(k)$. By iteratively replacing the blob on the right-hand side by the entire right-hand side, it is easy to see that the diagrammatic equation includes the effects of an arbitrary number of intermediate contact interactions. Using the Feynman rules the LSE becomes
%-------------------------------
\begin{align}
f^0(k)
&= -\frac{g}{4\pi}
-f^0(k)\frac{g}{2\pi^2} I(k^2/M),
\label{eq:LSE_contact}
\end{align}
%-------------------------------
where the function $I(E)$ is 
%------------------------------
\begin{align}
I(E) = \int\limits_0^\Lambda dq\frac{q^2}{q^2 - ME - i0^+} = -\frac{\pi}{2}\sqrt{-ME-i0^+} + \Lambda.
\label{eq:Iintegral}
\end{align}
%------------------------------
In the definition of $I(E)$, we introduced an upper cutoff $\Lambda$ in the momentum magnitude $q$. In the end we must take the limit $\Lambda\to\infty$.  Though $I(E)$ is ill-defined in this limit, the amplitude $f^0(k)$ is nevertheless well defined (as we will show). 

Solving Eq.~\eqref{eq:LSE_contact} for $f^0(k)$, we find
%-------------------------------
\begin{align}
f^0(k)
&= \left(\frac{-4\pi}{g} - \frac{2\Lambda}{\pi} - ik \right)^{-1},
\label{eq:f0_EFT_sol1}
\end{align}
%-------------------------------
where we have used $\sqrt{-k^2-i0^+} = -ik$ for real-valued $k$. Inserting
%-----------------------------
\begin{equation}
g = \frac{4\pi}{1/a-2\Lambda/\pi}
\label{eq:g}
\end{equation}
%-----------------------------
into Eq.~\eqref{eq:f0_EFT_sol1}, the cutoff dependence cancels, and we obtain the universal form of the scattering amplitude $f_0(k)$ in Eq.~\eqref{fk}. Comparing with Eq.~\eqref{eq:ERE_ti}, we see that the scattering length is indeed $a$ and that the effective range is 0. 
%-------------------------------
\begin{align}
f^0(k)
&= \frac{1}{-1/a - ik}.
\label{eq:f0_EFT_sol2}
\end{align}
%-------------------------------
In Chapter~\ref{Chap:Applications} we consider the case of a time-periodic contact interaction controlled with a time periodic scattering length $a(t)$. We use the effective field theory outlined above to describe the time-dependent contact interaction by replacing $a$ with $a(t)$ in Eq.~\eqref{eq:g}.

%%% Local Variables:
%%% mode: latex
%%% TeX-master: "../../HudsonPhDThesis"
%%% End:

%% file: Chapters/Floquet/Floquet.tex
% !TEX root = ../HudsonPhDThesis.tex
\cleardoublepage

\chapter{Floquet theory}
\label{Chap:Floquet}

Floquet theory, named for the French mathematician Gaston Floquet, provides a natural framework for dealing with Hamiltonians that are periodic in time. The application of Floquet theory to the solution of Schr{\" o}dinger's equation was developed in the pioneering work of Jon Shirley \cite{Shirley_1963, Shirley_1965} and later extended to the case of multiple modes \cite{many_mode_floquet, many_mode_floquet_2, many_mode_floquet_3, many_mode_floquet_4}. Here we introduce the basic concepts of single-mode Floquet theory with a focus on those ideas that will be most relevant in our later discussions. For more thorough reviews, see Refs.~\cite{Chu85a, Joachain_2014}.

\section{Formalism}
Consider a system with a Hamiltonian of the form
%-------------------------------
\begin{equation}
H(t) = H_0 + V(t),
\end{equation}
%-------------------------------
where $H_0$ is the kinetic energy operator and $V(t) = V(t+T)$ for some period $T$. We will sometimes refer to systems described by Hamiltonians of this form as \textit{Floquet systems}. The periodicity of $V(t)$ allows us to decompose $V(t)$ into its harmonic components:
%-------------------------------
\begin{equation}
V(t) = \sum\limits_{n=-\infty}^{\infty} V_n e^{-in\Omega t},
\end{equation}
%-------------------------------
where $\Omega = 2\pi/T$ is the fundamental angular frequency. Using the identity
%-------------------------------
\begin{align}
\frac{1}{T}\int\limits_0^T dt e^{i(m-n)\Omega t} = \delta_{m,n},
\label{eq:diracdelta}
\end{align}
%-------------------------------
where $\delta_{m,n}$ is the Kronecker delta function, one can easily check that the harmonic components are
%-------------------------------
\begin{equation}
V_n = \frac{1}{T}\int\limits_{0}^{T}d\tau V(\tau)e^{in\Omega \tau}.
\label{eq:Vn}
\end{equation}
%-------------------------------

Let $\ket{\Psi_\alpha(t)}$ be a solution to the time-dependent Schr{\"o}dinger equation (TDSE) with time-periodic hamiltonian $H(t)$:
%-------------------------------
\begin{equation}
\left[H(t) -i\partial_t\right] \ket{\Psi_\alpha(t)} = 0.
\label{eq:TDSE}
\end{equation}
%-------------------------------
Here $\alpha$ indexes the set of solutions. The solution of interest is selected by the relevant boundary condition. For example, in scattering problems the particular solution is chosen to match the known incoming wavefunction. According to \textit{Floquet's theorem} \cite{Arfken_2005}, Eq.~\eqref{eq:TDSE} has solutions of the form 
%-------------------------------
\begin{equation}
\ket{\Psi_\alpha(t)} = e^{-i\epsilon_\alpha t}\ket{\Phi_\alpha(t)},
\label{eq:Psi_Ansatz}
\end{equation}
%-------------------------------
for some real parameter $\epsilon_\alpha$ termed the \textit{quasienergy}, and where $\ket{\Phi_\alpha(t)}$ is periodic:
%-------------------------------
\begin{equation}
\ket{\Phi_\alpha(t)} = \ket{\Phi_\alpha(t+T)}.
\end{equation}
%-------------------------------
This periodicity implies that $\ket{\Phi_\alpha(t)}$ can be expanded as a Fourier series:
%-------------------------------
\begin{equation}
\ket{\Phi_\alpha(t)}=\sum\limits_{n=-\infty}^{\infty} \ket{\phi_\alpha^n} e^{-in\Omega t}.
\label{eq:phiModeExpansion}
\end{equation}
%-------------------------------
Substituting Eq.~\eqref{eq:Psi_Ansatz} into the TDSE gives the \textit{Floquet eigenvalue equation} (FEE):
%-------------------------------
\begin{equation}
\mathcal{H}(t)\ket{\Phi_\alpha(t)} = \epsilon_\alpha \ket{\Phi_\alpha(t)},
\label{eq:FEE}
\end{equation}
%-------------------------------
where we have defined the \textit{Floquet Hamiltonian} $\mathcal{H}(t)$
%-------------------------------
\begin{equation}
\mathcal{H}(t) = H(t) -i\partial_t,
\end{equation}
%-------------------------------
As the name suggests, the FEE is an eigenvalue equation for the eigenvalues $\epsilon_\alpha$ and the eigenstates $\ket{\Phi_\alpha(t)}$. We make this connection concrete by noting that the Hilbert space for the hermitian operator $\mathcal{H}(t)$ is the product space $\mathcal{R}\otimes\mathcal{T}$, where $\mathcal{R}$ is the space of square-integrable functions in configuration space and $\mathcal{T}$ is the space of periodic functions with period $T$. Thus we see that for Floquet systems the problem of determining the time-evolution of a state vector through configuration space can be reexpressed as a stationary eigenvalue problem on $\mathcal{R}\otimes\mathcal{T}$. In effect, this allows us to use the machinery of time-independent quantum mechanics to solve the time-dependent problem!

From the FEE in Eq.~\eqref{eq:FEE}, we see that the state vector
%-------------------------------
\begin{equation}
\ket{\Phi_{\alpha,m}(t)} = e^{-im\Omega t} \ket{\Phi_\alpha(t)},
\end{equation}
%-------------------------------
for any integer $m$ yields the same solution $\ket{\Psi_{\alpha}(t)}$ but with $\epsilon_\alpha$ replaced by $\epsilon_\alpha + m\Omega$. Thus we see that the quasienergy $\epsilon_\alpha$ associated with a solution $\ket{\Psi_\alpha(t)}$ is only uniquely defined modulo the oscillation frequency $\Omega$, hence the term \textit{quasienergy}. This arbitrariness is a necessary consequence of the fact that $H(t)$ does not have general time-translational invariance but only invariance under translations by integer multiples of the periodicity $T$. This property of the Floquet solutions implies that the set of quasienergies $\{\epsilon_\alpha\}$ can be mapped into the range $[0, \Omega)$, often termed the \textit{first Brillouin zone}.

To determine the eigenvalues $\epsilon_\alpha$ and the eigenstates $\ket{\Phi_\alpha(t)}$ for the Floquet system, we insert the mode expansion of $\ket{\Phi_\alpha(t)}$, Eq.~\eqref{eq:phiModeExpansion}, into the FEE, Eq.~\eqref{eq:FEE}, giving
%-------------------------------
\begin{align}
\sum\limits_m\left[(H_0-m\Omega)e^{-im\Omega t}-\sum\limits_n V_n e^{-i(m+n)\Omega t}\right]
\ket{\phi_\alpha^{m}} = \epsilon_\alpha\sum\limits_m e^{-im\Omega t} \ket{\phi_\alpha^m}.
\end{align}
%-------------------------------
Multiplying this equation by $e^{i k\Omega t}$ for integer $k$ and using the identity in Eq.~\eqref{eq:diracdelta}, we obtain the time-independent eigenvalue equations for the frequency components $\ket{\phi_\alpha^n}$:
%-------------------------------
\begin{align}
\sum\limits_n\left[(H_0-k\Omega)\delta_{k,n} + V_{k-n}\right] \ket{\phi_\alpha^n}
= \epsilon_\alpha \ket{\phi_\alpha^k}
.
\label{eq:floquet_eigenvalue_index}
\end{align}
%-------------------------------
In the case that only $V_0\neq 0$ (a time-independent potential), the modes decouple and Eq.~\eqref{eq:floquet_eigenvalue_index} is equivalent to the normal eigenvalue problem in time-independent quantum mechanics. On the other hand, if $V_n\neq 0$ for some $n\neq 0$, Eq.~\eqref{eq:floquet_eigenvalue_index} is an infinite set of coupled equations. To solve this system of equations, we define the matrix $\mathcal{F}$ with matrix elements
%-------------------------------
\begin{align}
(\mathcal{F})_{mn} = (H_0-m\Omega)\delta_{m,n} + V_{m-n}
\end{align}
%-------------------------------
and the vector $\bm{\phi}_\alpha$ with elements
%-------------------------------
\begin{align}
(\bm{\phi}_\alpha)_m = \ket{\phi_\alpha^m}.
\end{align}
%-------------------------------
The FEE can then be written
%-------------------------------
\begin{align}
\mathcal{F}\bm{\phi}_\alpha = \epsilon_\alpha\bm{\phi}_\alpha.
\label{eq:Floquet_characteristic_eqn}
\end{align}
%-------------------------------
In principle the quasienergies are then determined by the condition
%-------------------------------
\begin{align}
\det(\mathcal{F} - \epsilon_\alpha I)=0,
\end{align}
%-------------------------------
where $I$ is the identity matrix. Once the quasienergies are known, the eigenvectors are determined by solving Eq.~\eqref{eq:Floquet_characteristic_eqn} for the components of $\bm{\phi}_\alpha$. 

Since $\mathcal{F}$ is an infinite-dimensional matrix, we must in practice truncate to a submatrix with indices constrained to the range $m,n \in [N_\mr{min},N_\mr{max}]$ for some negative integer $N_\mr{min}$ and positive integer $N_\mr{max}$. This truncation is motivated by conservation of probability which implies that the elements $(\bm{\phi}_\alpha)_n$ decrease to zero for large $|n|$. This truncation is justified {\it a posteriori} by checking that observables calculated using a choice of $N_\mr{min}$ and $N_\mr{max}$ are not strongly sensitive to small changes to $N_\mr{min}$ or $N_\mr{max}$.

\section{Example: Rabi Oscillation}
As an illustration of Floquet theory, we now consider the weak driving of a two-level system near resonance. Let the Hamiltonian for the two-level system be
%-------------------------------
\begin{align}
H(t) = H_0 + U\cos(\Omega t),
\label{eq:RabiHam}
\end{align}
%-------------------------------
where
%-------------------------------
\begin{align}
H_0 &= E_1\ketbra{1}{1} + E_2\ketbra{2}{2}, \nonumber \\
U &= u\left(\ketbra{1}{2}+\ketbra{2}{1}\right).
\end{align}
%-------------------------------
The states $\ket{1}$ and $\ket{2}$ are a basis for the quantum state of the 2-level system. They are eigenstates of $H_0$ with energies $E_1$ and $E_2$. The harmonic components of the potential are 
%-------------------------------
\begin{align}
V_n = \begin{cases} 
      U/2 & n= \pm 1 \\
      0 & \mr{else}. 
   \end{cases}
\end{align}
%-------------------------------
The $\mathcal{F}$ matrix has the form
%-------------------------------
\begin{align}
\mathcal{F} = 
\begin{blockarray}{cccccccc}
	 &	\cdots &	n=-2 &	-1 &	0 &	1 & 2 & \cdots\\
\begin{block}{c(ccccccc)} 
 \vdots &\ddots 	&\vdots		&\vdots		&\vdots		&\vdots		&\vdots		& \iddots \\
 m=-2 	&\cdots 	&H_0+2\Omega&V_{-1}		& 0			&0 			&0 			& \cdots \\
 -1 	&\cdots 	&V_1		&H_0+\Omega	&V_{-1}		& 0			&0 			& \cdots \\
 0 	&\cdots 	&0 			&V_1		&H_0		&V_{-1}		&0 			& \cdots \\
 1 	&\cdots 	&0 			&0			&V_1		&H_0-\Omega	&V_{-1}		& \cdots \\
 2 	&\cdots 	&0 			&0 			&0			&V_1		&H_0-2\Omega& \cdots \\
 \vdots &\iddots 	&\vdots		&\vdots		&\vdots		&\vdots		&\vdots		& \ddots \\
\end{block}
\end{blockarray}
\label{eq:Fmatrix}
\end{align}
%-------------------------------
When expanded in the $\ket{1}, \ket{2}$ basis, each element of $\mathcal{F}$ in Eq.~\eqref{eq:Fmatrix} is a $2\times 2$ matrix. The full matrix elements are
%-------------------------------
\begin{align}
(\mathcal{F})_{mn,ij} = \bra{i}(\mathcal{F})_{mn}\ket{j}
\end{align}
%-------------------------------
For example, consider the $m,n \in [-1,0]$ submatrix of $\mathcal{F}$ expanded in the $\ket{1}, \ket{2}$ basis:
%-------------------------------
\begin{align}
\mathcal{F}_{m,n \in [-1,0]} \overset{\ket{1},\ket{2}}{\longrightarrow}
\begin{blockarray}{ccccc}
	 &	n,j=-1,1 &	-1,2 &	0,1 &	0,2 \\
	\begin{block}{c(cccc)}
	m,i=-1,1 	&E_1+\Omega 	&0			& 0			&u/2 	\\
	-1,2  		&0				&E_2+\Omega	&u/2		& 0		\\
	0,1  		&0				&u/2		&E_1		& 0		\\
	0,2  		&u/2			& 0			& 0 		&E_2	\\
	\end{block}
\end{blockarray}
\label{eq:Fsubmatrix}
\end{align}
%-------------------------------
Similarly, the Floquet eigenstates $\ket{\phi_\alpha^m}$ can be expanded in the $\ket{1}, \ket{2}$ basis:
%-------------------------------
\begin{align}
\ket{\phi_\alpha^m} = \ket{1}\phi_{\alpha,1}^m + \ket{2}\phi_{\alpha,2}^m.
\end{align}
%-------------------------------

We consider the case of near-resonance driving, i.e. 
%-------------------------------
\begin{align}
\Omega = \Delta E + \delta,
\end{align}
%-------------------------------
where $\Delta E = E_2 - E_1$ and the detuning is $\delta \ll \Delta E$. Near resonance, then, $\mathcal{F}_{-1-1,11}$ nearly equals $\mathcal{F}_{00,22}$. Moreover, these matrix elements are directly coupled by the off-diagonal elements $\mathcal{F}_{-10,12}$ and $\mathcal{F}_{0-1,21}$. This situation is directly analogous to degenerate perturbation theory in time-independent quantum mechanics, the only difference being that the effective degeneracy is induced by the oscillating field. As in degenerate perturbation theory, even for small $u$ the degenerate or nearly degenerate subspace must be exactly diagonalized. Once this is done, the remaining elements of $\mathcal{F}$ can be taken into account perturbatively. Here we simply extract the leading order result. To do so, we solve the truncated Floquet eigenequation
%-------------------------------
\begin{align}
\begin{pmatrix}
E_1+\Omega - \epsilon_\alpha	& u/2 \\
u/2 		& E_2	-\epsilon_\alpha
\end{pmatrix}
\begin{pmatrix}
\phi_{\alpha,1}^{-1} \\
\phi_{\alpha,2}^{0}
\end{pmatrix}
=
\begin{pmatrix}
0\\
0
\end{pmatrix}.
\label{eq:FEE_truncated}
\end{align}
%-------------------------------
The quasienergies are
%-------------------------------
\begin{align}
\epsilon_\pm = E_2 + \frac{\delta \pm \sqrt{\delta^2+u^2}}{2}.
\label{eq:Rabi_eigenval}
\end{align}
%-------------------------------
The eigenvectors are
%-------------------------------
\begin{align}
\begin{pmatrix} \phi_{\pm,1}^{-1} \\ \phi_{\pm,2}^{0} \end{pmatrix}
= \frac{1}{\sqrt{1+b_\pm^2}}
\begin{pmatrix} 1 \\ b_\pm \end{pmatrix},
\label{eq:Rabi_eigenvec}
\end{align}
%-------------------------------
where 
%-------------------------------
\begin{align}
b_\pm = \frac{\delta\mp \sqrt{\delta^2+u^2}}{2}.
\end{align}
%-------------------------------

We can now determine the time-evolution of the two level system. We start by writing down the time evolution of the wavefunctions $\ket{\Psi_\pm(t)}$ by inserting the results from Eqs.~\eqref{eq:Rabi_eigenval} and \eqref{eq:Rabi_eigenvec} into Eq.~\eqref{eq:Psi_Ansatz}:
%-------------------------------
\begin{align}
\ket{\Psi_\pm(t)} = e^{-i\epsilon_\pm t}
\left(\ket{1}\phi_{\pm,1}^{-1}e^{-i\Omega t} + \ket{2}\phi_{\pm,2}^{0} \right).
\end{align}
%-------------------------------
The general solution has the form
%-------------------------------
\begin{align}
\ket{\Psi(t)} = A\ket{\Psi_+(t)}+B\ket{\Psi_-(t)}.
\end{align}
%-------------------------------
The coefficients $A$ and $B$ are determined by the normalization
%-------------------------------
\begin{align}
\langle \Psi(t)\ket{\Psi(t)} = 1
\end{align}
%-------------------------------
and the boundary condition
%-------------------------------
\begin{align}
\ket{\Psi(t_0)} = \ket{\Psi_0}.
\end{align}
%-------------------------------
Letting $t_0=0$ and $\ket{\Psi_0}=\ket{1}$, we find
%-------------------------------
\begin{align}
A = \frac{\sqrt{1+b_+^2}}{1-b_+/b_-}, \hspace{1cm}
B = \frac{\sqrt{1+b_-^2}}{1-b_-/b_+}.
\end{align}
%-------------------------------
The probability that the system is in level $i$ at time $t$ is
%-------------------------------
\begin{align}
P_i(t) = |\langle i \ket{\Psi(t)}|^2.
\end{align}
%-------------------------------
We arrive at the familiar result for Rabi oscillation in two-level systems:
%-------------------------------
\begin{align}
P_1(t) = 1-\frac{u^2}{\Omega_R^2}\sin^2\left(\frac{\Omega_R}{2}t\right), \hspace{1cm}
P_2(t) = \frac{u^2}{\Omega_R^2}\sin^2\left(\frac{\Omega_R}{2}t\right),
\label{eq:RabiP12}
\end{align}
%-------------------------------
where $\Omega_R$ is the \textit{Rabi frequency}:
%-------------------------------
\begin{align}
\Omega_R = \sqrt{\delta^2+u^2}.
\end{align}
%-------------------------------

Starting with the Hamiltonian in Eq.~\eqref{eq:RabiHam}, the usual procedure for calculating the Rabi oscillation formula in Eq.~\eqref{eq:RabiP12} is to use the rotating wave approximation to isolate the dominant mode. In the Floquet approach the same effect is achieved by diagonalizing the (nearly) degenerate Floquet submatrix in Eq.~\eqref{eq:FEE_truncated}. Unlike the rotating wave approximation, the Floquet approach provides a natural framework for incorporating corrections to the leading order results. In particular, let $\mathcal{F}_0$ be the degenerate submatrix of $\mathcal{F}$ and let $\mathcal{F}_1 = \mathcal{F} - \mathcal{F}_0$. Analogously to degenerate perturbation theory, one first explicitly diagonalizes $\mathcal{F}_0$, and then incorporates the effects of $\mathcal{F}_1$ perturbatively \cite{Shirley_1965, longabstract}.

%%%%%%%%%%%%%%%%%%%%%%%%%%%%%%%%%%%%%%%%%%
%%% Local Variables:
%%% mode: latex
%%% TeX-master: "../../HudsonPhDThesis"
%%% End:

%% file: Chapters/Formalism/Formalism.tex
% !TEX root = ../HudsonPhDThesis.tex
\cleardoublepage

\chapter{Floquet scattering theory}
\label{Chap:Formalism}

The scattering of neutral alkali-metal atoms in the presence of a time-periodic magnetic field can be expressed in terms of the quantum mechanical scattering of particles with a short-range, time-periodic interaction potential. For sufficiently low-energy scattering, the scatterers ``resolve'' the angular  frequency $\Omega$ of the periodic potential and thus only undergo transitions to states whose energies differ from the scattering energy by integer multiples of $\hbar \Omega$ \cite{Chu85a}, in the same spirit as Fermi's Golden Rule. The purpose of this chapter is to derive a formalism for determining the probability amplitudes for scattering into these energy components. In this chapter we derive an extension of Floquet theory to the problem of scattering from short-range, time-periodic potentials.

Floquet scattering theory has previously been used to accurately describe laser-assisted electron-atom collisions \cite{Mason1993, Joachain1994, Joachain_2014}. In that case, it is assumed that the scattering electron is in a light-dressed state composed of several energy components coupled by the laser. If the frequency of the laser nearly corresponds to an internal transition of the atom, the scattering observables are resonantly modified. In this chapter, we instead consider the scattering of two neutral atoms whose short-range interaction potential is modulated periodically in time. In the weak-driving case, two-body transition rates can be expressed in terms of transition matrix elements of Tan's contact \cite{MagnetoAssoc2015, Mohapatra:pra:2015}. The alternative approach presented here allows us to incorporate reactions involving the exchange of arbitrarily many quanta of the oscillating field.

Let $H_0$, $V(t)$, and $|\Psi(t)\rangle$ be the kinetic energy operator, the time-dependent interaction potential, and the wavevector describing the 2-body system, respectively. In principle, we should solve the time-dependent Schr{\"o}dinger equation (TDSE)
%-------------------------------
\begin{equation}
(i\partial_t -H_0)|\Psi(t)\rangle =  V(t)|\Psi(t)\rangle.
\label{eq:TDSE_2}
\end{equation}
%-------------------------------
The solution for $\ket{\Psi(t)}$ must satisfy the asymptotic boundary condition (identical to Eq.~\eqref{eq:ScatBC_ti}):
%-------------------------------
\begin{equation}
\ket{\Psi(t)} \overset{t\rightarrow -\infty}{\longrightarrow} e^{-i(k^2/M) t}\ket{\bm{k}}.
\label{eq:ScatBC}
\end{equation}
%-------------------------------

By using the boundary condition in Eq.~\eqref{eq:ScatBC}, we are tacitly assuming that the desired solution does not depend upon the width of the incident physical wave packet, as is always the case for stationary potentials. This is not guaranteed for a short-range, time-periodic potential, however, since the time-scale associated with $V(t)$ generally introduces sensitivity to the window of time over which the scattering event occurs. This time window, in turn, depends upon the width of the wave-packet. To avoid this complication, we will assume that the incident wave ``feels'' many cycles of the interaction potential, such that observables are insensitive to the window of time over which the scattering occurs. The number of cycles ``felt'' by the scattering wavefunction is inversely related to the width, $\Delta E$ of the wave packet in energy space. So, concretely, we require $\Omega / \Delta E \gg 0$. The width $\Delta E$ depends on the process that generates the physical free states. For asymptotic states belonging to a homogeneous ideal gas, we can take the limit $\Delta E = 0$. 

In the following sections, we will develop the formal methods necessary to describe scattering from short-range time-periodic potentials. Many of our results will closely resemble the results from multichannel scattering theory. This connection goes deep. The time-periodic potential causes transitions between states whose energies differ by $n\Omega$ for some integers $n$. If the incoming state has energy $k^2/M$, the outgoing state will have components with energies $k^2/M + n\Omega$ for some integers $n$. The connection to multichannel scattering is this: the outgoing component with energy $k^2/M + n\Omega$ relative to the threshold of the incoming channel can be expressed as a state with energy $k^2/M$ in a channel whose threshold energy is $n\Omega$ relative to the threshold of the incoming channel. As we will see, these insights emerge naturally as consequences of Floquet's theorem, Eq.~\eqref{eq:Psi_Ansatz}.

\section{Floquet Lippmann-Schwinger equation}
We begin our discussion of Floquet scattering theory by deriving an analog of the Lippmann-Schwinger equation from stationary scattering theory which we term the \textit{Floquet Lippmann-Schwinger equation}. This equation will prove very useful in the development of the scattering formalism and as a starting point for actual calculations.

As in Eq.~\eqref{eq:Psi_Ansatz}, we expect the time-dependent Schr{\"o}dinger equation in Eq.~\eqref{eq:TDSE} to have solutions of the form
%-------------------------------
\begin{equation}
\ket{\Psi(t)} = e^{-i\epsilon t}\ket{\Phi(t)} =e^{-i\epsilon t}\sum\limits_{n=-\infty}^{\infty} \ket{\phi^n} e^{-in\Omega t}.
\label{eq:psiAnsatz2}
\end{equation}
%-------------------------------
Inserting this into the TDSE and projecting out the $m^\mr{th}$ Floquet mode gives
%-------------------------------
\begin{align}
(\epsilon + m\Omega - H_0)\ket{\phi^m}=
\sum\limits_{n=-\infty}^\infty V_{m-n}\ket{\phi^n}.
\label{eq:SEomega}
\end{align}
%-------------------------------
We can immediately write down a formal solution to Eq.~\eqref{eq:SEomega} that also satisfies the boundary condition in Eq.~\eqref{eq:ScatBC}:
%-------------------------------
\begin{equation}
\ket{\phi^m} =\delta_{m,0}\ket{\bm{k}} + G_0(\epsilon+m\Omega)\sum\limits_{n=-\infty}^\infty V_{m-n}\ket{\phi^n}.
\label{eq:FLSE}
\end{equation}
%-------------------------------
$G_0(E)$ is the noninteracting Green's function given in Eq.~\eqref{eq:Green}.
By acting the operator $G_0^{-1}(\epsilon + m\Omega)$ on both sides of Eq.~\eqref{eq:FLSE}, we recover Eq.~\eqref{eq:SEomega} if
%-------------------------------
\begin{equation}
\epsilon = \frac{k^2}{M}.
\end{equation}
%-------------------------------

Equation \eqref{eq:FLSE} is the Floquet Lippmann-Schwinger equation (FLSE). It is directly analogous to the familiar Lippmann-Schwinger equation (LSE), Eq.~\eqref{eq:LSE}, that we derived in the context of time-independent scattering theory. The main difference between the FLSE and the LSE is the sum over Floquet modes that appears in the second term on the right hand side of \eqref{eq:FLSE}. The summation over Floquet components appears because incident particles with center of mass energy $k^2/M$ can absorb or emit quanta from the oscillating potential, exiting with energy $k^2/M+n\Omega$ for any integer $n$. The coupling between Floquet components is controlled by the potential components $V_n$. If $V_n = 0$ for every $n\neq 0$ (i.e. if the potential is time-independent), then Eq.~\eqref{eq:FLSE} reduces to the LSE, Eq.~\eqref{eq:LSE}. Whereas the normal LSE is a single integral equation, the FLSE is an infinite set of coupled integral equations. 

The FLSE fully encodes the information from both the time-dependent Sch{\"o}dinger equation and the boundary condition in the asymptotic past. Like the LSE, the FLSE is not a solution in the practical sense because the Floquet component $\ket{\phi^n}$ depends upon all Floquet components $\ket{\phi^k}$ (including itself) operated upon by $G_0(\epsilon+n\Omega)V_{n-k}$. As with the normal LSE, however, the FLSE naturally leads to methods for calculating the desired physical quantities. 

\section{Asymptotic wavefunction}
As with time-independent scattering theory, we can extract scattering observables from knowledge about the wavefunction in the region where the distance between the particles is much larger than the range of the potential. Using the FLSE, we can extract the Floquet component wavefunctions $\phi^n(\bm{r}) = \langle \bm{r}\ket{\phi^n}$ in the region of large separation.

Projecting the Floquet Lippmann-Schwinger equation into position space gives
%-------------------------------
\begin{align}
\phi^m(\bm{r}) 
&= e^{i\bm{k}\cdot\bm{r}}\delta_{m,0} 
+ \int d^3r'G_0(\bm{r},\bm{r'};k^2/M+n\Omega)\sum\limits_{n=-\infty}^\infty \langle\mbf{r'}|V_{m-n}\ket{\phi^n}.
\label{eq:FLSER}
\end{align}
%-------------------------------
$G_0(\bm{r},\bm{r'}; E)$ is the free position-space Green's function defined in Eq.~\eqref{eq:G0r}. In the region where the particle separation $r=|\bm{r}|$ is much greater than the range of the potential, Eq.~\eqref{eq:FLSER} takes the form
%-------------------------------
\begin{align}
\phi^m(\bm{r}) 
&\underset{r\rightarrow\infty}{\longrightarrow} e^{i\bm{k}\cdot\bm{r}}\delta_{m,0}  -\frac{M}{4\pi}\frac{e^{ik_m r}}{r} \int d^3r'e^{-i k_m\bm{\hat{r}}\cdot\bm{r}'}\sum\limits_{n=-\infty}^\infty \bra{\bm{r'}}V_{m-n}|\phi^n\rangle \nonumber \\
&= e^{i\bm{k}\cdot\bm{r}}\delta_{m,0} -\frac{M}{4\pi}\frac{e^{ ik_m r}}{r} \sum\limits_{n=-\infty}^\infty \langle k_m\mbf{\hat{r}}|V_{m-n}|\phi^n\rangle,
\label{eq:asymptoticphi}
\end{align}
%-------------------------------
where 
\begin{equation}
k_n = \sqrt{k^2 +  n M \Omega}.
\label{eq:kn}
\end{equation}
To obtain Eq.~\eqref{eq:asymptoticphi} we have dropped all terms that fall faster than $1/r$. This equation manifestly satisfies the asymptotic boundary condition in Eq.~\eqref{eq:ScatBC}, since the only incoming plane-wave component equals the one specified by the boundary condition. All other components are spherical outgoing waves. Note also that if $k^2/M+n\Omega$ is negative, $k_n$ is imaginary and the second term on the right-hand side of Eq.~\eqref{eq:asymptoticphi} falls off exponentially fast with $r$. Such ``bound states'' contribute nothing to the asymptotic wavefunction.

A comparison of the form of Eq.~\eqref{eq:asymptoticphi} with the form of the asymptotic wavefunction in time-independent scattering theory, Eq.~\eqref{eq:asymptoticphi_ti_std}, allows us to identify the Floquet scattering amplitudes:
%-------------------------------
\begin{equation}
f_m(\bm{p},\mbf{k})=-\frac{M}{4\pi}\sum\limits_{n=-\infty}^\infty \bra{\bm{p}}V_{m-n}|\phi^n\rangle,
\label{eq:amplitude_deff}
\end{equation}
%------------------------------- 
where the dependence on the incident relative momentum $\bm{k}$ is hidden in the Floquet components $\ket{\phi^n}$. The amplitude $f_n(\bm{p},\mbf{k})$ is the scattering amplitude for incident scatterers with relative momentum $\bm{k}$ and center of mass energy $k^2/m$ to exit with momentum $\bm{p}$ and energy $k^2/M + n\Omega$. Note that the asymptotic wavefunctions only depend on amplitudes that satisfy the `on-shell' condition $p^2/M = k^2/M+ n\Omega$.

\section{Floquet scattering cross section}
The quantum mechanical scattering from a short-range, time-periodic potential is fully characterized by the Floquet scattering amplitudes $f_n(k_n\bm{\hat p},\mbf{k})$ defined in Eq. \eqref{eq:amplitude_deff}. In particular, we are interested in the differential cross section $d\sigma_n(k)/d\Omega$ for the incoming scatterers with energy $k^2/M$ to exit in the $n^{\mathrm{th}}$ Floquet channel with energy $k^2/M+n\Omega$ into the solid angle $d\Omega$. The cross section into the solid angle $d\Omega$ is:
%-------------------------------
\begin{equation}
d\sigma_n = \frac{R_n d\Omega}{|\bm{j}_{\mathrm{inc}}|},
\label{eq:dsigmadomega_1}
\end{equation}
%-------------------------------
where $R_n$ is the differential rate of probability flow into the $n^{\mathrm{th}}$ Floquet channel, and $|\bm{j}_{\mathrm{inc}}|$ is the magnitude of the incident particle current density. $R_n d\Omega$ equals the dot product of the vector area element $r^2 d\Omega \,\hat{\bm{r}}$ with the scattered current density $\bm{j}_{\mathrm{sc},n}$ into the $n^{\mathrm{th}}$ Floquet channel and into the solid angle $d\Omega$. Dividing out an overall factor of $d\Omega$ this gives
%-------------------------------
\begin{equation}
R_n = \bm{j}_{\mathrm{sc},n}\cdot \bm{\hat r} r^2.
\label{eq:R_n1}
\end{equation}
%-------------------------------
To determine $\bm{j}_{\mathrm{sc},n}$, we consider the asymptotic Floquet component wavefunction
%-------------------------------
\begin{align}
\phi^n(\bm{r}) 
\underset{r\rightarrow\infty}{\longrightarrow} e^{i\bm{k}\cdot\bm{r}}\delta_{n,0} + \phi^n_{\mathrm{sc}}(\bm{r}),
\end{align}
%-------------------------------
where, comparing with Eqs.~\eqref{eq:asymptoticphi} and \eqref{eq:amplitude_deff},
%-------------------------------
\begin{align}
 \phi^n_{\mathrm{sc}}(\bm{r}) = \frac{e^{ ik_n r}}{r}f_n(k_n\bm{\hat r},\mbf{k}).
\end{align}
%-------------------------------
The scattered current density at large separations $r$ is
%-------------------------------
\begin{align}
\bm{j}_{\mathrm{sc},n}
&\underset{r\rightarrow\infty}{\longrightarrow} \frac{1}{iM}\big[{\phi^n_{\mathrm{sc}}(\bm{r})}^*\nabla\phi^n_{\mathrm{sc}}(\bm{r})-\phi^n_{\mathrm{sc}}(\bm{r})\nabla{\phi^n_{\mathrm{sc}}(\bm{r})}^*\big]
\nonumber\\
&=\frac{2k_n}{M}\frac{\bm{\hat r}}{r^2}\big|f_n(k_n\bm{\hat r},\mbf{k})\big|^2 + \mathcal{O}\left(1/r^3\right).
\end{align}
%-------------------------------
Inserting this result into Eq.~\eqref{eq:R_n1}, we find
%-------------------------------
\begin{equation}
R_n = \frac{2k_n}{M}\big|f_n(k_n\bm{\hat r},\mbf{k})\big|^2.
\label{eq:R_n2}
\end{equation}
%-------------------------------
The magnitude of the incident current density is simply
%-------------------------------
\begin{equation} 
|\bm{j}_{\mathrm{inc}}|= \frac{2k}{M}.
\label{eq:jinc}
\end{equation}
%-------------------------------
Inserting $R_n$ and $|\bm{j}_{\mathrm{inc}}|$ into Eq.~\eqref{eq:dsigmadomega_1}, we obtain the differential cross section for the incident wave to scatter into the $n^{\mathrm{th}}$ Floquet channel:
%-------------------------------
\begin{equation}
\frac{d\sigma_n}{d\Omega} = \frac{k_n}{k}\big|f_n(k_n\bm{\hat r},\mbf{k})\big|^2.
\label{eq:dsigmadomega_2}
\end{equation}
%-------------------------------
This result differs from the result from time-independent scattering theory by the ratio of the magnitudes of the outgoing to incoming momentum. This result is identical in form to the standard result for time-independent multichannel scattering (see e.g. \cite{taylor_scattering_thy}). The total cross section for scattering into all possible Floquet modes is
%-------------------------------
\begin{equation}
\sigma_{\mr{tot}}(\bm{k}) ={\sum_n}'\sigma_n(\bm{k}),
\label{eq:totalcross}
\end{equation}
%-------------------------------
where the prime on the sum indicates that the summation is restricted to values of $n$ such that $k^2/M + n\Omega > 0$. This restriction comes from the observation made following Eq.~\eqref{eq:asymptoticphi} that states for other values of $n$ are exponentially suppressed at large separations $r$. 

\section{Integral equation for the Floquet scattering amplitudes}
The Floquet scattering amplitudes defined in Eq.~\eqref{eq:amplitude_deff} depend on the unknown Floquet components $\ket{\phi^n}$, which are fully determined by the Floquet Lippmann-Schwinger Equation, Eq.~\eqref{eq:FLSE}. Since the scattering observables only depend on the amplitudes, it is convenient to convert the FLSE for the components $\ket{\phi^n}$ into a similar equation for $f_m(\bm{p},\mbf{k})$. By premultiplying Eq.~\eqref{eq:FLSE} with $-(M/4\pi)\bra{\bm{p}}V_{n-m}$ and summing over $m$, we find an integral equation for the scattering amplitudes:
%-------------------------------
\begin{align}
f_n(\mbf{p},\mbf{k})
&= -\frac{M}{4\pi}\langle \mbf{p}|V_n|\mbf{k}\rangle -\frac{M}{4\pi}\sum\limits_{m,j=-\infty}^\infty \bra{\bm{p}}V_{n-m}G_0(k^2/M+m\Omega)
V_{m-j}\ket{\phi^j}\nonumber \\
&= -\frac{M}{4\pi}\langle \mbf{p}|V_n|\mbf{k}\rangle +\sum\limits_{m=-\infty}^\infty \int \frac{d^3q}{(2\pi)^3}
\langle \mbf{p}|V_{n-m}|\mbf{q}\rangle G_0(q, k^2/M+m\Omega) f_m(\mbf{q},\mbf{k}).
\label{eq:FLSEf}
\end{align}
%-------------------------------
$G_0(q, E)$ is the free momentum-space Green's function defined in Eq.~\eqref{eq:Gmomentum}. Equation \eqref{eq:FLSEf} can be used as the starting point for exact or approximate calculations of the Floquet scattering amplitudes. Once the amplitudes $f_n(\bm{p},\bm{q})$ are known, the `on-shell' amplitudes that appear in Eq.~\eqref{eq:asymptoticphi} are obtained by setting $\mbf{p}=k_n\hat{\mbf{r}}$ where $\mbf{k}$ is the relative momentum of the incoming particles and $k_n$ is defined in Eq.~\eqref{eq:kn}. For $\bm{p}=k_n\hat{\bm{r}}$, the energy of the outgoing particles differs from the energy of the incoming particles by an integer multiple of $\Omega$.

\section{Partial wave expansion}
As discussed in Sec.~\ref{chap:scat::sec:theory::subsec:partialwave}, the conditions of ultracold atomic gas experiments are such that we can usefully perform a partial-wave expansion, keeping only the lowest few terms. This is because at very low temperatures, only a small number of length scales contribute to observable phenomena. Thus, the ``high-energy'' details of the interaction potential can be ignored. 

At low energies, the scattering potentials of alkali atoms are effectively spherically symmetric. If follows that the harmonic components of the potential $\langle\mbf{p}|V_n|\bm{q}\rangle$ and the Floquet scattering amplitudes $f_n(\mbf{p},\mbf{q})$ depend only on $\bm{\hat p}\cdot\bm{\hat q}$ but not upon the rotation angle around the axis of the incoming relative momentum. The amplitudes and potential components therefore have expansions of the form
%-------------------------------
\begin{align}
f_n(\mbf{p},\mbf{q})
 &= 4\pi\sum\limits_{l,m}f_n^{l}(p,q)Y_{lm}^*(\Omega_{\bm{p}})Y_{lm}(\Omega_{\bm{q}}),
\nonumber\\
\langle\mbf{p}|V_n|\bm{q}\rangle
 &=4\pi\sum\limits_{l,m}V_n^{l}(p,q)Y_{lm}^*(\Omega_{\bm{p}})Y_{lm}(\Omega_{\bm{q}}),
\label{eq:sphericalHarmonics}
\end{align}
%-------------------------------
where the $l$ sum runs from $0$ to $\infty$ and the $m$ sum runs from $-l$ to $l$. We have assumed that the expansion coefficients $f_n^l$ and $V_n^l$ do not depend upon the $m$ quantum number. This is equivalent to our assumption that the interaction potential is spherically symmetric. With this assumption the $m$ sum can be performed analytically using the addition theorem for spherical harmonics given in Eq.~\eqref{eq:addition_theorem}. Using this identity, Eqs.~\eqref{eq:sphericalHarmonics} simplify to 
%-------------------------------
\begin{align}
f_n(\mbf{p},\mbf{q})
 &= \sum\limits_{l}f_n^{l}(p,q)(2l+1)P_l(\bm{\hat p}\cdot\bm{\hat q}),
\nonumber\\
\langle\mbf{p}|V_n|\bm{q}\rangle
 &=\sum\limits_{l}V_n^{l}(p,q)(2l+1)P_l(\bm{\hat p}\cdot\bm{\hat q}).
\label{eq:LegendreExapansion}
\end{align}
%-------------------------------
Using the orthogonality relation in Eq.~\eqref{eq:Pcompleteness}, the partial wave components are
%-------------------------------
\begin{align}
f_n^l(p,q)
 &=\frac{1}{2}\int\limits_{-1}^1 d(\bm{\hat p}\cdot\bm{\hat q}) P_l(\bm{\hat p}\cdot\bm{\hat q})f_n(\bm{p},\bm{q}),\nonumber\\
V_n^l(p,q)
 &=\frac{1}{2}\int\limits_{-1}^1 d(\bm{\hat p}\cdot\bm{\hat q}) P_l(\bm{\hat p}\cdot\bm{\hat q})\langle\mbf{p}|V_n|\mbf{k}\rangle.
\label{eq:partialWaveComponents}
\end{align}
%-------------------------------
Notice that the $s$-wave component ($l=0$) is the simple angle average of the original function since $P_0(x)=1$. 

The angular momentum components of the wavefunction scatter independently. This allows us to re-express the FLSE in terms of the individual angular momentum components.  Inserting the spherical harmonic expansions of $f_n(\mbf{p},\mbf{q})$ and  $\langle\mbf{p}|V_n|\bm{q}\rangle$, Eq.~\eqref{eq:sphericalHarmonics}, into the FLSE, Eq.~\eqref{eq:FLSE}, and projecting out the $l^{\mathrm{th}}$ angular momentum component, we obtain the FLSE for the partial wave Floquet amplitudes $f_n^l(p,k)$:
%-------------------------------
\begin{align}
f_n^l(p,k)
&= -\frac{M}{4\pi}V_n^l(p,k)
+\frac{1}{2\pi^2}\sum\limits_{m=-\infty}^\infty \int\limits_0^\infty dqq^2
\frac{V_{n-m}^l(p,q)}{k^2/M+m\Omega - q^2/M + i0^+} f_m^l(q,k).
\label{eq:FLSEpartialwave}
\end{align}
%-------------------------------
The left-hand side and the first term on the right-hand side follow directly using the orthogonality relation for Legendre polynomials, Eq.~\eqref{eq:Pcompleteness}. To arrive at the second term on the right-hand side, we first used the orthogonality relation for the spherical harmonics in Eq.~\eqref{eq:Ycompleteness} to integrate over the $\bm{q}$ angles. We then apply the addition theorem in Eq.~\eqref{eq:addition_theorem} and finally the orthogonality relation for the Legendre polynomials, Eq.~\eqref{eq:Pcompleteness}. For the full derivation, see Appendix \ref{App:partial_wave}.

\section{Effective range expansion}
For low-energy scattering, the $s$-wave ($l=0$) Floquet scattering amplitudes defined in Eq.~\eqref{eq:partialWaveComponents} can be expressed as explicit functions of the scattering momentum $k$ by using the effective range expansion, giving
%-------------------------------
\begin{equation}
f^0_n(k)^{-1}+ik = -\frac{1}{a_n} + \frac{1}{2}{r_e}_nk^2  + \mathcal{O}(k^4),
\label{eq:ERE}
\end{equation}
%-------------------------------
where $a_n$ and ${r_{e}}_n$ are the frequency-dependent $s$-wave Floquet scattering length and effective range for scattering into the $n^{\mathrm{th}}$ Floquet level. In practice, the first $N$ resonance parameters can be extracted by fitting a polynomial of order $N$ in $k^2$ to $f_n(k)^{-1}+ik$ at small values of $k$.  The higher order terms in the effective range expansion can be calculated in this fashion, but they are rarely needed. In general, $a_n$ and ${r_e}_n$ will be complex functions of $\Omega$ because of the coupling between different Floquet modes. 

The benefit of the effective range expansion is that it makes the momentum dependence completely explicit. Low-energy scattering is then completely characterized by the Floquet effective range expansion coefficients $a_n$, ${r_e}_n$, \textit{etc}. The low-energy cross section into the $n^{\mathrm{th}}$ Floquet mode is
%-------------------------------
\begin{equation}\label{eq:crosssec_ERE}
\sigma_n(k) = \frac{4\pi|a_n|^2}{|1 + ia_nk -\tfrac{1}{2}a_n{r_e}_nk^2+\mathcal{O}(k^4)|^2}.
\end{equation}
%-------------------------------

% The effective range expansion is only useful for values of $k^2$ well within the radius of convergence of $f^0_n(k)^{-1} +ik$ such that we only need to keep a small number of effective range parameters $a_n,\,{r_e}_n,\, etc$. However, in the context of multichannel problems, such as the Floquet scattering problem, this radius of convergence becomes extremely small as two coupled channels approach degeneracy. In Floquet theory, such effective degeneracy occurs when, for example, the oscillation energy scale $\Omega$ is tuned near the transition energy between the incoming scattering state and a bound state supported by the scattering potential in the adiabatic limit. Thus, as we will see concretely in Chapter \ref{chap:atomic_scattering}, the effective range expansion is of little practical use when exploring resonance phenomena.

\section{Floquet \textit{S}-matrix.}
In the context of Floquet scattering theory, we can define an $S$-matrix which relates the amplitude for an incoming state in the $m^{\mathrm{th}}$ Floquet channel to the amplitude for an outgoing state in the $n^{\mathrm{th}}$ Floquet channel. Thus the $S$-matrix in Floquet scattering theory is directly analogous to the $S$-matrix in multichannel time-independent scattering theory (see e.g. \cite{taylor_scattering_thy}).

We label the asymptotic state with momentum $\bm{p}$ in channel $n$ as $\ket{\bm{p}}\otimes\ket{n} \equiv \ket{\bm{p};n}$. This state is an eigenstate of the asymptotic channel Hamiltonian
%------------------------------
\begin{equation}
H_0^n = \frac{\bm{P}^2}{M} + n\Omega,
\end{equation}
%------------------------------
where $\bm{P}$ is the momentum operator. The normalization of the asymptotic states is
%------------------------------
\begin{equation}
  \braket{\bm{p'};n'}{\bm{p};n} = (2\pi)^3\delta(\bm{p}'-\bm{p})\delta_{n'n}.
  \label{eq:pnnorm}
\end{equation}
%------------------------------
In terms of these asymptotic states, the completeness relation for the entire Hilbert space can be expressed as
%------------------------------
\begin{equation}
1 = \sum\limits_{n=-\infty}^\infty \int\frac{d^3p}{(2\pi)^3}\ketbra{\bm{p};n}{\bm{p};n}.
\end{equation}
%------------------------------

The $S$-matrix elements for transitioning between our asymptotic states is directly related to the Floquet scattering amplitude. Following the pattern of Eq.~\eqref{eq:Sf1}, the relationship between the $S$-matrix elements and the scattering amplitude is
%------------------------------
\begin{align}
  \matel{\bm{p'};n'}{\bm{p};n}{S} &= (2\pi)^3\delta(\bm{p'}-\bm{p})\delta_{n',n}\nonumber\\
  &+\frac{i}{2\pi M}\delta\left[E' - E - (n'-n)\Omega\right]f_{n'n}(p'_{n'}\hat{\bm{p}}',p_{n}\hat{\bm{p}}),
  \label{eq:Sf2}
\end{align}
%------------------------------
where $E'=p'^2/M+n'\Omega$ and $E=p^2/M+n\Omega$ are the outgoing and incoming energies, respectively. The amplitude $f_{n'n}(p'_{n'}\hat{\bm{p}}',p_{n}\hat{\bm{p}})$ is the scattering amplitude defined in Eq.~\eqref{eq:amplitude_deff} but extended to the case that the incoming state is not in the $m=0$ channel. Moving the $\delta$-function term to the left-hand side and using the normalization in Eq.~\eqref{eq:pnnorm}, Eq.~\eqref{eq:Sf2} can be expressed as
%------------------------------
\begin{align}
  \matel{\bm{p}';n'}{\bm{p};n}{(S-1)} = \frac{i}{2\pi M} & \delta\left[E' - E - (n'-n)\Omega\right]
  \nonumber \\
  &\times
  f_{n'n}(p'_{n'}\hat{\bm{p}}',p_{n}\hat{\bm{p}}).
\label{eq:Stof}
\end{align}
%------------------------------

Since for describing cold atoms we are often interested in the partial wave amplitudes $f_{nm}^l(p,q)$, it is convenient to define the partial wave $S$-matrix element $S_{nm}^l(E)$ as
%------------------------------
\begin{equation}
\matel{E'l'm';n'}{Elm;n}{S} = \delta\left[E'-E-(n'-n)\Omega\right]\delta_{l'l}\delta_{m'm}S_{nn'}^l(E).
\label{eq:Spartial}
\end{equation}
%------------------------------
The states $\ket{Elm;n}$ are the simultaneous eigenstates of $H_0^n$, $L^2$, and $L_z$. The `on-shell' condition enforced by the energy $\delta$-functions in Eqs.~\eqref{eq:Stof} and \eqref{eq:Spartial} requires that
%------------------------------
\begin{equation}
E' - E = (n'-n)\Omega.
\end{equation}
%------------------------------
As we expect, the energy difference between the incoming and outgoing state must equal the energy exchanged with the oscillating field, represented by the Floquet-channel labels $n'$ and $n$.

The momentum dependence of the left- and right-hand sides
of Eq.~\eqref{eq:Stof} can be expanded into partial-wave components using the
completeness relation in Eq.~\eqref{eq:Elmcompleteness} followed by Eq.~\eqref{eq:Elmexpansion}. By using the definition of $S_{nn'}^l(E)$ in Eq.~\eqref{eq:Spartial}, we extract the relationship between the partial-wave Floquet scattering amplitude and the partial-wave $S$-matrix element (see Appendix \ref{App:multi_channel}):
%------------------------------
\begin{equation}
f_{n'n}^l(p'_{n'},p_n) = \frac{S_{n'n}^l(E)-\delta_{n'n}}{2i\sqrt{p'p}}.
\label{eq:ftoS_partial}
\end{equation}
%------------------------------
This result is identical in form to the result from multichannel scattering theory, where the analogs of $n'$ and $n$ are channel labels \cite{taylor_scattering_thy}.

%%%%%%%%%%%%%%%%%%%%%%%%%%%%%%%%%%%%%%%%%%
%%% Local Variables:
%%% mode: latex
%%% TeX-master: "../../HudsonPhDThesis"
%%% End:

%% file: Chapters/PeriodicPotentials/PeriodicPotentials.tex
% !TEX root = ../HudsonPhDThesis.tex
\cleardoublepage

\chapter{Square-well model}
\label{chap:atomic_scattering}

Having worked out the formalism for binary scattering from short-range, time-periodic potentials in Chapter \ref{Chap:Formalism}, we are now ready to roll up our sleeves and actually calculate low-energy scattering observables such as the total scattering cross section $\sigma_{\mr{tot}}(\bm{k})$ defined in Eq.~\eqref{eq:totalcross}. We restrict our discussion to $s$-wave scattering. We also assume that the low-energy observables are sufficiently described in terms of the time-dependent $s$-wave scattering length
%------------------------------
\begin{equation}
a(t) = \abar + \atil \cos(\Omega t).
\label{eq:aoft}
\end{equation}
%------------------------------
We postpone our discussion of the physical origin of $a(t)$ until Chapter \ref{Chap:Applications}. The assumption that the properties of the physical system depend only upon $a(t)$ will hold if the effective range of the potential $r_e$ is much smaller than $a(t)$ and if the scattering momentum $k$ is much smaller than the momentum scale $\hbar/r_e$ associated with range of the potential.

We solve this problem in two ways. In this chapter, we model the interaction potential using a square well with oscillating depth, carefully choosing the parameters of the square-well potential to reproduce the low-energy observables of a two-atom system with time-oscillating scattering length $a(t)$ and a very small effective range $r_e/\abar\ll 1$. We explicitly solve the time-dependent Schr{\"o}dinger equation for this system and extract information about scattering. In the next chapter, we solve the same problem using the Floquet Lippmann-Schwinger formalism derived in Section \ref{Chap:Formalism} for a contact interaction potential with time-dependent scattering length $a(t)$ and (by definition) zero effective range $r_e=0$. 

This duplication of effort serves three purposes. First, it allows us to verify that the results from the Floquet scattering formalism derived in Chapter \ref{Chap:Formalism} agree with the results from the more straightforward approach of solving the time-dependent Schr{\" o}dinger equation. Second, it demonstrates some of the power of the Floquet formalism, since, as we will find, the calculation is much simpler using the latter approach, yielding exact, analytic results in the zero-range limit.

\section{Square well potential with oscillating depth}
\label{chap:periodic::sec:square}
%%%%%%%%%%%%%%%%%%%%%%%%%%%%%%%%%%%%%%%%%%%%%%%%%%%%%%
\begin{figure}[h]
\centering
\includegraphics[width=0.6\textwidth]{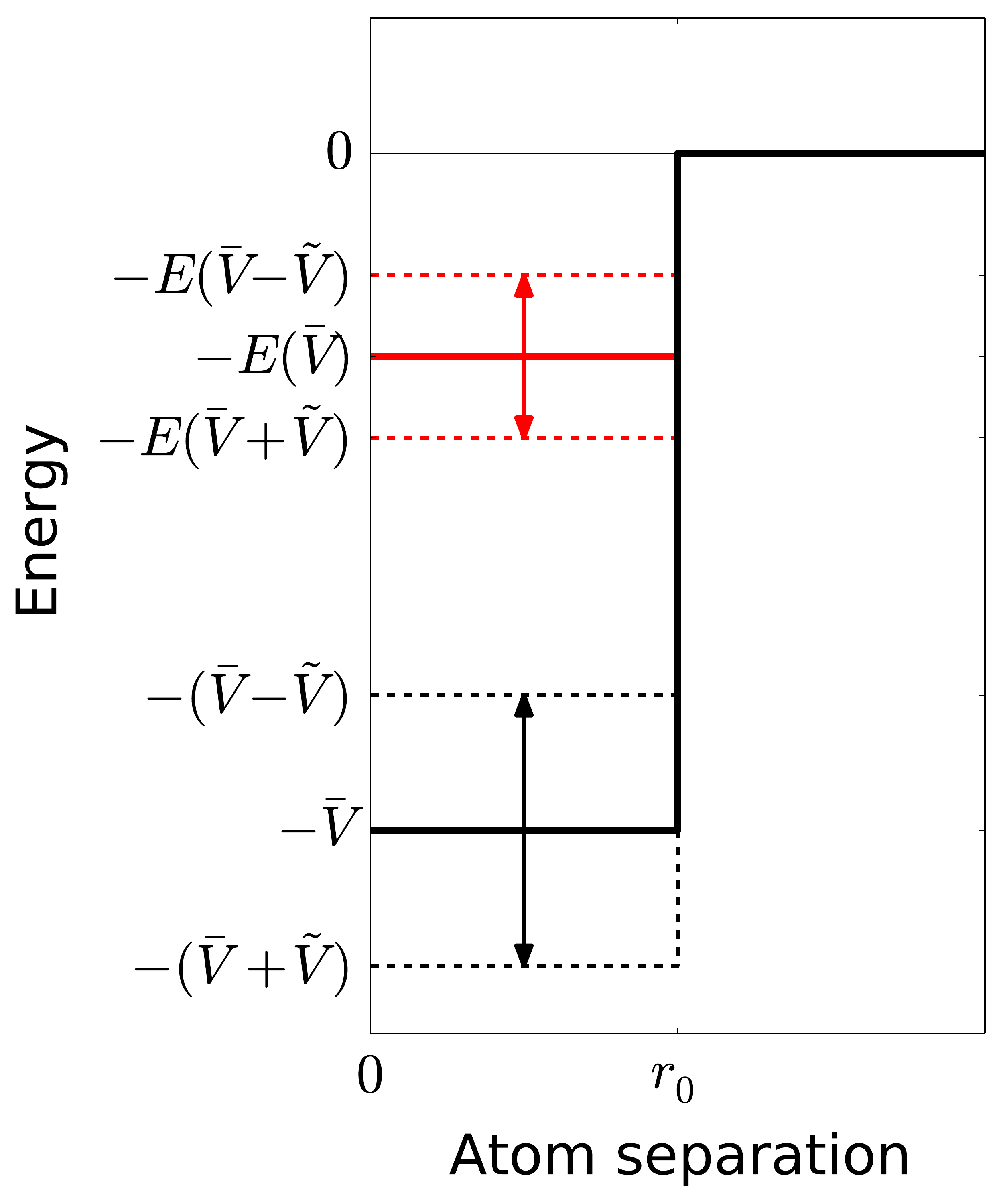}
\caption[The square-well interaction potential with oscillating depth.]{The square-well interaction potential with oscillating depth. In the adiabatic limit, the modulation in the depth of the potential, represented by the black double arrow, results in a modulation of the energy of the bound state, represented by the red double arrow. The black solid and dashed lines in the region $r<r_0$ represent the central and extremal values of the potential depth. The red solid and dashed lines represent the central and extremal values of the adiabatic bound-state energy.}
\label{fig:square_well}
\end{figure}
%%%%%%%%%%%%%%%%%%%%%%%%%%%%%%%%%%%%%%%%%%%%%%%%%%%%%%

Consider the binary collision of atoms interacting through a square-well potential with time-modulated depth (See Figure \ref{fig:square_well}). For atomic separations $r$ greater than the potential radius $r_0$, the atoms are non-interacting. For $r<r_0$, the atoms feel a potential $-[\Vbar + \Vtil\cos(\Omega t)]$. The time-averaged depth of the potential $\Vbar$ and the range $r_0$ are chosen such that the potential supports a bound state with binding energy $E_B$ when the oscillation amplitude $\Vtil$ is set to zero. We extract the $s$-wave scattering properties for atoms with mass $M$ from the solution to the time-dependent Schr{\"o}dinger equation for $u(r,t)=rR(r,t)$, where $R(r,t)$ is the radial wave-function for the separation of the atom pair:
%------------------------------------
\begin{align}
i\frac{d}{dt}u(r,t)&=-\frac{1}{M}\frac{\partial^2}{\partial r^2}u(r,t)
-\left[\Vbar+\Vtil\cos(\Omega t)\right]\theta(r_0-r)u(r,t).
\label{eq:single_channel_model}
\end{align}
%------------------------------------
The distribution $\theta(x)$ is the Heaviside theta function. Here and below, $\hbar$ is set to $1$. 

\section{Solving the time-dependent Schr{\"o}dinger equation}
Equation \eqref{eq:single_channel_model} can be solved by using Floquet's theorem as in Ref.~\cite{Reichl_1999}. Here we outline the derivation. For a detailed derivation, see Appendix \ref{App:square_well}. As discussed in Section \ref{Chap:Floquet}, Floquet's theorem asserts that the Schr{\"o}dinger equation with a time periodic potential, such as Eq.~\eqref{eq:single_channel_model}, has solutions of the form
%------------------------------------
\begin{equation}\label{eq:floquet_theorem}
u(r,t)=e^{-i\epsilon_\mathrm{F}t}\Phi(r,t),
\end{equation}
%------------------------------------
where $\epsilon_\mathrm{F}$ is the Floquet quasienergy, and $\Phi(r,t)$ has the same periodicity as the time-dependent part of the potential: 
%------------------------------------
\begin{equation}
\Phi(r,t)=\Phi(r,t+2\pi/\Omega). 
\end{equation}
%------------------------------------
The wavefunction for $r>r_0$ contains an incoming mode with energy $k^2/M$, where $k$ is the small relative momentum of the atom pair. The requirement that $u(r,t)$ be continuous at $r=r_0$ implies that $\epsilon_\mathrm{F}=k^2/M+j\Omega$, where $j$ is any integer (taken to be zero below for simplicity). Letting $\Phi (r,t)=g(r)f(t)$ and substituting Eq.~\eqref{eq:floquet_theorem} into
\eqref{eq:single_channel_model}, the Schr{\"o}dinger equation separates into equations for $g(r)$ and $f(t)$ that can be solved analytically. We then obtain the general solution for $u(r,t)$:
%------------------------------------
\begin{equation}\label{eq:usol} 
 u(r,t) = \sum\limits_{n=-\infty}^{\infty} \\
  \begin{cases}
   2ia_n \sin(q_nr)\exp\left[{-i(k_n^2/M)t+i\Vtil\sin(\Omega t)/\Omega}\right]
    & r < r_0, \\
   (A^\mathrm{out}_n e^{ik_n r}+A^\mathrm{in}_n e^{-ik_n r})\exp\left[{-i(k_n^2/M)t}\right]
   & r \geq r_0,
  \end{cases}
\end{equation}
%------------------------------------
where $k_n=[k^2+nM\Omega]^{1/2}$ and $q_n=[k^2+M(\Vbar+n\Omega)]^{1/2}$. The coefficients $a_n$ are determined by the boundary conditions at $r=r_0$. The solution for $r>r_0$ is a superposition of freely propagating modes. The coefficients $A^\mathrm{out}_n$ and $A^\mathrm{in}_n$ represent the amplitudes of outgoing and incoming modes, respectively. The $n=0$ mode corresponds to the low-energy scattering state with energy $k^2/M$. The additional so-called {\it Floquet modes} with energies that differ by integer multiples of $\Omega$ are necessary to satisfy the boundary conditions at $r=r_0$. Physically speaking, an incident wave with energy $k^2/M$ can absorb (emit) quanta with energy $\Omega$ from (to) the oscillating field. As a result, the outgoing wave is a superposition of modes with energies that differ from $k^2/M$ by integer multiples of $\Omega$. The negative-energy modes are exponentially damped for $r>r_0$, and they do not propagate. 

Both $u(r,t)$ and $\partial u(r,t)/\partial r$ must be continuous at $r=r_0$. Using the expansion
%--------------------------------
\begin{equation}
\exp\left[i\frac{\Vtil}{\Omega}\sin(\Omega t)\right]=\sum\limits_{j=-\infty}^{\infty}J_j(\Vtil/\Omega)\exp\left[-ij\Omega t\right],
\label{eq:besselexp}
\end{equation}
%--------------------------------
where $J_n(x)$ is the $n^\mathrm{th}$ Bessel function of the first kind, these boundary conditions can be used to eliminate the $a_n$ coefficients and find a relationship between the amplitudes of the incoming and outgoing modes:
%--------------------------------
\begin{equation}\label{eq:S_matrix}
A_n^{\mathrm{out}}=\sum\limits_{m=-\infty}^{\infty}S^{0}_{nm}(k)A_m^{\mathrm{in}},
\end{equation}
%--------------------------------
where
%--------------------------------
\begin{equation}\label{eq:S_matrix_def}
S^{0}_{nm}(k)=\sum\limits_{j=-\infty}^\infty (W_-)_{nj}(W_+^{-1})_{jm}.
\end{equation}
%--------------------------------
$W_+$ and $W_-$ are the infinite-dimensional matrices with matrix elements
%----------------------------
\begin{align}\label{eq:M1M2}
(W_\pm)_{jn} &= \frac{e^{\pm ik_jr_0}}{k_j}
\left[(k_j\mp q_n)e^{iq_nr_0}
-
(k_j \pm q_n)e^{-iq_nr_0}\right]J_{j-n}(\Vtil/\Omega).
\end{align}
%--------------------------------

$S^{0}(k)$ is the $l=0$ partial wave $S$-matrix element defined in Eq.~\eqref{eq:Spartial}. We will  assume that all of the incoming flux is in the $m=0$ channel with energy $k^2/M$. Note that some of the elements $S^0_{nm}(k)$ correspond to either incoming or outgoing Floquet channels with negative energy and are therefore not true $S$-matrix elements in the usual sense. All scattering observables can be calculated from $S^{0}(k)$. The $s$-wave Floquet scattering amplitudes $f_n^0(k)$, defined in Eq.~\eqref{eq:sphericalHarmonics}, can be extracted from $S^{0}(k)$ using Eq.~\eqref{eq:ftoS_partial}:
%-------------------------------
\begin{equation}
f^0_{n}(k)=\frac{S^0_{n0}(k)-\delta_{n,0}}{2ik}.
\label{eq:fnk_square}
\end{equation}
%-------------------------------
With this result in hand, we can calculate the cross section into the $n^{\mathrm{th}}$ Floquet level, $\sigma_n(k)$, using Eq.~\eqref{eq:dsigmadomega_2}, giving
%-------------------------------
\begin{equation}
\sigma_n(k)=4\pi\frac{k_n}{k}|f^0_n(k)|^2.
\label{eq:sigma_n_square}
\end{equation}
%-------------------------------
The total cross section is then obtained by summing over positive-energy Floquet channels as in Eq.~\eqref{eq:totalcross}. The negative-energy Floquet channels correspond to resonance states with wavefunctions that are exponentially suppressed at large separation.

In practice, one must truncate the infinite sum that appears in the definition of the partial-wave $S$-matrix element, Eqs.~\eqref{eq:S_matrix_def}, to a finite range $j\in[-j_{\mathrm{max}},j_{\mathrm{max}}]$. This truncation is guaranteed to converge for sufficiently large values of $j_{\mathrm{max}}$, even for large driving amplitudes $\Vtil$. This is because the coupling between Floquet modes separated by energy $n\Omega$ is numerically suppressed by a factor of $n!$ for large $n$. For weakly-driven systems, convergence can often be achieved with fewer than six Floquet modes. For strongly-driven systems, one may need to include dozens or even hundreds of Floquet modes to reach convergence. 

\section{Atomic scattering in the zero-range limit}
As discussed in Chapter \ref{chap:exper::sec:tuning}, it is possible to control the effective $s$-wave scattering length $a$ by applying an external magnetic field. By applying an oscillating magnetic field with frequency $\Omega$, one can induce the effectively time-periodic scattering length, Eq.~\eqref{eq:aoft}. We would like to choose the parameters of the square-well model to reproduce the physics of the atoms scattering with an effective scattering length $a(t)$. After performing this matching, we can make quantitative statements about the physical system using the results from the square-well model. 

We assume that the effective range is zero so that the properties of the two-atom system only depend upon the periodic function $a(t)$. We also assume that $a(t)$ is positive so that in the adiabatic limit ($\Omega\rightarrow 0$) the physical system supports a bound state with energy $-1/(Ma(t)^2)$. Assuming finally that $\atil\ll \abar$, we can expand the energy of the bound state as a power series in $\atil$ keeping only the first time-dependent term:
%-------------------------
\begin{equation}
-\frac{1}{Ma(t)^2} =-\frac{1}{M\abar^2}\left[1 + \frac{2\atil}{\abar}\cos(\Omega t) + \mathcal{O}\left((\atil/\abar)^2\right)\right].
\label{eq:abinding}
\end{equation}
%-------------------------
On the other hand, the square-well model in the adiabatic limit supports a bound state with binding energy $E(V(t))$, as represented by the horizontal red lines in Figure \ref{fig:square_well}. Though the bound-state energy $E$  depends upon the range $r_0$, in the following discussion, we will hold $r_0$ fixed and only vary the depth of the potential. The $s$-wave bound-state energies $E$ in the spherical well with depth $V$ and range $r_0$ are the roots of the transcendental equation
%-------------------------
\begin{equation}
\sqrt{\frac{V-|E|}{|E|}} = - \tan\sqrt{(V-|E|)Mr_0^2},
\end{equation}
%-------------------------
where, since $E$ is the energy of a bound state, $E = -|E|$. This equation can be solved numerically to determine $E(V)$. For small $\Vtil$, we can expand $E(V(t))$ as a power series in $\Vtil$ keeping only the first time-dependent term:
%-------------------------
\begin{equation}
E(V(t)) = E(\Vbar) +\frac{\partial E(\Vbar)}{\partial \Vbar}\Vtil\cos(\Omega t) + \mathcal{O}(\Vtil^2).
\label{eq:squarebinding}
\end{equation}
%-------------------------
We now choose the parameters of the square-well model to reproduce the physics of the zero-range atomic system with time-dependent scattering length $a(t)$. By comparing the expansions in Eqs.~\eqref{eq:abinding} and Eq.~\eqref{eq:squarebinding} and remembering that the effective range is zero, we obtain the following matching criteria:
%-------------------------
\begin{align}
\lim_{r_0\rightarrow 0}E(\Vbar,r_0) &= -\frac{1}{M\abar^2}, \nonumber \\
\lim_{r_0\rightarrow 0}\frac{\partial E(\Vbar,r_0)}{\partial \Vbar}\Vtil &= -\frac{2\atil}{M\abar^3}.
\label{eq:matching}
\end{align}
%-------------------------
These conditions determine $\Vbar$ and $\Vtil$; however, the solution is ill-defined because in the limit $r_0\rightarrow 0$, the potential depth $\Vbar$ must approach $\infty$ as $\sim 1/(Mr_0^2)$ in order to sustain the bound state. In practice, we can simply fix the value of $r_0$ and tune the value of $\Vbar$ so that $\abar \gg r_0$. Using this procedure, the ratio of $\abar/r_{0}$ becomes extremely sensitive to the choice of $\Vbar$ as the ratio becomes large. This has no physical consequences, because the physics is only sensitive to the value of $\abar$ in this limit. However, from a numerical point of view, this sensitivity places a practical limit on the maximum value of $\abar/r_0$ that can be encoded in the parameters of the square-well model.

\section{Induced scattering resonances}
%%%%%%%%%%%%%%%%%%%%%%%%%%%%%%%%%%%%%%%%%%%%%%%% 
\begin{figure}
\centering
\includegraphics[width=0.85\textwidth]{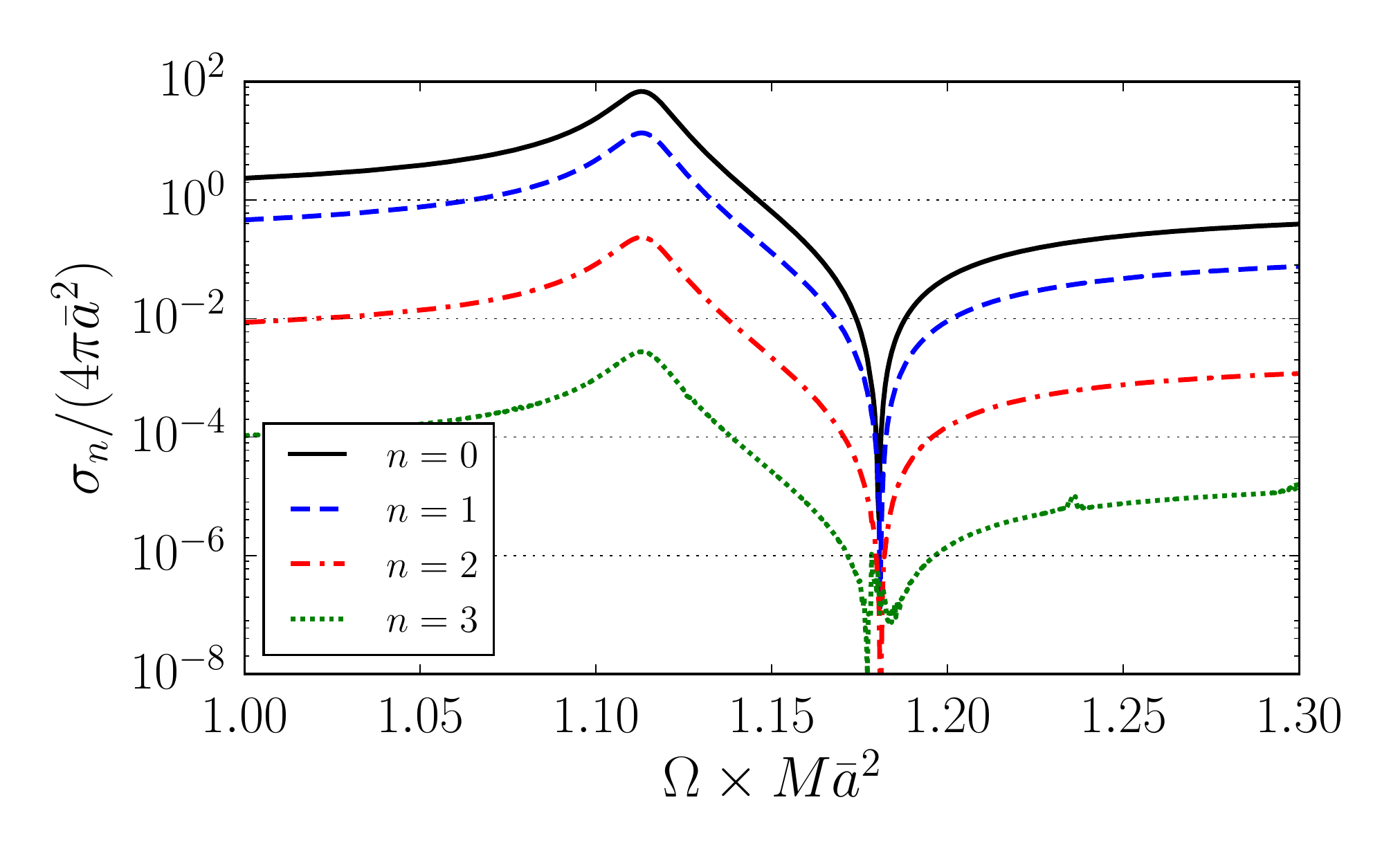}
\caption[Frequency-dependent Floquet-component cross sections, square-well model]{The Floquet-channel cross sections $\sigma_n$, $n\in\{0,1,2,3\}$, from the square-well model as functions of $\Omega$ for $\atil/\abar = 0.4$ and $k\abar = 0.1$. The vertical axis is scaled by the value of the total cross section in the absence of modulation. The horizontal axis is scaled by the binding energy of the resonance molecule in the absence of modulation. For this choice of parameters, the $n=0$ mode is the largest. The kinks in $\sigma_3$ are numerical artifacts.}
\label{fig:sigmaChannel_square}
\includegraphics[width=0.85\textwidth]{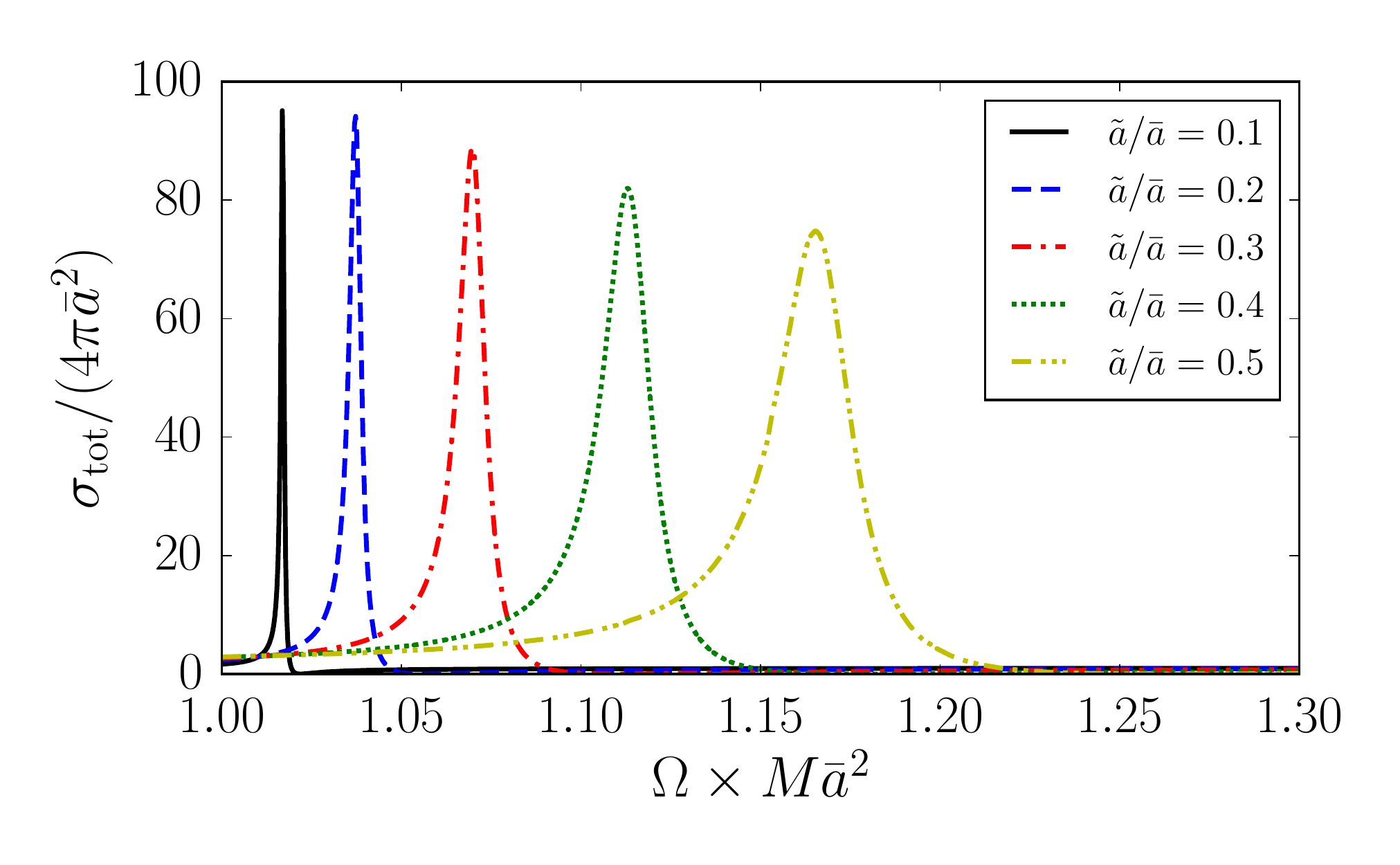}
\caption[Frequency-dependent total cross section, square-well model]{The total cross-section $\sigma_{\mathrm{tot}}$ from the square-well model as a function of $\Omega$ for left-to-right increasing values of $\atil / \abar$ and $k\abar=0.1$. Larger values of $\atil$ correspond to broader and slightly shorter resonances. The position of the peak grows quadratically with $\atil/\abar$.}
\label{fig:sigmaTotal_square}
\end{figure}
%%%%%%%%%%%%%%%%%%%%%%%%%%%%%%%%%%%%%%%%%%%%%%%%%%%%%%%%%% 
%%%%%%%%%%%%%%%%%%%%%%%%%%%%%%%%%%%%%%%%%%%%%%%%%%%%%%%%%% 
\begin{figure}
\centering
%\includegraphics[width=0.85\textwidth]{Chapters/PeriodicPotentials/Figures/sigmaKDisc2.pdf}
%\caption[Frequency-dependent total cross section for different momenta]{The total cross-section $\sigma_{\mathrm{tot}}$ as a function of $\Omega-k^2/M$ for three small values of the scattering momentum $k$ (shown in the plot legend) and for $\atil / \abar=0.4$. The axis dimensions are identical to those in Fig.~\ref{fig:sigmaChannel_square}. The horizontal dotted line shows the value of the total cross section in the absence of modulation.}
%\label{fig:sigmaKDisc_square}
%
\includegraphics[width=0.85\textwidth]{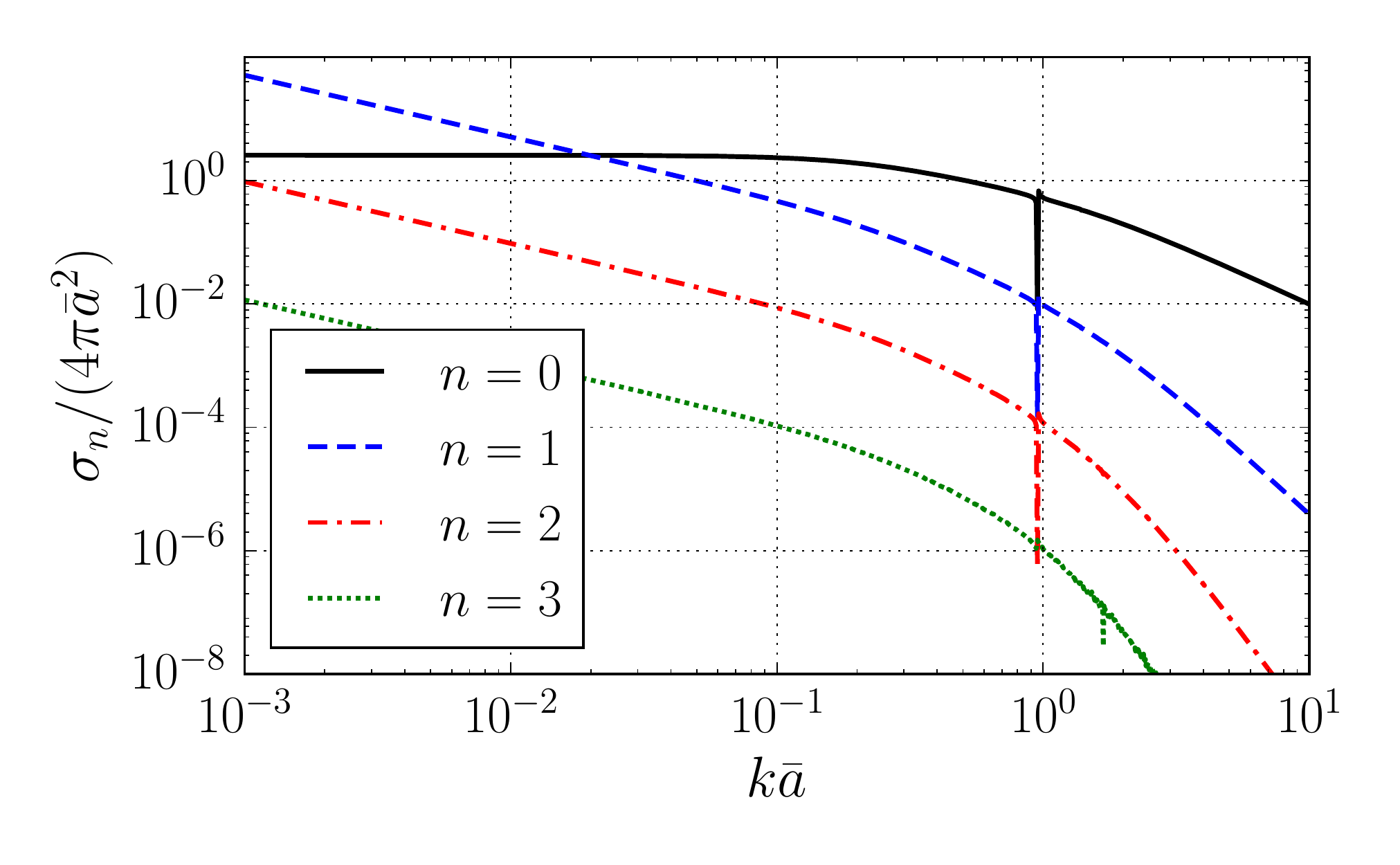}
\caption[Momentum-dependent Floquet-component cross sections, square-well model]{The Floquet-channel cross-sections $\sigma_n$, $n\in\{0,1,2,3\}$, from the square-well model as functions of momentum $k$ for $\atil/\abar = 0.4$ and $\Omega\times M \abar^2 = 1$ (chosen arbitrarily). The $n > 0$ modes grow as $k^{-1}$ as $k\to 0$. A narrow sub-harmonic resonance is visible in each channel near $k\abar=1$. The noise in $\sigma_3$ for $k\abar \gtrsim 1$ is a numerical artifact.}
\label{fig:sigmaKCont_square}
\end{figure}
%%%%%%%%%%%%%%%%%%%%%%%%%%%%%%%%%%%%%%%%%%%%%%%%%%%%%%%%%% 

With the choice of potential parameters specified by the matching procedure in Eqs.~\eqref{eq:matching}, the square-well model maps onto the physical system with the time-dependent scattering length $a(t)$ in Eq.~\eqref{eq:aoft}. We perform the matching procedure numerically for $\abar/r_0 = 100$ and for several different values of $\atil/\abar$. We then extract the Floquet scattering amplitudes $f^0_n(k)$ using Eq.~\eqref{eq:fnk_square}. Using these amplitudes, we then calculate the Floquet component cross sections $\sigma_n(k)$ using Eq.~\eqref{eq:sigma_n_square} as well as the total cross section $\sigma_{\mathrm{tot}}(k)$.

Figures \ref{fig:sigmaChannel_square} and \ref{fig:sigmaTotal_square} explore the the $\Omega$-dependence of the zero-range scattering cross sections in detail. These figures demonstrate that the scattering cross sections are dramatically altered by the application of the oscillating scattering length $a(t)$ in Eq.~\eqref{eq:aoft}. Figure \ref{fig:sigmaChannel_square} shows the Floquet-channel cross-sections $\sigma_n$, $n\in\{0,1,2,3\}$, as functions of $\Omega$ for scattering momentum $k\abar=0.1$ and driving strength $\atil/\abar=0.4$. For this choice of parameters, the $n=0$ cross section obtains a maximum value that is nearly two orders of magnitude larger than its value in the absence of modulation. As is clear from the plot, the $n>0$ cross-sections have the same shape as the $n=0$ resonance. For $|\atil/\abar|\ll 1$, the ratio $\sigma_n/\sigma_{n-1}$ is proportional to $(\atil/\abar)^2$, leading to the suppression of scattering into higher Floquet modes. All component cross sections are dramatically suppressed for a value of $\Omega$ slightly less than $1.2/(M\abar^2)$. Figure \ref{fig:sigmaTotal_square} shows the total cross section as a function of $\Omega$ for five different values of the driving amplitude $\atil/\abar$ from $0.1$ to $0.5$. Larger values of $\atil$ correspond to broader and slightly-shorter asymmetric resonances. 

Figure \ref{fig:sigmaKCont_square} shows the momentum dependence of the Floquet-component cross sections at a fixed value of $\Omega$. This figure demonstrates that $\sigma_0$ approaches a constant as $k\to 0$ and that the $n>0$ cross sections grow as $k^{-1}$ as $k\to 0$. At sufficiently small $k$, this enhancement overcomes the $(\atil/\abar)^{2n}$ suppression, resulting in large cross sections for scattering into higher Floquet modes. This enhancement results from the ratio $k_n/k$ in Eq.~\eqref{eq:sigma_n_square} becoming very large at small values of $k$ for $n>0$. Also evident in Fig.~\ref{fig:sigmaKCont_square} is a narrow sub-harmonic resonance in each Floquet channel near $k\abar = 1$. At this value of $k$, the energy of the scattering atoms is larger than the dimer energy by $2\Omega$.

\section{Effective range expansion}
%%%%%%%%%%%%%%%%%%%%%%%%%%%%%%%%%%%%%%%%%%%%%%%%%%%%%%%%%% 
\begin{figure}
\centering
\includegraphics[width=0.9\textwidth]{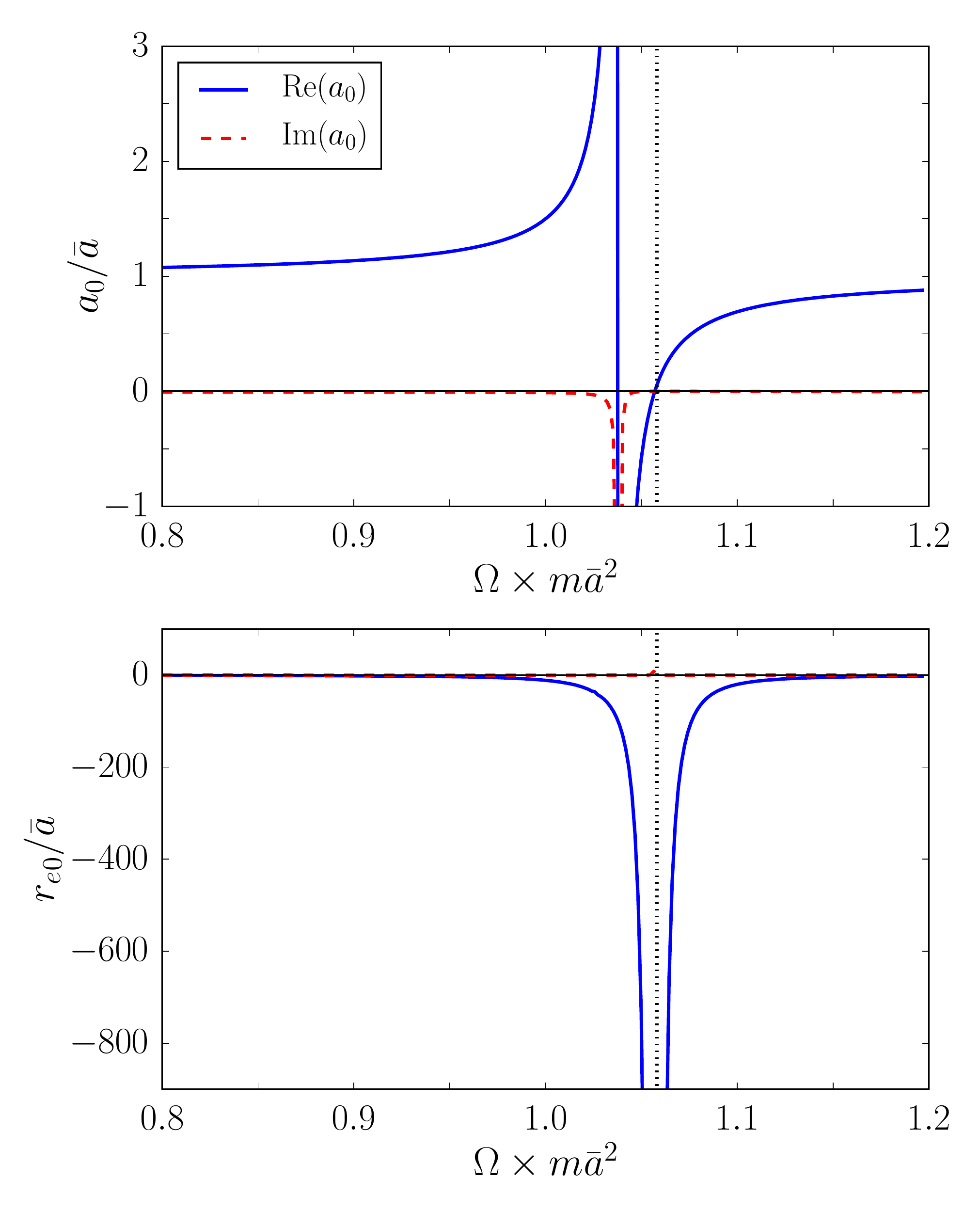}
\caption[The effective range parameters $a_0$ and ${r_e}_0$.]{\textbf{Top Panel}: The real part (solid line) and imaginary part (dashed line) of the effective scattering length $a_0$ as a function of the oscillation frequency $\Omega$. The dotted line marks the position of the zero crossing in the real part of $a_0$. \textbf{Bottom panel}: the real part (solid line) and imaginary part (dashed line) of ${r_e}_0$. The magnitude of the real part of ${r_e}_0$ becomes very large and negative at the zero crossing in the real part of $a_0$ (marked by the vertical dotted line) .}
\label{fig:a0}
\end{figure}
%%%%%%%%%%%%%%%%%%%%%%%%%%%%%%%%%%%%%%%%%%%%%%%%%%%%%%%%%%  

As discussed in Sec.~\ref{chap:scat::sec:thy::subsec:ERE}, the low-energy $s$-wave observables for the Floquet system can be described in terms of a few frequency-dependent effective-range parameters $a_n,\,{r_e}_n,$ \textit{etc.}, where $n$ labels the outgoing Floquet channel. Subsequent terms in the effective range expansion are suppressed by higher powers of $k^2$. For the effective range expansion to be useful, the largest relevant momentum scale $k_{\mathrm{max}}$ should be well within the radius of convergence the expansion, such that only a small number of effective range parameters are needed
to describe the physics. 

Using Eq.~\eqref{eq:ERE}, we calculate the parameters $a_n$ and ${r_e}_n$ for $n=0$. We focus on the $n=0$ Floquet channel because of our observation from Figs.~\ref{fig:sigmaChannel_square} and \ref{fig:sigmaKCont_square} that for some values of the scattering momentum, $\sigma_0$ is the largest Floquet-component cross section. Figure \ref{fig:a0} shows $a_0$ and ${r_e}_0$ as functions of $\Omega$. The real part of $a_0$ is resonantly enhanced for a value of $\Omega$ close to $1/(M\abar^2)$. The shape of this resonance as a function of $\Omega$ is very similar to the shape of a magnetic Feshbach resonance as a function of the magnetic field shown in Fig.~\ref{fig:Feshbach}. The magnitude of the imaginary part of ${a}_0$ becomes very large within a narrow range of frequencies surrounding the resonance. The real part of ${r_e}_0$ becomes extremely large at the value of $\Omega$ where the real part of $a_0$ equals zero. We discuss the range of validity of the effective range expansion in Chapter~\ref{Chap:Applications}.

\begin{figure}[t]
\centering
\includegraphics[width=0.85\textwidth]{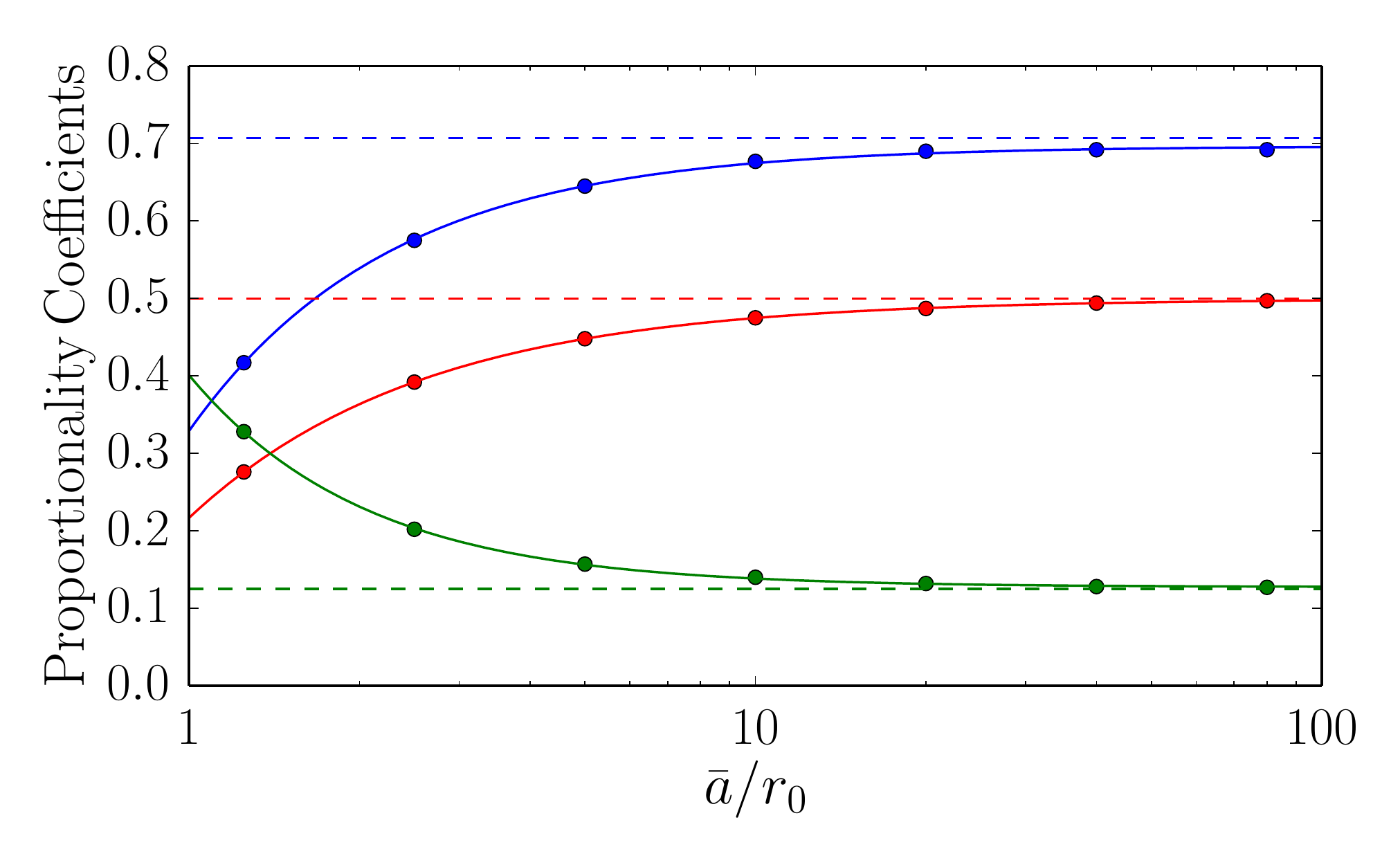}
\caption[Universal limit of the resonance parameter coefficients]{Numerical results for the coefficients of $\alpha^2$ for the dimensionless resonance parameters $\Omega_0\times M\abar^2$, $\delta\times M\abar^2$, and $\gamma\abar$ (top to bottom) as functions of $\abar/r_0$ along with power-law fits to guide the eye. In the large-$\abar$ limit, the coefficients approach the universal numbers given in Eqs.~\eqref{eq:universal}, represented by the dashed lines.}
\label{fig:universallimit}
\end{figure}

The functional forms of $a_0$ and ${r_{e}}_0$ are well parametrized by the expressions
%------------------------------
\begin{align}
\frac{\abar}{a_0} &= \frac{\Omega - \Omega_0}{\Omega - \Omega_0 - \delta}+i\gamma\abar ,
\nonumber \\
\frac{{r_e}_0}{\abar} & = \frac{1/(M\abar^{2})^2}{\left(\Omega - \Omega_0 - \delta\right)^2}\,\alpha^2,
\label{eq:a0r0_square}
\end{align}
%------------------------------
where $\alpha \equiv \atil/\abar$. We have defined the resonance parameters
%------------------------------
\begin{align}
\Omega_0\cdot M\abar^2 &= 1 + 0.69\,\alpha^2,
\nonumber \\ 
\delta \cdot M\abar^2 &= 0.50\,\alpha^2,
\nonumber \\
\gamma  \abar & = 0.13\,\alpha^2.
\label{eq:universal}
\end{align}
%------------------------------
The numbers $0.69$, $0.50$, and $0.13$ were determined as best-fit parameters in fitting Eq.~\eqref{eq:a0r0_square} to numerical results for $a_0$. Note that
%------------------------------
\begin{align}
\frac{1}{a_0} & \underset{\alpha\to 0}{\longrightarrow}  \frac{1}{\abar}
\nonumber \\
{r_s}_0 & \underset{\alpha\to 0}{\longrightarrow} 0,
\end{align}
%------------------------------
exactly as we expect, since taking $\alpha\to 0$ is equivalent to having a stationary, zero-range interaction potential with scattering length $\abar$. The results in Eq.~\eqref{eq:universal} are universal in the sense that any system with $r_0\ll \abar$ will have these effective frequency-dependent range parameters. Figure \ref{fig:universallimit} plots the dependence of the coefficients of $\alpha^2$ in Eq.~\eqref{eq:universal} as functions $\abar/r_0$. For $\abar/r_0=10$ the coefficients are already within a few percent of their universal values.

%%%%%%%%%%%%%%%%%%%%%%%%%%%%%%%%%%%%%%%%%%%%%%%%%%%%%%%%%% 

%% file: Chapters/Applications/Applications.tex
\cleardoublepage

\chapter{Zero-range model}
\label{Chap:Applications}

In the last chapter, we used a square-well model to derive the scattering properties of atoms with the time-dependent scattering length $a(t) = \abar + \atil \cos(\omega t)$ and zero range. We obtained numerical results for the universal resonance parameters associated with a Floquet resonance (see Eq.~\eqref{eq:universal}). We now solve the same problem using an explicitly zero-range model. Instead of explicitly solving the Schr{\"o}dinger equation as we did in Chap.~\ref{chap:atomic_scattering}, we use the Floquet Lippmann-Schwinger Equation (FLSE) derived in Sec.~\ref{Chap:Formalism}. We find excellent agreement between the predictions of the zero-range model and the square-well model in the zero-range limit. In addition, we obtain analytic results for the resonance parameters associated with a Floquet resonance, determined numerically in Chapter~\ref{chap:atomic_scattering}.

\section{Time-periodic contact potential}
Consider the binary collision of atoms interacting through a zero-range (``contact'') potential with time-modulated strength. This problem can be formulated in terms of the effective field theory for atoms with large scattering length outlined in Sec.~\ref{chap:scat::sec:EFT}. The quantum field operator for the contact interaction is given in Eq.~\eqref{eq:L}. To obtain the contact interaction for particles with the time-dependent scattering length $a(t)$ given in Eq.~\eqref{eq:aoft}, we simply substitute $a(t)$ for $a$ in the definition of the bare coupling constant $g$, given in Eq.~\eqref{eq:g}.

The FLSE for $s$-wave scattering depends upon the matrix elements of the $s$-wave component of the harmonic potential components, $V^0_n(p,q)$, defined in Eq.~\eqref{eq:FLSEpartialwave}. To calculate the Floquet components $V^0_n(p,q)$, we first need to evaluate $\matel{\bm{p}}{\bm{q}}{V_n}$, where $V_n$ are the harmonic components of the potential defined in Eq.~\eqref{eq:Vn}. To determine the matrix elements of the harmonic components, we first calculate the expectation value $\matel{\bm{p}}{\bm{q}}{V(t)}$ and then calculate the harmonic components of the result. Using the effective field theory discussed in Sec.~\ref{chap:scat::sec:EFT}, we can easily evaluate this matrix element, giving
%-------------------------------
\begin{equation}
\matel*{\bm{p}}{\bm{q}}{V(t)} = \matel*{\bm{p}}{\bm{q}}{\frac{g(t)}{M}\psi_1^\dagger \psi_2^\dagger \psi_2 \psi_1} = \frac{g(t)}{M}.
\label{eq:potential_matrix_elements_time_dependent}
\end{equation}
%-------------------------------
The time-dependent coupling is
%-------------------------------
\begin{equation}
g(t) = \frac{4\pi}{1/a(t)-2\Lambda/\pi}.
\label{eq:goft}
\end{equation}
%-------------------------------
The asymptotic states $\ket{\bm{p}}$ and $\ket{\bm{q}}$ are the non-interacting eigenstates with no initial-state or final-state interactions. The effects of interactions are taken into account through the structure of the FLSE.

To treat the case where $1/a(t)$ has small deviations from $1/\abar$, we expand $g(t)$ in powers of $1/a(t)-1/\abar$. Following Ref.~\cite{Mohapatra:pra:2015}, we drop terms of order $(1/a(t)-1/\abar)^2$ and higher because these terms are suppressed by higher powers of $1/\Lambda$. We then expand $1/a(t)-1/\abar$ in powers of $\atil/\abar$. Keeping terms up to order $\atil/\abar$ we find
%------------------------------
\begin{equation}
g(t)=\bar g\left[1 + \frac{\bar g}{4\pi\abar}\frac{\atil}{\abar}\cos(\omega t) + \mathcal{O}\big((\atil/\abar)^2\big) \right].
\label{eq:gexp}
\end{equation}
%------------------------------
We can now read off the Fourier components of $g(t)$:
%------------------------------
\begin{align}
g_0 =\bar g,~~~
g_{\pm 1} = \frac{\bar g^2}{8\pi\abar}\,\frac{\atil}{\abar}, 
\label{eq:gns}
\end{align}
%------------------------------
with all other components equal to zero. It also follows directly that $V^0_n(p,q)$ = $g_n/M$. 

\section{Solving the Lippmann-Schwinger equation}

%%%%%%%%%%%%%%%%%%%%%%%%%%%%%%%%%%%%%%%%%%%%%%%% 
\begin{figure}
\centering
\includegraphics[width=0.85\textwidth]{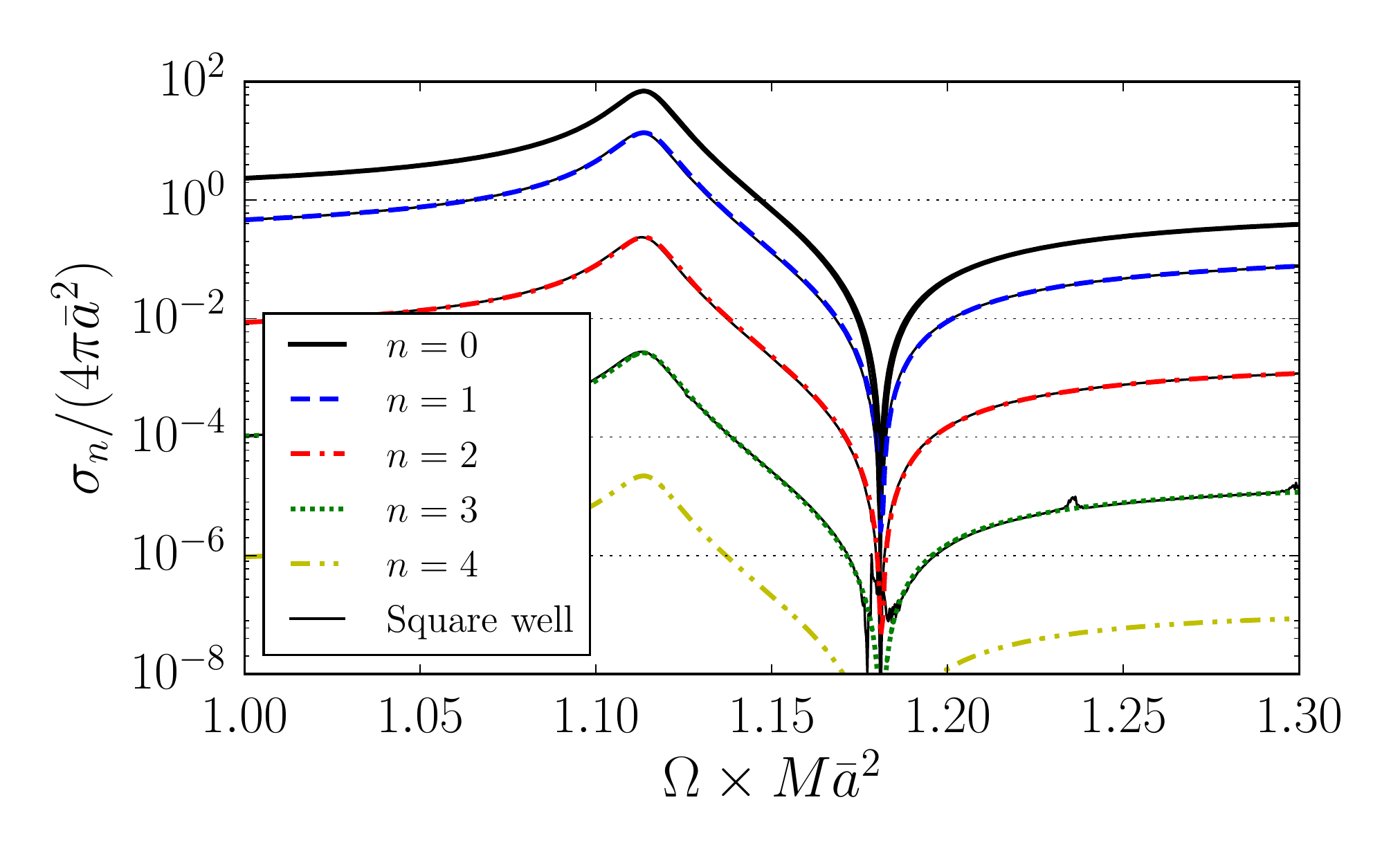}
\caption[Frequency-dependent Floquet-component cross sections, zero-range model]{The Floquet-channel cross sections $\sigma_n$, $n\in\{0,1,2,3,4\}$, from the zero-range model as functions of $\Omega$ for $\atil/\abar = 0.4$ and $k\abar = 0.1$. The thin black curves (nearly indistinguishable from each $\sigma_n$, $n\in\{0,1,2,3\}$) are the corresponding results from the square-well model.}
\label{fig:sigmaChannel}
\includegraphics[width=0.85\textwidth]{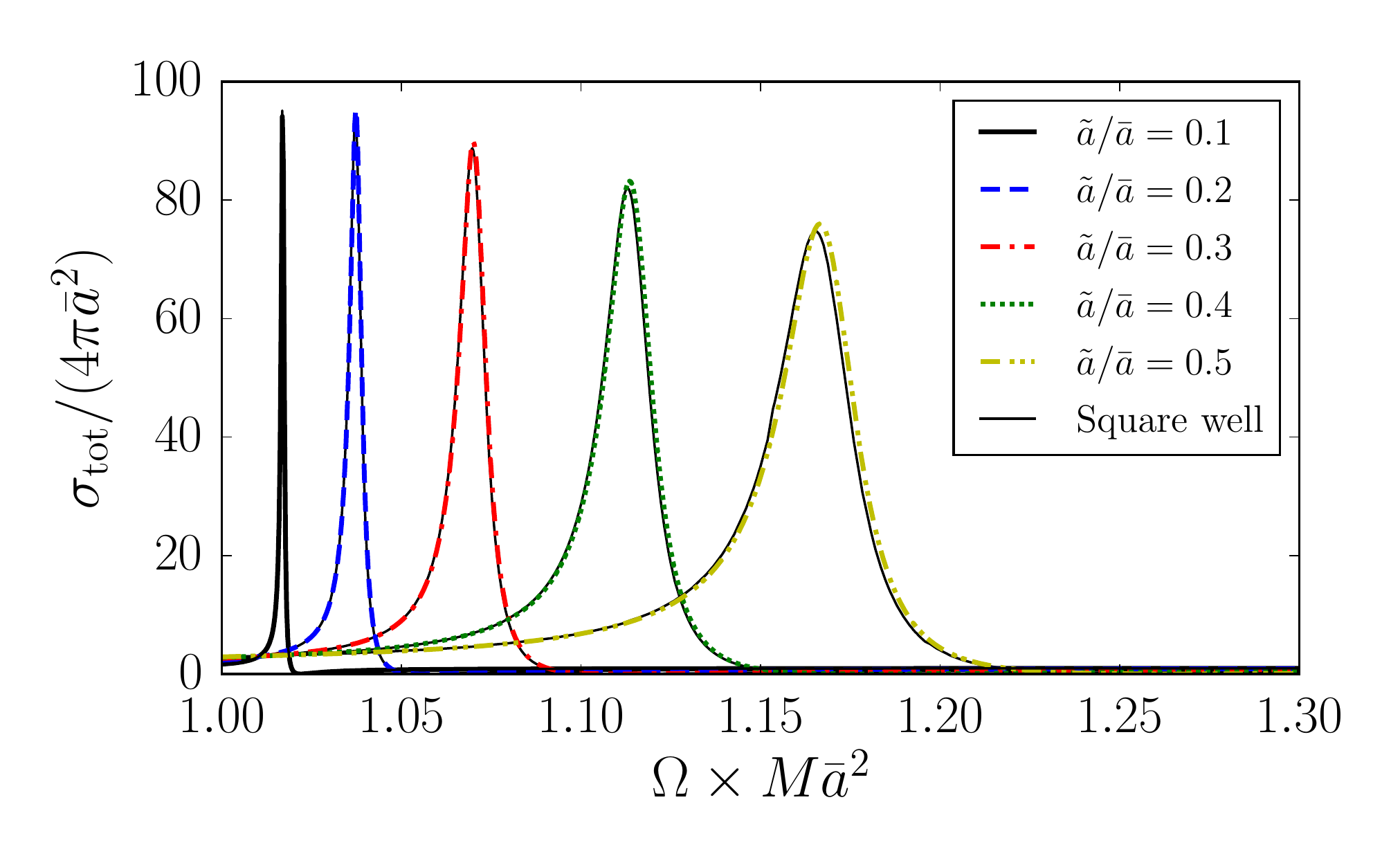}
\caption[Frequency-dependent total cross section, zero-range model]{The total cross-section $\sigma_{\mathrm{tot}}$ from the zero-range model as functions of $\Omega$ for left-to-right increasing values of $\atil / \abar$ and $k\abar=0.1$. The thin black curves (nearly indistinguishable from the zero-range results) are the corresponding results from the square-well model.}
\label{fig:sigmaTotal}
\end{figure}
%%%%%%%%%%%%%%%%%%%%%%%%%%%%%%%%%%%%%%%%%%%%%%%%%%%%%%%%%% 
%%%%%%%%%%%%%%%%%%%%%%%%%%%%%%%%%%%%%%%%%%%%%%%%%%%%%%%%%% 
\begin{figure}
\centering
\includegraphics[width=0.85\textwidth]{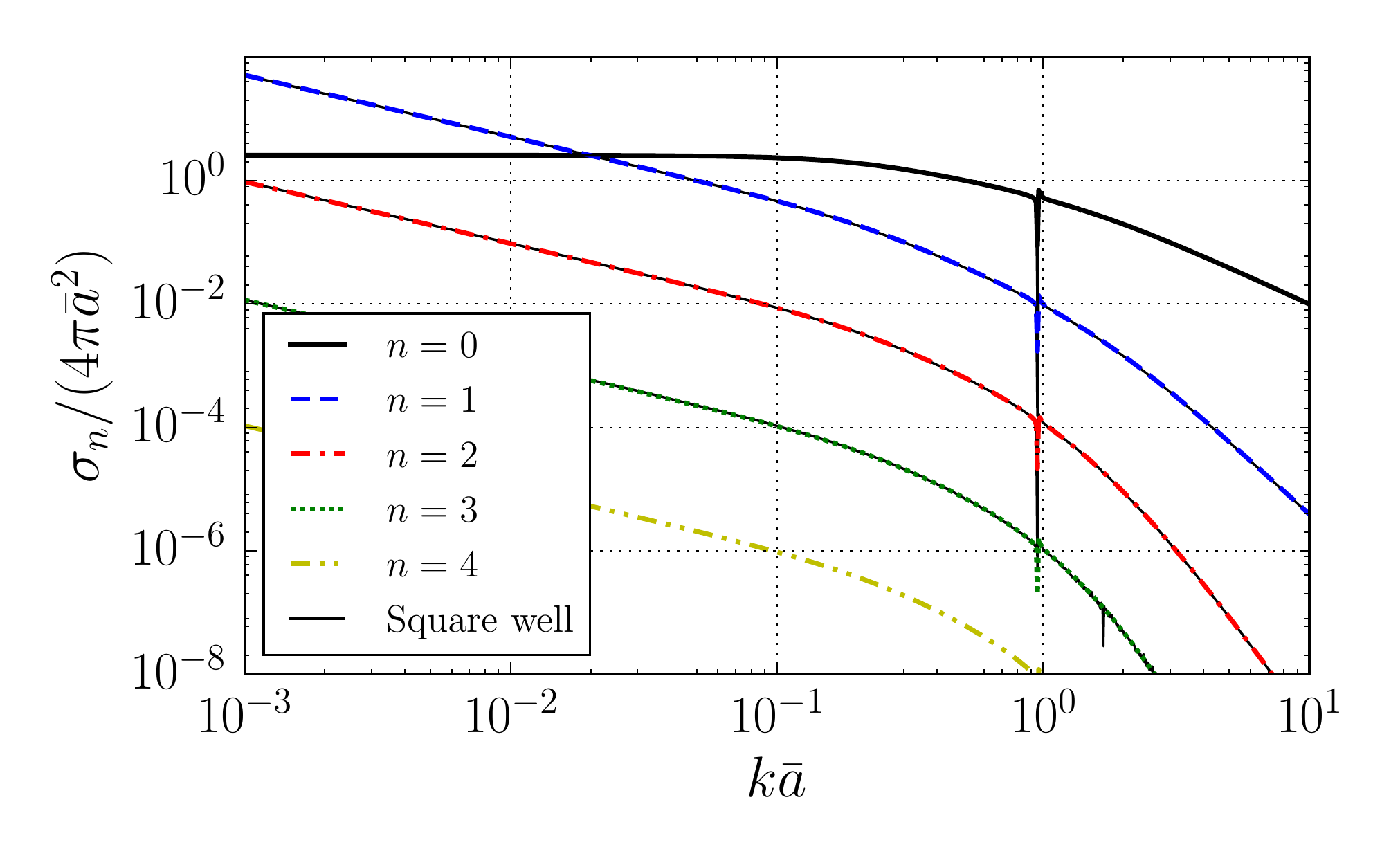}
\caption[Momentum-dependent Floquet-component cross sections, zero-range model]{The Floquet-channel cross-sections $\sigma_n$, $n\in\{0,1,2,3,4\}$, from the zero-range model as functions of momentum $k$ for $\atil/\abar = 0.4$ and $\Omega\times M \abar^2 = 1$ (chosen arbitrarily). The thin black curves (nearly indistinguishable from each $\sigma_n$, $n\in\{0,1,2,3\}$) are the corresponding results from the square-well model.}
\label{fig:sigmaKCont}
\end{figure}
%%%%%%%%%%%%%%%%%%%%%%%%%%%%%%%%%%%%%%%%%%%%%%%%%%%%%%%%%% 

For contact interactions, the potential matrix elements $V_n^l(p,q)$ are independent of $p$ and $q$. This significantly simplifies the partial wave FLSE, Eq.~\eqref{eq:FLSEpartialwave}, which becomes
%-------------------------------
\begin{align}
f_n^0(p,k)
&= -\frac{g_n}{4\pi}
-\sum\limits_{j=-\infty}^\infty \frac{g_{n-j}}{2\pi^2} \int\limits_0^\Lambda dqq^2
\left(q^2 - k^2 - jM\Omega - i0^+\right)^{-1} f_j^0(q,k),
\label{eq:FLSE_contact}
\end{align}
%-------------------------------
where we have introduced a cutoff $\Lambda$ in the $q$-integral for consistency with our use of cutoff regularization in defining the contact potential in Eq.~\eqref{eq:goft}. The amplitudes $f_n^0(p,k)$ are finite in the physical limit $\Lambda\rightarrow\infty$. Since the right-hand side depends upon $k$ but not $p$, the amplitude $f_n^0(p,k)$ is independent of $p$. We can then relabel the amplitude as $f_n^l(k)\equiv f^l_n(p,k)$. It follows that the amplitudes $f_j(q,k)$ which appear inside the $q$-integral on the right-hand side of Eq.~\eqref{eq:FLSE_contact} are actually independent of $q$ and can be pulled outside of the integral. The FLSE then reduces to 
%-------------------------------
\begin{align}
\sum\limits_{j=-\infty}^\infty\left[\delta_{n,j}+\frac{g_{n-j}}{2\pi^2} I(k_j^2/M)\right] f^0_j(k)
&= -\frac{g_n}{4\pi},
\label{eq:FLSE_contact_2}
\end{align}
%-------------------------------
where $k_j = \sqrt{k^2 + jM\Omega}$, and $I(E)$ is the integral defined in Eq.~\eqref{eq:Iintegral}. We recover the solution to the time-independent, zero-range LSE in Eq.~\eqref{eq:f0_EFT_sol2} by setting $g_n=0$ for $n\neq 0$ and $g_0=g$, where $g$ is the bare coupling defined in Eq.~\eqref{eq:g}.

We see from Eq.~\eqref{eq:FLSE_contact_2} that the FLSE reduces to a set of coupled linear equations for the scattering amplitudes $f^0_n(k)$. Defining the $W$-matrix with elements
%------------------------------
\begin{align}
W_{nj} = \delta_{n,j}+\frac{g_{n-j}}{2\pi^2}I(k_j^2/M),
\label{eq:Wmat}
\end{align}
%------------------------------
we can write the FLSE in a compact form using matrix notation:
%------------------------------
\begin{align}
W\bm{f} = -\frac{1}{4\pi}\bm{g}
\label{eq:fsolution}
\end{align}
%------------------------------
where $\bm{f}$ and $\bm{g}$ are vectors with elements
%------------------------------
\begin{align}
(\bm{f})_n &= f_n^0(k) \nonumber \\
(\bm{g})_n &= g_n
\label{eq:fgelements}
\end{align}
%------------------------------
To solve the FLSE, we must replace the infinite-dimensional matrix $W$ and the infinite-dimensional vectors $\bm{f}$ and $\bm{g}$ with the truncated objects $W_N$, $\bm{f}_N$, and $\bm{g}_N$, where the subscript $N$ denotes that the absolute values of all indices are constrained to be less than or equal to the positive integer $N$. $W_N$ is a $(2N+1)\times(2N+1)$ dimensional matrix; $\bm{f}_N$ and $\bm{g}_N$ are $2N+1$ dimensional vectors. 

The truncation approximation introduced above relies upon the convergence condition
%------------------------------
\begin{align}
(\bm{f}_N)_n \underset {N\to\infty}{\longrightarrow} (\bm{f})_n.
\end{align}
%------------------------------
The minimum value of $N$ required to reach convergence up to a specified error depends upon the ratio $\alpha \equiv \atil/\abar$. This is because $\alpha$ controls the strength of the coupling between Floquet channels which, in turn, controls the number of elements $(\bm{f})_n$ that are nonzero to the specified accuracy. For $\alpha=0.1$ and $N=5$, the relative error in the total zero-energy cross section $\sigma_{\mathrm{tot}}(0)$, Eq.~\eqref{eq:totalcross}, is less than 1 part in $10^6$ for values of $\Omega$ near $1/(M\abar)^2$. For $\alpha = 10$, we require $N=150$ to achieve the same level of accuracy. Since the $W$-matrix Eq.~\eqref{eq:Wmat} is tridiagonal, it is computationally practical to solve for $\bm{f}_N$ even for very large values of $N$. This computational efficiency was not present in our calculation of the scattering amplitudes using the square-well model in Chapter~\ref{chap:atomic_scattering}.

We now compare the results for zero-range scattering using Floquet theory with those we obtained using the square-well model in Chapter~\ref{chap:atomic_scattering}. We do this by solving Eq.~\eqref{eq:fsolution} for the Floquet-component scattering amplitudes. Using these amplitudes we re-generate the results from the square-well model displayed in Figs.~\ref{fig:sigmaChannel_square}, \ref{fig:sigmaTotal_square}, and \ref{fig:sigmaKCont_square}. The results are displayed in Figures \ref{fig:sigmaChannel}, \ref{fig:sigmaTotal}, and \ref{fig:sigmaKCont}. These figures demonstrate the excellent agreement between the predictions of the zero-range model and the square-well model in the zero-range limit. We also note that the zero-range model is numerically much better behaved than the square-well model, as evidenced by the absence of numerical artifacts in $\sigma_3$ (Figs.~\ref{fig:sigmaChannel} and \ref{fig:sigmaKCont}). In the square-well model, due to numerical constraints, we were only able to calculate the first three Floquet-component cross sections with reasonable accuracy. In contrast, using the zero-range model, we can accurately calculate at least ten Floquet-component cross sections. This advantage of the zero-range model is suggested by the inclusion of $\sigma_4$ in Figs.~\ref{fig:sigmaChannel} and \ref{fig:sigmaKCont}.

\section{Analytic results for the weakly driven system}
For $\alpha \lesssim 0.05$ and $k\abar \gtrsim 0.001$, the total cross section is approximated within a few percent by the truncated solution for $f_0^0(k)\approx(\bm{f}_N)_0$ with $N=2$. In this case, it is possible to obtain a useful analytic approximation of the scattering amplitude $f^0_0(k)$ at low momentum. The other amplitudes $f_n^0(k)$ for $n\neq 0$ are irrelevant at such small values of $\alpha$ and for the specified range of momenta. After solving for $f_0^0(k)$, we can perform the effective range expansion
%------------------------------
\begin{align}
[f_0^0(k)]^{-1}+ik = -1/a_0 + \frac{1}{2}{r_e}_0k^2 + \mathcal{O}(k^4).
\label{eq:universalERE}
\end{align}
%------------------------------
We then simply read off the Floquet effective range parameters $1/a_0$, ${r_e}_0$, \ldots. Even for $N=2$, the resulting expressions for $1/a_0$ and ${r_e}_0$ are quite complicated. However, for small $\alpha$ and for values of $\Omega$ near $1/(M\abar^2)$, the expressions reduce to the parametrizations in Eq.~\eqref{eq:a0r0_square} (see Appendix \ref{App:analytic_parametrization}):
%------------------------------
\begin{align}
\frac{\abar}{a_0} &= \frac{\Omega - \Omega_0}{\Omega - \Omega_0 - \delta}+i\gamma\abar ,
\nonumber \\
\frac{{r_e}_0}{\abar} & = \frac{1/(M\abar^{2})^2}{\left(\Omega - \Omega_0 - \delta\right)^2}\,\alpha^2.
\label{eq:a0r0_zero}
\end{align}
The zero-range results for the resonance parameters are
%------------------------------
\begin{align}
\Omega_0\cdot M\abar^2 &= 1 + \frac{\sqrt{2}}{2}\,\alpha^2,
\nonumber \\ 
\delta \cdot M\abar^2 &= \frac{1}{2}\,\alpha^2,
\nonumber \\
\gamma  \abar & = \frac{1}{8}\,\alpha^2.
\label{eq:universal_2}
\end{align}
%------------------------------
The imaginary part of ${r_{e}}_0$ equals zero up to order $\alpha^4$. The analytic results for the coefficients of $\alpha^2$ given in Eq.~\eqref{eq:universal_2} agree quantitatively with the numerical results in Eq.~\eqref{eq:universal}. 

We obtain an analytic approximation for the momentum-dependent cross section by inserting the results for $a_0$ and ${r_e}_0$ in Eqs.~\eqref{eq:a0r0_zero} and \eqref{eq:universal_2} into the expression for $\sigma_0(k)$ in Eq.~\eqref{eq:crosssec_ERE}. In the vicinity of the resonance, the resulting expression is a very good approximation to the numerically ``exact'' total cross section (calculated from $\bm{f}_N$ for $N=100$) for $\alpha \lesssim 0.05$ and $0.001 \lesssim k\abar\lesssim 1$. For larger values of $k$, higher terms in the effective range expansion in Eq.~\eqref{eq:universalERE} are no longer suppressed, and one must use the full momentum-dependent amplitude. For smaller values of $k$, higher Floquet components contribute, and the total cross-section is no longer well approximated by $f_0^0(k)$.

As can be seen in Fig.~\ref{fig:a0}, the effective range ${r_e}_0$ becomes much larger than the scattering length $a_0$ in the vicinity of the zero crossing in the real part of $a_0$. Using the analytic expressions in Eq.~\eqref{eq:a0r0_zero}, it is easy to show that the effective range diverges at the zero crossing, signaling the breakdown of the effective range expansion near the zero crossing. The product $a_0{r_s}_0$ also diverges at the zero crossing in the real part of $a_0$. Thus the low-energy $s$-wave cross section in Eq.~\eqref{eq:crosssec_ERE} is strongly suppressed near the zero crossing.
% %------------------------------
% \begin{align}
% \frac{a_0{r_e}_0}{\abar^2} & \underset{\Omega\to\Omega_0+\delta}{\longrightarrow}  \frac{2}{M\abar^2}\frac{1}{\Omega - \Omega_0 - \delta}.
% \end{align}
% %------------------------------
% This ratio becomes infinity large at the zero crossing. 

%%% Local Variables:
%%% mode: latex
%%% TeX-master: "../../HudsonPhDThesis"
%%% End:

%% file: Chapters/Conclusion/Conclusion.tex
% !TEX root = HudsonPhDThesis.tex
% !TEX encoding = UTF-8 Unicode
% !TEX spellcheck = en_US
\cleardoublepage

\chapter{Summary}
\label{Chap:Final}

The properties of atomic systems depend upon the strength of the interatomic interactions. The ability to experimentally adjust these interactions has led to unprecedented breakthroughs in few- and many-body physics. At low scattering energy, the interaction strength is parametrized by the $s$-wave scattering length $a$. In most current experiments, the scattering length is controlled by exploiting a magnetic Feshbach resonance (MFR), where an external constant magnetic field aligned along the spin-quantization axis of the atoms induces a resonant coupling between the scattering atoms. The scattering length diverges when the magnetic field $B$ is tuned near the resonant value $B_0$ (See Fig.~\ref{fig:Feshbach}). It follows from Eq.~\eqref{eq:crosssec_ERE_ti} that the atomic cross section becomes extremely large in the vicinity of the resonance.

In this thesis, we have explored a new mechanism for controlling the strength of interatomic interactions termed modulated magnetic Feshbach resonance (MMFR). Like MFR, this mechanism involves the application of an external magnetic field aligned along the spin-quantization axis of the atoms. Unlike MFR, the amplitude of the applied field is periodic in time. This periodic field can inject or remove energy into the two-atom system. The atomic cross section becomes extremely large when the frequency is tuned near the transition frequency between the scattering atoms and a bound state in the same channel. 

In Chapter \ref{Chap:Formalism} we developed an extension to Floquet theory to describe the scattering of particles with short-range, time-periodic interactions. The culmination of this formalism is captured by the Floquet Lippmann-Schwinger equation (FLSE) Eq.~\eqref{eq:FLSEf}. This equation encodes all information about the quantum-mechanical scattering of particles from a short-range, time-periodic potential. All scattering observables can be deduced from the Floquet scattering amplitudes obtained by solving the FLSE. Though we later focus on the case of a zero-range potential, the Floquet scattering developed in this thesis applies to any short-ranged potential.

After deriving the scattering theory for short-range, time-periodic potentials, we calculated the scattering properties of particles with zero-range, time-periodic potentials. We did this in two ways. In Chapter \ref{chap:atomic_scattering}, we explicitly solved the time-dependent Schr{\" o}dinger equation for atoms with a square-well interaction potential with oscillating depth. By carefully tuning the parameters of the square-well model, we mapped this problem onto the problem of particles scattering with a zero-range, time-periodic potential. We showed that the resulting total cross section can be made thousands of times larger than the cross section in the absence of the oscillating field (see Fig.~\ref{fig:sigmaTotal_square}). We also showed that in the weak-driving limit, low energy scattering is well described by the effective range expansion. We found simple analytic parametrizations of the frequency-dependent scattering length and effective range (see Eq.~\eqref{eq:a0r0_square}), and we numerically determined the parametrization coefficients (see Eq.~\eqref{eq:universal}). 

In Chapter \ref{Chap:Applications}, we used the Floquet scattering formalism developed in Chapter \ref{Chap:Formalism} to calculate the scattering properties of particles with an explicitly zero-range time-periodic potential. The resulting cross sections agree with the cross section calculated using the square-well model to within the numerical accuracy. We obtained analytic results for the coefficients in the parametrization of the frequency-dependent scattering length and effective range (see Eq.~\ref{eq:universal_2}). These analytic results demonstrate explicitly that those coefficients are universal numbers. The analytic results for the coefficients agree with the numerical results in Chapter \ref{chap:periodic::sec:square} to within the numerical accuracy.

This thesis introduced Floquet scattering theory for atoms with short-range time-periodic potentials. This theory, in turn, provided the theoretical basis for MMFR -- a new technique for manipulating interatomic interactions. This discussion is by no means complete. We limited our calculations to the case of a zero-range scattering potential. Moreover, we were completely silent on the application of these ideas to 1D and 2D systems, which are currently of great theoretical and experimental interest. In addition to these theoretical extensions of Floquet scattering theory, we have also not fully explored the scope of possible applications of MMFR. For example, MMFR can be used to selectively transfer atoms between energy levels that differ by quanta of the field. It may be possible to exploit this fact to remove energy from a gas of atoms, effectively cooling it. The oscillating field can be used to coherently control the state of a few-body system, leading to possible applications in coherent quantum control. It is my hope that these and other exciting possibilities will be explored in future literature. 

%%% Local Variables:
%%% mode: latex
%%% TeX-master: "../../HudsonPhDThesis"
%%% End:

%% file: Appendices/PartialWaveIntegralEqn.tex
% !TEX root = ../HudsonPhDThesis.tex

\cleardoublepage
\chapter{Integral equation for the partial-wave scattering amplitudes}
\label{App:partial_wave}

Here we provide the details in the derivation of Eq.~\eqref{eq:FLSEpartialwave_ti}. The same steps apply equally well to the derivation of Eq.~\eqref{eq:FLSEpartialwave}.

To begin, we insert the spherical harmonic expansions in Eqs.~\eqref{eq:sphericalHarmonics_ti} into the LSE for the scattering amplitudes in Eq.~\eqref{eq:LSEf}. This gives
%-------------------------------
\begin{align}
4\pi\sum\limits_{lm}&f^l(p,k)Y^{*}_{lm}(\Omega_{\bm{p}})Y_{lm}(\Omega_{\bm{k}})\nonumber\\
  &=
  -m\sum\limits_{lm}V^l(p,k)Y^*_{lm}(\Omega_{\bm{p}})Y_{lm}(\Omega_{\bm{k}}) \nonumber\\
  &+(4\pi)^2\sum\limits_{lm}\sum\limits_{l'm'} \int\limits \frac{d^3q}{(2\pi)^3}
  V^l(p,q)G_0(q,k^2/m)f^{l'}(q,k)\nonumber\\
&~~~~~~\times  Y^*_{lm}(\Omega_{\bm{p}})Y_{lm}(\Omega_{\bm{q}})Y^*_{l'm'}(\Omega_{\bm{q}})Y_{l'm'}(\Omega_{\bm{k}}).
\label{eq:applse1}
\end{align}
%-------------------------------
Using the identity in Eq.~\eqref{eq:Ycompleteness} to integrate over $\Omega_{\bm{q}}$ in the last term gives
%-------------------------------
\begin{align} 4\pi\sum\limits_{lm}&f^l(p,k)Y^{*}_{lm}(\Omega_{\bm{p}})Y_{lm}(\Omega_{\bm{k}})\nonumber\\
  &=
  -m\sum\limits_{lm}V^l(p,k)Y^*_{lm}(\Omega_{\bm{p}})Y_{lm}(\Omega_{\bm{k}}) \nonumber\\
  &+\frac{2}{\pi}\sum\limits_{lm} \int\limits_0^\infty dq q^2
  V^l(p,q)G_0(q,k^2/m)f^{l}(q,k)Y^*_{lm}(\Omega_{\bm{p}})Y_{lm}(\Omega_{\bm{k}}).
\label{eq:applse2}
\end{align}
%-------------------------------
Performing the sums over $m$ and dividing by an overall constant gives
%-------------------------------
\begin{align} \sum\limits_{l}&f^l(p,k)(2l+1)P_l(\hat p \cdot \hat k)\nonumber\\
  &=
  -\frac{m}{4\pi}\sum\limits_{l}V^l(p,k)(2l+1)P_l(\hat p \cdot \hat k) \nonumber\\
  &+\frac{1}{2\pi^2}\sum\limits_{l} \int\limits_0^\infty dq q^2
  V^l(p,q)G_0(q,k^2/m)f^{l}(q,k)(2l+1)P_l(\hat p \cdot \hat k).
\label{eq:applse3}
\end{align}
%-------------------------------
Finally, using Eq.~\eqref{eq:Pcompleteness} to project out a particular partial wave, we obtain Eq.~\eqref{eq:FLSEpartialwave_ti}. The derivation of Eq.~\eqref{eq:FLSEpartialwave} proceeds in a parallel fashion.

%%% Local Variables:
%%% mode: latex
%%% TeX-master: "../HudsonPhDThesis"
%%% End:

%% file: Appendices/SquareWellWF.tex
% !TEX root = ../HudsonPhDThesis.tex

\cleardoublepage
\chapter{Scattering wavefunction for a particle incident on a well with oscillating depth}
\label{App:square_well}

Here we provide the details in the derivation of Eq.~\eqref{eq:usol}. 

\section{Solving for the wavefunction}
Inserting Eq.~\eqref{eq:floquet_theorem} into the radial Schr{\"o}dinger equation, Eq.~\eqref{eq:single_channel_model}, we obtain
%------------------------------------
\begin{align}
i\frac{d}{dt}\Phi(r,t)+\epsilon_F\Phi(r,t)&=-\frac{1}{M}\frac{\partial^2}{\partial r^2}\Phi(r,t)
-\left[\Vbar+\Vtil\cos(\Omega t)\right]\theta(r_0-r)\Phi(r,t).
\label{eq:app_box_1}
\end{align}
%------------------------------------
Examining the form of Eq.~\eqref{eq:app_box_1} for $r<r_0$, we see that the dependencies upon $r$ and $t$ are separable. Inserting the ansatz $\Phi(r,t) = g(r)f(t)$, we obtain the equations 
%------------------------------------
\begin{align}
  \frac{i}{f(t)}\frac{d}{dt}f(t) + \Vtil \cos(\Omega t) &= C, \\
 - \frac{1}{M}\frac{1}{g(r)}\frac{\partial^2 }{\partial r^2}g(r) - \Vbar -\epsilon_F &= C,
\label{eq:app_box_2}
\end{align}
%------------------------------------
for some constant $C$.

We solve for $f(t)$ subject to the arbitrary boundary condition $f(0)=1$. This choice has no affect on the scattering observables. We find
%------------------------------------
\begin{align}
  f(t) = \exp\left[-iCt + i\Vtil\sin(\Omega t) / \Omega\right].
\label{eq:app_box_3}
\end{align}
%------------------------------------
The constant $C$ is constrained by the requirement that $f(t)$ be periodic with fundamental frequency $\Omega$ as implied by Floquet's theorem. It follows that $C$ can take the values $n\Omega$ for any integer $n$. There is therefore a ladder of solutions for $f(t)$ indexed by $n$:
%------------------------------------
\begin{align}
  f_n(t) = \exp\left[-in\Omega t + i\Vtil\sin(\Omega t) / \Omega\right].
\label{eq:app_box_4}
\end{align}
%------------------------------------

We solve for $g(r)$ subject to the physical constraint that $g(0) = 0$. If $g(0)\neq 0$, the radial wavefunction $R(r,t)$ would diverge at least as fast as $1/r$ as $r\to 0$. Since $g(r)$ depends on $C$, there is a ladder of solutions for $g(r)$ indexed by $n$. Solving Eq.~\eqref{eq:app_box_2} with $C=n\Omega$, we obtain
%------------------------------------
\begin{align}
  g_n(r) = 2ia_n \sin\left(r\sqrt{M(\Vbar + \epsilon_F + n\Omega)}  \right),
\label{eq:app_box_5}
\end{align}
%------------------------------------
where the $a_n$ coefficients will be determined by the boundary
conditions at $r=r_0$. The full solution inside the box is
%------------------------------------
\begin{align}
  \Phi(r,t) \overset{r<r_0}{=} 2i \sum\limits_{n=-\infty}^\infty &a_n \sin\left(r\sqrt{M(\Vbar +
  \epsilon_F + n\Omega)}  \right) \nonumber \\
&\times \exp\left[-i(\epsilon_F+n\Omega) t + i\Vtil\sin(\Omega t) / \Omega\right].
\label{eq:app_box_6}
\end{align}
%------------------------------------
Using the identity in Eq.~\eqref{eq:besselexp}, this can be rewritten
%------------------------------------
\begin{align}
  \Phi(r,t) \overset{r<r_0}{=} 2i \sum\limits_{n}\sum\limits_{m} &a_n J_m(\Vtil/\Omega) \sin\left(r\sqrt{M(\Vbar +
  \epsilon_F + n\Omega)}  \right) \nonumber \\
&\times \exp\left[-i(\epsilon_F+(n+m)\Omega) t\right]\nonumber \\
=2i\sum\limits_{n}\sum\limits_{m} &a_n J_{m-n}(\Vtil/\Omega) \sin\left(r\sqrt{M(\Vbar +
  \epsilon_F + n\Omega)}  \right) \nonumber \\
&\times \exp\left[-i(\epsilon_F+m\Omega) t\right].
\label{eq:app_box_6_3}
\end{align}
%------------------------------------
The last line follows by shifting $m\to m-n$ in the infinite sum over the $m$ index.

The wavefunction for $r\geq r_0$ is composed of an incoming mode with energy $k^2/M$ and outgoing modes with energies $k^2/M + n\Omega$ for integers $n$ such that $k^2/M +n\Omega > 0$. The outgoing wavefunction also contains negative-energy modes. These are exponentially suppressed for $r\geq r_0$, and they contribute zero outgoing flux. The ladder of outgoing modes is necessary to match the boundary conditions at $r=r_0$. To perform the matching at the boundary, we will also include a ladder of incoming modes with energy $k^2/M + n\Omega$. After the matching has been performed, we can safely ignore all but the $n=0$ incoming mode. The general solution outside the well is 
%------------------------------------
\begin{align}
  \Phi(r,t) \overset{r\geq r_0}{=} \sum\limits_{m}
  (A^\mathrm{out}_m e^{ik_m r}+A^\mathrm{in}_m e^{-ik_m
  r})\exp\left[{-i(k_m^2/M)t}\right]
\label{eq:app_box_7_2}
\end{align}
%------------------------------------
where $k_m\equiv \sqrt{k^2/M + m \Omega}$.

\section[Matching at $r=r_0$]{Matching at \texorpdfstring{$\bm{r=r_0}$}{$r=r_0$}}

The wavefunction must be continuous at $r=r_0$. This gives the equation
%------------------------------------
\begin{align}
\sum\limits_{n}\sum\limits_{m} &2ia_n J_{m-n}(\Vtil/\Omega) \sin\left(r_0\sqrt{M(\Vbar +
  \epsilon_F + n\Omega)}  \right)\exp\left[-i(\epsilon_F+m\Omega) t\right]
\nonumber \\
&=\sum\limits_{m}(A^\mathrm{out}_m e^{ik_m r_0}+A^\mathrm{in}_m e^{-ik_m
  r_0})\exp\left[{-i(k_m^2/M)t}\right].
\label{eq:app_box_6_2}
\end{align}
%------------------------------------
This equation must hold for all times $t$. This implies that $\epsilon_F=k^2/M + j\Omega$ for any integer $j$. The choice of $j$ is completely arbitrary. We choose $j=0$. Making this insertion and using the identity in \eqref{eq:diracdelta} to project out the $j\Omega$ mode, Eq.~\eqref{eq:app_box_6_2} becomes
%------------------------------------
\begin{align}
\sum\limits_{n} &2ia_n J_{j-n}(\Vtil/\Omega) \sin(r_0q_n)
=A^\mathrm{out}_j e^{ik_j r_0}+A^\mathrm{in}_j e^{-ik_j
  r_0},
\label{eq:app_box_7}
\end{align}
%------------------------------------
where $q_n\equiv \sqrt{k_n^2+M\Vbar}$. The derivative of the wavefunction must also be continuous at $r=r_0$. This leads to the equation (easily derived from Eq.~\eqref{eq:app_box_7})
%------------------------------------
\begin{align}
\sum\limits_{n} &2a_n J_{j-n}(\Vtil/\Omega) \cos(r_0q_n)\frac{q_n}{k_j} 
=-A^\mathrm{out}_j e^{ik_j r_0} + A^\mathrm{in}_j e^{-ik_j
  r_0}.
\label{eq:app_box_8}
\end{align}
%------------------------------------
Adding and subtracting Eqs.~\eqref{eq:app_box_7} and \eqref{eq:app_box_8}, we obtain the equations
%------------------------------------
\begin{align}
A^\mathrm{in}_j  &= \frac{1}{2}\sum\limits_{n}(W_{+})_{jn}  a_n ,
\label{eq:app_box_9_1}\\
A^\mathrm{out}_j  &= \frac{1}{2}\sum\limits_{n}(W_{-})_{jn}  a_n ,
\label{eq:app_box_9_2}
\end{align}
%------------------------------------
where $W_\pm$ is defined in Eq.~\eqref{eq:M1M2}. Using
Eq.~\eqref{eq:app_box_9_1} to eliminate $a_n$ in Eq.~\eqref{eq:app_box_9_2} in favor of $A^\mathrm{in}_n$, we obtain the results in Eqs.~\eqref{eq:S_matrix} and \eqref{eq:S_matrix_def}.

%%% Local Variables:
%%% mode: latex
%%% TeX-master: "../HudsonPhDThesis"
%%% End:

%% file: Appendices/MultiChannel.tex
% !TEX root = ../HudsonPhDThesis.tex

\cleardoublepage
\chapter{Partial-wave Floquet S-matrix}
\label{App:multi_channel}

Here we derive the relationship between the partial-wave $S$-matrix
elements and the partial-wave Floquet scattering elements given in
Eq.~\eqref{eq:ftoS_partial}. 

Starting with Eq.~\eqref{eq:Stof}, we insert complete sets of states
%------------------------------
\begin{align}
1 = \sum\limits_{lm}\sum\limits_{n}\int dE \ket{Elm;n}\bra{Elm;n}
\label{eq:app_multi_1}
\end{align}
%------------------------------
before and after the operator $S-1$. The left-hand side becomes
%------------------------------
\begin{align}
  \matel{\bm{p}';n'}{\bm{p};n}{(S-1)} &= \sum\limits_{l'm'j'}\sum\limits_{lmj}\int dE' \int dE \braket{\bm{p}';n'}{E'l'm';j'} \braket {Elm;j}{\bm{p};n}
\nonumber \\
&\times \delta_{l'l}\delta_{m'm} \delta(E' - E - (n'-n)\Omega)
\nonumber \\
&\times \left(S^l_{j'j}(E) - \delta_{j'j}\right).
\label{eq:app_multi_2}
\end{align}
%------------------------------
The innter product $\braket{\bm{p};n}{Elm;j}$ has an energy delta function as a factor (see Ref.~\cite{taylor_scattering_thy}, Chapter 6):
%------------------------------
\begin{align}
\braket{\bm{p};n}{Elm;j} = \frac{1}{\sqrt{Mp}}\delta(p^2/M+n\Omega - E)\delta_{nj}Y_{lm}(\Omega_{\bm{p}})
\label{eq:app_multi_3}
\end{align}
%------------------------------
We evaluate both energy integrals in Eq.~\eqref{eq:app_multi_2}, giving
%------------------------------
\begin{align}
  \matel{\bm{p}';n'}{\bm{p};n}{(S-1)} &= \frac{1}{M\sqrt{p'p}}\sum\limits_{lm}
 Y^*_{lm}(\Omega_{\bm{p}}) Y_{lm}(\Omega_{\bm{p'}}) 
\nonumber\\
&\times \delta(p'^2/M - p^2/M)
\nonumber\\
&\times\left[S^l_{n'n}(p^2/M+n\Omega) - \delta_{n'n}\right].
\label{eq:app_multi_4}
\end{align}
%------------------------------
We can now use the addition theorem in Eq.~\eqref{eq:addition_theorem} to perform the sum over $m$, giving
%------------------------------
\begin{align}
  \matel{\bm{p}';n'}{\bm{p};n}{(S-1)} &= \frac{1}{4\pi M\sqrt{p'p}}\sum\limits_{l}
 (2l+1)P_l(\hat{\bm{p'}}\cdot\hat{\bm{p}})
\nonumber\\
&\times \delta(p'^2/M - p^2/M)
\nonumber\\
&\times\left[S^l_{n'n}(p^2/M+n\Omega) - \delta_{n'n}\right].
\label{eq:app_multi_5}
\end{align}
%------------------------------
We now expand the right-hand side of Eq.~\eqref{eq:Stof} using Eq.~\eqref{eq:LegendreExapansion} giving
%------------------------------
\begin{align}
\frac{i}{2\pi M} &\delta\left(p'^2/M - p^2/M\right)
  f_{n'n}(p'_{n'}\hat{\bm{p}}',p_{n}\hat{\bm{p}}) 
\nonumber \\ 
&=\frac{i}{2\pi M} \sum\limits_l (2l+1) P_l(\hat{\bm{p}}'\cdot\hat{\bm{p}})
\delta\left(p'^2/M - p^2/M\right) f_{n'n}^l(p'_n,p_n).
\label{eq:app_multi_6}
\end{align}
%------------------------------
After using Eq.~\eqref{eq:Pcompleteness} to project out a particular partial-wave component, we can compare the right-hand sides of Eqs.~\eqref{eq:app_multi_5} and \eqref{eq:app_multi_6}, leading to the final result in Eq.~\eqref{eq:ftoS_partial}.

%%% Local Variables:
%%% mode: latex
%%% TeX-master: "../HudsonPhDThesis"
%%% End:

%% file: Appendices/AnalyticParametrization.tex
% !TEX root = ../HudsonPhDThesis.tex

\cleardoublepage
\chapter{Analytic parametrization of frequency-dependent effective range parameters}
\label{App:analytic_parametrization}

Here we derive the analytic parameterizations of the frequency-dependent effective range parameters $a_0$ and $r_0$ given in Eqs.~\eqref{eq:a0r0_zero}. We solve Eq.~\eqref{eq:fsolution} with $\bm{W}$, $\bm{f}$ and $\bm{g}$ replaced by $\bm{W}_N$, $\bm{f}_N$ and $\bm{g}_N$ with $N=2$ as discussed in Chapter \ref{Chap:Applications}. This results in a system of five coupled equations for the components of $\bm{f}_2$. We solve this system analytically and then take the limit $\Lambda\to\infty$, yielding a finite result for $\bm{f}_2$. We do not copy the solution here because it is quite complicated and not very informative. After solving for the components of $\bm{f}_2$, we use $(\bm{f}_2)_0\approx f_0^0(k)$ along with Eq.~\eqref{eq:universalERE} to calculate the frequency-dependent effective range parameters $a_0$ and $r_0$. Specifically,
%------------------------------
\begin{align}
  \frac{\abar}{a_0} &= -\abar(f_0^0(k) + ik)|_{k=0}, \nonumber \\
  \frac{r_0}{\abar} &= \frac{2}{\abar} \left.\frac{d}{d (k^2)}(f_0^0(k) + ik)\right|_{k=0}.
\end{align}
%------------------------------

The expressions for $\abar/a_0$ and $r_0/\abar$ found using the procedure described above are too complicated to be usefully presented here. Of importance is the fact that they are analytic functions of the dimensionless variables $m\abar^2\Omega$ and $\alpha^2$. We simultaneously expand about $\alpha^2=0$ and about the dimer pole at $m\abar^2\Omega = 1$. For the frequency dependent scattering length, performing the multivariate expansion to next-to-leading order gives
%------------------------------
\begin{align}
  \frac{\abar}{a_0} &= \frac{m\abar^2\Omega - \left(1+\frac{\sqrt{2}}{2}\alpha^2\right)}{m\abar^2\Omega -\left(1+\frac{\sqrt{2}}{2}\alpha^2\right) - \frac{1}{2}\alpha^2} + \frac{i}{8}\alpha^2,
  \label{eq:app:a0expansion}
\end{align}
%------------------------------
where the first (second) term corresponds to the (next-to-)leading order term in the multivariate expansion. For weak, near-resonant driving, the second term is numerically small compared to the first. However, we retain this term because it captures the important fact that the frequency-dependent scattering length is complex-valued. For $m\abar^2\Omega-1\ll \alpha^2$, the second term has the additional real-valued contribution $\alpha^2/8$. Since this term is numerically much smaller than the real contribution from the first term, we retain only the imaginary term. This imaginary term is insensitive to the ratio $(m\abar^2\Omega-1)/\alpha^2$ for a wide range of values of that ratio. Similarly, performing the multivariate expansion of the frequency-dependent effective range gives
%------------------------------
\begin{align}
  \frac{r_0}{\abar} &= \frac{\alpha^2}{\left[m\abar^2\Omega-\left(1+\frac{\sqrt{2}}{2}\alpha^2\right)-\frac{1}{2}\alpha^2 \right]^2}.
  \label{eq:app:r0expansion}
\end{align}
%------------------------------
We omit the second term in the multivariate expansion because it is numerically small compared to the first term.

Comparing Eqs.~\eqref{eq:app:a0expansion} and ~\eqref{eq:app:r0expansion} with Eqs.~\eqref{eq:a0r0_zero}, we can read off the resonance parameters in Eq.~\eqref{eq:universal_2}.

%%% Local Variables:
%%% mode: latex
%%% TeX-master: "../HudsonPhDThesis"
%%% End: